\documentclass[12pt]{article}
\usepackage{amsfonts}
\usepackage{amssymb}
\usepackage{pstricks-add} 
\usepackage{fancyhdr}
\usepackage{hhline} 

\def\Ne{{\mathbb{N}}}

\def\Qe{{\mathbb{Q}}}
\def\Re{{\mathbb{R}}}
\def\Ce{{\mathbb{C}}}

\def\eqalign#1{\matrix{#1}}
\def\build#1_#2^#3{\mathrel{
\mathop{\kern 0pt#1}\limits_{#2}^{#3}}}

\def\epsilon{\varepsilon}
\def\di{\displaystyle}
\def\ass{\longmapsto }
\def\bar#1{\overline{#1}}
\def\norm#1{\|#1\|}
\def\binomial#1#2{\pmatrix{#1\cr #2\cr}}
\def\diamant{\ifmmode \diamond\enspace
\else \par$\diamond\enspace$ \fi}

\def\it#1{\textit{#1}}
\def\bf#1{\textbf{#1}}

\title{Modelling of consciousness\\
 and interpretation of quantum mechanics
}
\author{\'Eric Merle\\
mathematics teacher in french preparatory classes,\\
Lyc\'ee Louis-le-Grand, 
Paris, France.}
\date{July 17, 2018}

\begin{document}

\maketitle 

\section*{Abstract}I start from the fundamental principles of non-relativistic quantum mechanics, without probability, and interpret them using the notion of coexistence: a quantum state can be read, not uniquely, as a coexistence of other quantum states, which are pairwise orthogonal. 
In this formalism, I prove that a conscious observer is necessarily a physical object that can memorize local events by setting one of its parts in an exactly specified constant quantum state
 (hypotheses H1, H2 and H3). 
Then I define the probability of a future event as the proportion of initial observers, all identical, who will actually experience that event. 
It then becomes possible to establish the usual results of quantum mechanics. 
Furthermore, I detail the link between probabilities and relative frequencies. Additionally, I study the biological feasibility of this modelling of observer's mind. \par
The second part of this paper completes the neuronal description of  the mind functions, based on current neuroscientific knowledge.  It provides a model that is compatible with the assumptions of the first part and consistent with our daily conscious experience. In particular, it develops 
 a model of self-consciousness based on an explicit use of the random component of neuron behaviour; according to the first part, that random is in fact the coexistence of a multiplicity of possibilities. So, when the mind measures the random part of certain neurons in the brain, he goes himself within each of these possibilities. The mind has a decision-making component that is active in this situation, appearing then as the cause of the choice of this possibility among all the others. This models the self-consciousness which then  ensures the unity of our conscious experience by equating this experience with ``what the ego is conscious about''.\par
The conclusion details the points that remain to be developed. 

\newpage 

\tableofcontents

\newpage

\section*{Introduction :}
\addcontentsline{toc}{section}{Introduction}

This introduction is informal. It presents without technicality the ideas that are more rigorously developed in this  article.\par
I make the assumption that our universe is exactly governed by non-relativistic quantum mechanics.  So we explore what this theory tells about our universe and ourselves.\par

Paragraphs I.1.1 and I.1.2 reveal the lack of an acceptable definition of probabilities, even though they are used by most interpretations of quantum mechanics. The notion of probability is however too complex to be a primitive notion, not defined from other concepts. Therefore, I start only from non-relativistic quantum mechanics deprived of any use of probabilities.\par

Paragraphs I.1.3 to I.1.7 develop an interpretation of this mechanics based on the primitive notions of existence, coexistence, presence and time\footnote{
Time is a notion whose general relativity highlights all the complexity. However, in order to avoid the use of quantum gravity, I am forced here to consider time as a primitive notion.}.  Quantum mechanics imposes strong constraints on its own interpretation. This helps to clarify and better understand these primitive concepts. \par

The experimental observation of interference phenomena leads us to accept that one object can live at the same time in several states. Its quantum state can then be read as a coexistence of this object in several states. This a priori baroque idea is in fact consistent with our daily experience. Indeed, we only make imprecise measurements on the objects which surround us. We can therefore conceive that each object is in fact in a state of coexistence, each state within the coexistence being compatible with our observations. For example, the water molecules contained in this bottle that I observe do not conform to a precise dynamic of vertiginous complexity where each water molecule vibrates and turns and collides with other molecules along inextricable trajectories. No, this bottle of water is in a state of indeterminate coexistence of all the possibilities of determined states that are compatible with my observations.\par
Within a coexistence, experimental results indicate that the different states do not have the same presence, except in a particular case of coexistence that I call uniform. It is only for this kind of coexistence that it makes sense to 
count the different states.\par

I prove that a non-uniform coexistence of states can be reduced to a uniform coexistence, in which each initial state is multiplied into several similar states. For each state present in the non-uniform coexistence, the proportion of states thus multiplied relative to all states follows a variant of the Born rule.\par\medskip
However,  to actually prove the Born rule, we must first define probabilities. They appear 
when an observer measures a state of coexistence and thus enters that coexistence.
He does not select one of the possibilities present in coexistence because he enters them all, but in each possibility his consciousness is modified differently by his observation, so that his consciousness, in each coexisting state, experiments only one of these possibilities.  This is Everett's original approach [T.D.1, pages 150-151].\par\medskip

What is new in this article is the precise modelling of consciousness and the fact that it allows a natural definition of probabilities.\par\medskip

Chapter I.2 deals with modelling the consciousness of an observer. It is the state of a physical object that I call the observer's mind.\par
 According to the interpretation of quantum mechanics without probability developed in I.1, every system of the universe exists intrinsically in a state given by its wave function and it collects information from its environment by entanglement during interactions. As Bergson explains, this kind of ``immediate consciousness'' is only unconsciousness if it retains nothing of its past, if it constantly forgets itself. The mind of a conscious observer is therefore necessarily a physical object of the universe in interaction with its environment that can  preserve for some time the memory of its past conscious experiences, that is to say a
part of his quantum state itself. It is therefore necessarily a physical object that can store a local event by placing one of its parts in an exactly defined and constant quantum state.\par

We can thus model the observer's mind by a collection of elementary components, which I call c-bits (``c'' for ``conscious''), whose quantum state during a waking period can only take two values, necessarily mutually orthogonal. Each c-bit is a degree of freedom of a system of a few particles. In I.4, I develop the possibility that a c-bit is the spin of a set of 4 spin $\frac12$ particles or atoms. Each c-bit is housed in the heart of a protein, called c-protein, itself located in a neuron, called a c-neuron. The c-protein protects the integrity of the c-bit and controls its evolution according to the activity of the c-neuron.\par
A local event to the individual translates into a certain pattern of neuronal activity in the brain. That involves certain c-neurons whose  c-bits  have their state modified accordingly, thus memorizing the local event. 
The set of conscious quantum states of the mind is a finite pairwise orthogonal set of states.\par\medskip

When a conscious observer measures the state of an object, chapter I.3 defines the probability that he will obtain a certain result: before the measurement, the object is in a state of coexistence of different possible results. During the measurement, the measuring device and then the observer himself are taken into this coexistence. It can be reduced to a uniform coexistence which is interpreted as follows: the observer's mind before the measurement is in a given state $mind$, which can be equated with a multitude of observers in the same state $mind$, then at the end of the measurement, among this multitude, some of these observers have experienced the $i$th result, which they memorize by setting their mind in the state $mind_i$. 
The probability of the $i$th result is then just
 the number of observers aware of that result  divided by the total number of initial observers.
All observers with the same conscious experience $mind_i$ can be grouped into a single observer, but according to a coexistence which is no longer uniform. 
It is a natural definition of probabilities. Its precise statement requires the previous modelling of the observer's consciousness.\par
The remainder of I.3 establishes the classical results of probabilistic quantum mechanics. It explains 
the Young's double-slit experiment and the phenomenon of decoherence. It also studies the relationship between probability and relative frequency for experiments that can be repeated many times.\par \bigskip

At a metaphysical level, according to each mind, there is the mind itself and the rest of the universe. According to each mind,  the whole universe is a coexistence within which each universe is characterized  by the state $mind$ of the mind  and where the rest of the universe is itself in a state of coexistence, that of all states that are compatible with  the state $mind$.   Such an interpretation of the universe in terms of coexistence is not unique. This explains why our own consciousness does not forbid the consciousness of others.  We perceive from others only the coexistence of their states of mind which are in coherence with our own current state of mind. Each mind is a particular and egocentric way of reading the universe.\par\medskip

The first part studies probabilities by using an initial modelling of the observer's consciousness. The second part reverses the roles. It seeks to develop the modelling of consciousness and in particular self-consciousness by using the definition of probabilities.\par\medskip

Chapter II.1 recalls the simplified operation of a neuron: it receives several electrical signals at its synapses and when their sum exceeds a certain threshold, it emits an action potential along its axon, which is connected to synapses of other neurons.\par
Some neurons, called sensory, interact with the external environment and translate this interaction into electrical signals. For example, the retina is covered with light-sensitive neurons.\par\medskip
Chapter II.2 sets out what is known about how
the information collected by sensory neurons flows through the brain.
It has for each sense a hierarchical system of neural maps that can recognize patterns of the environment.
The more one progresses in the hierarchy, the more complex the patterns are, like the presence of a smile on a face, and the more invariant they are, for example in relation to the size and orientation of the same smile.\par\medskip

Chapter II.3 focuses on neural architecture downstream of the upper sensory maps. 
A Hopfield network is a neuron network  that behaves like a content-addressable memory; it can learn certain activity patterns, so that when presented with any activity pattern, it converges to the nearest learned pattern. We assume that, for each sense, the upper maps are connected to Hopfield networks that learn their most common activity patterns. Thus, at the perception of a partially hidden glass of water, the upper visual maps are set in a specific activity configuration which is transmitted to a Hopfield network which then can reconstitute the whole glass if it is part of its learned configurations. I label these Hopfield networks as  imaginary maps because they augment reality. \par
BAMs (Bidirectional Associative Memory) consist of two layers of neurons. They can learn to associate an activity configuration of one layer with an activity configuration of the other layer.\par

We assume that the imaginary maps of the different senses are linked by BAMs. That allows for example to associate the configuration corresponding to seeing a glass with the hearing of the word ``glass''. \par
In short, sensations are shaped in high-level sensory maps. Thanks to the BAMs, they are augmented with the imagination of other sensations. \par
We assume that these sensory and imaginary maps are made of c-neurons that encode in their c-bits these real or imagined sensations into conscious perceptions; the consciousness of the mind \it{is} the state of all c-bits. This decorrelation between neuronal electrical activity and conscious perception helps to explain the phenomenon of delayed consciousness highlighted by Libet experiments. \par\medskip

However, the brain is not a passive organ merely analyzing sensory information.
Chapter II.4 focuses on the motor component of the brain.  The motor neurons in the spinal cord can control the contractions of our muscles to make movements. They are themselves controlled by a motor map housed in the brain.
We assume that it is connected by BAMs to the various imaginary maps. Then the imagination of a sensation can induce a suitable movement. Conversely, the activation of the motor map can induce the imagination of the associated movement \it{before} the sensation of the movement actually performed.\par\medskip

In paragraph II.1.3, I assume that some neurons, that I call q-neurons, have a random behavior; under certain conditions, they are set in a state of coexistence, where active states and rest states  coexist. This coexistence is entangled with the environment.\par
In Chapter II.6, we use these q-neurons to develop decision-making mechanisms.  A typical decision-making algorithm  is to imagine different possible decisions, to retain some optimal solutions and then  to choose randomly among these the actual decision.\par
These algorithms are triggered by the brain through the activation of neurons, which I label as decisional. They act on q-neurons that introduce a little randomness in the imagined movements, which are then possibly realized. \par
When a decisional neuron is activated, it leads to the creation of a coexistence of states in certain q-neurons, thus a coexistence of different possibilities of movement, real or only imagined. When these movements are perceived by sensory and imaginary maps, they are actually measured by the mind. The mind  itself then enters the different states of the coexistence. \par\medskip

In chapter II.7, I suppose that decisional neurons are c-neurons. The mind then has a decisional component. 
The conscious chronology of decision making leads the mind to perceive a decision making at the very moment when the imagination of the result of the decision appears in the mind, just before the sensations produced by its actual realization, in case of actual choice. 

For the mind, therefore, it is not a random behaviour, but a decision making,
and the perception of the decisional c-bits is related to the decision made. For the mind, the decisional c-bits encode the characteristics of the \it{cause} that made the decision, that is, of the ego.\par
The ego has a real existence for the mind, just as sensory perceptions do. However, the ego does not exist in itself, as the cause of a decision, because no decision is made. All possibilities coexist.
This makes the ego inaccessible to the brain's own analytical abilities.
This aspect of the human condition is not negative because it is partially in conformity with Buddhist wisdom.\par

As a cause, during the construction of the mind, the ego acquires a more and more central place, to the point that the content of the mind becomes ``what the ego is conscious about''. 
This is how self-consciousness ensures the unity of our conscious experience.\par
Moreover, the perception of an ego that is the cause of decisions imposes a sense of responsibility. This is probably a selective advantage. 

\vglue 5cm
\section*{Reading Advice}

Based on your knowledge and interests, the table on the next page shows how to read this article without losing time. 
I  define eight reader profiles:
\begin{itemize}
\item 
L1 = reader studying the different interpretations of quantum mechanics.   \item
L2 = Reader interested by a natural definition of probabilities within the framework of quantum mechanics, for which he seeks a coherent interpretation.  \item
L3 = Reader interested in a scientific definition of consciousness.  \item
L4 = Reader seeking to link consciousness and molecular biology.  \item
L5 = Reader seeking to connect consciousness and the notion of spin.  \item
L6 = Reader interested in a neuroscientific construction of consciousness.  \item
L7 = Reader interested in artificial intelligence and decision making algorithms.  \item
L8 = Philosopher.
\end{itemize}
\par\medskip

Each reader can just read the paragraphs checked in his column. When a cell contains the words  ``FP$k$'' or/and ``IP$k$'', the reader may just read  ``FP$k$'' or/and  ``IP$k$'' in the relevant paragraph. More complex indications are provided by numbers that refer to the bottom of the table. 
\newpage

\addcontentsline{toc}{section}{Reading advice}

\begin{minipage}{15cm}
{\setlength{\tabcolsep}{0.15 cm}
\begin{tabular}{|c||*{8}{c|}}
\hline
&L1&L2&L3&L4&L5&L6&L7&L8\\
\hhline{|=#*{8}{=|}}
I.1.1&$\times$&$\times$&$\times$&$\times$&$\times$&$\times$&$\times$&$\times$\\
\hline
I.1.2&$\times$&&&&&&&\\
\hline
I.1.3&$\times$&\scriptsize FP1$\!$\footnotemark[1], IP1&
\scriptsize FP1$\!$\footnotemark[1], IP1&\scriptsize FP1$\!$\footnotemark[1], IP1&\scriptsize FP1$\!$\footnotemark[1], IP1&
\scriptsize FP1$\!$\footnotemark[1], IP1&\scriptsize FP1$\!$\footnotemark[1], IP1&
$\times$\\
\hline
I.1.4&$\times$&\scriptsize IP2&\scriptsize IP2&\scriptsize IP2&\scriptsize IP2&\scriptsize IP2&\scriptsize IP2&\scriptsize IP2\\
\hline
I.1.5&\scriptsize FP2&$\times$&\scriptsize FP2&\scriptsize FP2&\scriptsize FP2&\scriptsize FP2&\scriptsize FP2&\scriptsize FP2\\
\hline
I.1.6&\scriptsize FP3, IP3&$\times$&\scriptsize FP3\footnotemark[2], IP3&\scriptsize FP3\footnotemark[2], IP3&\scriptsize FP3\footnotemark[2], IP3&\scriptsize FP3\footnotemark[2], IP3&\scriptsize FP3\footnotemark[2], IP3&\scriptsize FP3\footnotemark[2], IP3\\
\hline
I.7.1&$\times$&$\times$&$\times$&$\times$&$\times$&$\times$&$\times$&$\times$\\
\hline
I.7.2&\footnotemark[3]&$\times$&\footnotemark[3]&\footnotemark[3]&\footnotemark[3]&\footnotemark[3]&\footnotemark[3]&\footnotemark[3]\\
\hline
I.7.3&&$\times$&&&&&&\\
\hline
I.7.4&\footnotemark[4]&$\times$&\footnotemark[4]&\footnotemark[4]&\footnotemark[4]&\footnotemark[4]&\footnotemark[4]&\footnotemark[4]\\
\hline
I.7.5&\footnotemark[5]&$\times$&\footnotemark[5]&\footnotemark[5]&\footnotemark[5]&\footnotemark[5]&\footnotemark[5]&\footnotemark[5]\\
\hline
I.2.1&&&&&&&&$\times$\\
\hline
\scriptsize I.2.2 - I.2.4&$\times$&$\times$&$\times$&$\times$&$\times$&$\times$&$\times$&$\times$\\
\hline
I.2.5&&&&&&$\times$&&\\
\hline
I.2.6&$\times$&$\times$&$\times$&$\times$&$\times$&$\times$&$\times$&$\times$\\
\hline
I.2.7&&&&$\times$&$\times$&&&\\
\hline
I.3.1&$\times$&$\times$&$\times$&$\times$&$\times$&$\times$&$\times$&$\times$\\
\hline
\scriptsize I.3.2 - I.3.6&&$\times$&&&&&&\\
\hline
I.4&&&&&$\times$&&&\\
\hline
II.1.1&&&&$\times$&&$\times$&$\times$&$\times$\\
\hline
\scriptsize II.1.2 - II.3.5&&&&&&$\times$&$\times$&$\times$\\
\hline
II.3.6&&&&$\times$&&$\times$&&$\times$\\
\hline
\scriptsize II.4 - II.7.2&&&&&&$\times$&$\times$&$\times$\\
\hline
II.7.3&&&&&&&&$\times$\\
\hline
II.7.4&&&&&&$\times$&&$\times$\\
\hline
II.7.5&&&&&&&&$\times$\\
\hline
\end{tabular}}
\footnotetext{$1.$ 
You can limit yourself to finite-dimensional Hilbert spaces.}
\footnotetext{$2.$ 
If the notion of tensor product is not mastered, you can just admit \par
\quad - that $x\otimes y$ is linear with respect to $x$ and with respect to $y$, that is to say \par
\quad
$(\alpha x+\beta x')\otimes y=\alpha x\otimes y+\beta x'\otimes y$ and
$x\otimes (\alpha y+\beta y')=\alpha x\otimes y+\beta x\otimes y'$, \par
\quad
- that the Hermitian product of  $x\otimes y$ and $x'\otimes y'$ is $\langle x|x'\rangle.\langle y|y'\rangle$.}
\footnotetext{$3.$ 
Just read this paragraph from IP4 included.}
\footnotetext{$4.$ 
It is enough to read this paragraph until  ``a poorly specified existence''.}
\footnotetext{$5.$ 
Just read the beginning of this paragraph until ``called the presence of $obj_i$ in the\par
\quad\quad coexistence'' then just read 
IP5 and IP6.}
\end{minipage}
\newpage

\part{The Observer and Quantum Mechanics}

\section{Formalism and Interpretation\\ of Quantum Mechanics}

\subsection{The Problem of Probabilities}

\subsubsection{Quantum Mechanics and Probabilities}
Almost a century after its birth, quantum mechanics keeps a part of mystery. Its mathematical formalism is well established and the results it produces are in perfect agreement with experimental data. However the exact link between this formalism and the reality of the world in which we live remains poorly understood. We know that quantum mechanics is a very good model, but we do not know precisely what it models.
  It surely expresses something about the very nature of our universe, however that {something} is currently beyond our comprehension.
A coherent and understandable interpretation of quantum mechanics is lacking.\par\medskip

Let us take the example of the Stern-Gerlach experiment [J.D.C pg 57]; 
silver atoms are passed through an inhomogeneous magnetic field in the laboratory vertical direction. Each atom is then deflected up or down and both events occur with the probability $\frac12$. 
We know that this deflection depends only on the spin of the atoms, though if we impose them 
a same initial spin state in a horizontal direction, then they are still deflected up or down in an equiprobable manner [J.D.C pg 65].\par

More generally, quantum mechanics predictions always have a probabilistic nature, of the form  ``given that an object is in a certain initial state, if it is allowed to interact with a specific (measuring) device, then a particular result will be observed with a certain probability''.\par
In this context, a coherent and understandable interpretation of quantum mechanics must be accompanied by a coherent and understandable interpretation of the notion of  probability.

\subsubsection{The Frequentist Interpretation\label{fre}}
So, what do we mean by the phrase ``the silver atom is deflected up with a probability $\frac12$'', or more generally ``a certain object satisfies a given property with a probability $p$'', where $p\in[0.1]$?\par
The frequentist approach claims that in an experiment with a clearly defined protocol,
the  object \textbf{obj} satisfies the property  $P$ with a probability $p$ if and only if, 
when one repeats this experience $N$ times with $N$ large, then most often the property is satisfied $K$ times with $K$ close to $pN$. \par
This frequentist approach does not stand up to scrutiny [A.H], [D.W.2, pg 123-132]. 
Firstly, apart from the fact that some experiments are not reproducible, it is never possible to repeat an experiment in exactly the same manner. Certain parameters are necessarily modified and this is a first  imprecision source. In addition, the terms ``$N$ large'', ``most often'' and ``close to'' are explicitly imprecise. 
Indeed, for $N$ given, $K$ is most often close to, but different from $pN$, and this all the more frequently as $N$ is large. Thus, since $p$ is distinct from $\frac KN$, the latter fraction is not a definition of the probability $p$. To substitute these approximations with accuracies, we would have to repeat the experiment an infinite number of times, which is impossible here below.\par
 Lastly, such an interpretation provides no phenomenological explanation. 
When using a balanced die, before the roll, the probability of a ``3'' appearing is $\frac 16$, then after the roll, this probability becomes 0 or 1 depending on the result. What happened? 
 The frequentist approach does not answer this question.\par
In short, frequentism does not currently have a strictly admissible enunciation. Furthermore, even if one does exist, it would not have the interpretative strength expected.\par
Kolmogorov's axioms, which provide the mathematical definition of probability, do not better answer the previous question. Moreover as Hajek [A.H, pg 211] points out, the notion of mass also satisfies these axioms (by requiring the total mass to be equal to unity), therefore these axioms circumscribe the notion of probability 
without
characterizing it.

\subsubsection{Bayesian Interpretation}
If frequentist approaches are disqualified, to my knowledge,  there are only the
 subjective approaches left, which consist in deriving the notion of probability from that of information acquired by an observer. Saying that the object \textbf{obj} will verify the property $P$ with a probability  $p$ means that the observer making this prediction has more or less accurate information about \textbf{obj} and $P$.
When $p=1$ ou $p=0$, he is certain of the result, his information is complete. In other cases, his information only allows him to quantify his belief in a probabilistic form.  This evolves over time according to the knowledge newly obtained by the observer as determined by Bayes formulas (this is why this approach is often called Bayesian).  In particular, if the observer has statistical information about similar objects and properties, he can adjust $p$ according to relative frequencies.\par
To derive from this concept an observer independent notion of probability  [D.W.1, pg 132-144], one agrees that the objective probability of an event occurrence is the subjective probability for a rational observer who knows all about the dynamics and initial conditions of the object.\par
Such an approach therefore places the observer at the heart of the interpretation, an observer 
who can acquire information, deduce beliefs, an observer 
 capable of rationality.  \par\medskip
In short, to rigorously interpret probabilities,  
it is first necessary to model the observer and his consciousness. This last problem seems unreachable, firstly because it involves the higher brain functions  and thus the inextricable complexity of our about 100 billion neurons, and secondly because we do not know how to translate neuronal activities
in conscious properties, because we do not have the slightest draft of a model of our conscious perceptions.\par
It is a dead end in which all interpretations of quantum mechanics are stuck.  \par\medskip

\subsection{The Measurement Problem}

\subsubsection{Presentation of the Problem}
These interpretations are distinguished by their explanations of the measurement problem [A.N, pg 22-24], which I briefly recall by simplifying to the extreme and limiting myself to Stern-Gerlach experiment, with a silver atom 
whose spin is oriented in a horizontal direction; its state, also called its wave function, is denoted by $\uparrow\rangle_x$.\par
 According to the quantum mechanical  formalism, we have 
 $\uparrow\rangle_x=\di\frac1{\sqrt2}\Bigl(\uparrow\rangle_z+\downarrow\rangle_z\Bigr)$, where 
$\uparrow\rangle_z$ (resp.  $\downarrow\rangle_z$) means a spin oriented in the upward vertical direction
(resp. downward). Hence $\uparrow\rangle_x$ also means a superposition of a spin
$\uparrow\rangle_z$ and a spin $\downarrow\rangle_z$,   a superposition which each theory interprets differently.\par
  The system made up of the spin and the rest of the laboratory has as wave function\par
$\Big[\di\frac1{\sqrt2}\Bigl(\uparrow\rangle_z+\downarrow\rangle_z\Bigr), labo_{ini}\Big]$, where $labo_{ini}$ is the state of the  lab at the start of the experiment.
Then the atom passes through the magnetic field and a measuring device records its trajectory deflection; the  quantum mechanical formalism, without probability, asserts that the system is then in the state 
$\di\frac1{\sqrt2}\Bigl([\uparrow\rangle_z, labo_{up}]+[\downarrow\rangle_z,labo_{down}]\Bigr)$, where 
$labo_{up}$ (resp. $labo_{down}$) is the state of the   laboratory  when an upward deflection (resp. downward) is observed.
Nevertheless such a superposed state is never observed; in practice, the experienced state is either 
$[\uparrow\rangle_z, labo_{up}]$ or $[\downarrow\rangle_z, labo_{down}]$, equiprobably. 
We say that the superposition $\di\frac1{\sqrt2}\Bigl([\uparrow\rangle_z, labo_{up}]+[\downarrow\rangle_z,labo_{down}]\Bigr)$ collapsed into one of two wave functions $[\uparrow\rangle_z, labo_{up}]$ or  $[\downarrow\rangle_z, labo_{down}]$.\par\medskip

\subsubsection{Current Interpretations of Quantum Mechanics}
Several theories explain this collapse using the notion of probability, but without defining it:  

\begin{itemize}
	\item The Copenhagen interpretation assumes that the device measuring the trajectory deflection is macroscopic and as such is subject to classical mechanics and not quantum mechanics. 
	It cannot therefore be in a superposed state unlike microscopic objects. As in the end 
	our experience of the microscopic world systematically uses macroscopic objects, no interpretation of these superposed states is necessary.
	\item According to de Broglie-Bohm theory, the wave function does not model an object in our real universe, it is only a wave that guides that real object. The latter is never in superposition of states. However,  probabilities are used in this theory because we do not know the object initial position, we only have the probability distribution of the object presence in such or such place [C.C]. 
\item The GRW theory modifies the  quantum mechanical formalism by adding a term of spontaneous and probabilistic collapse of the wave function at the microscopic level [R.F]. 
\item According to Shan Gao [S.G.2], when the state of an object is a superposition of two states $state_1$ and $state_2$, it means that the real state of the object alternates between these two states, switching at each instant from the $state_1$ to the $state_2$ discontinuously and immediately. His model uses the probability that at time $t$ the object is in one of the two states.
\end{itemize}

These theories may provide a solution to the measurement problem, however they leave the problem of defining probabilities entirely unsolved. This is still the case with transactional interpretation 
[JG.C, paragraph 3.8] and the theory of consistent histories [RB.G].\par

The situation is perhaps even more serious. Indeed, as Andreas Albrecht and Daniel Philips [A.A.1] suggest, who notably take the example of coin toss, it is very plausible that the probabilistic nature of any phenomenon in the real world has its origin in the collapse of a wave function. Under these conditions, any theory that uses the notion of probability to explain the collapse phenomenon is marred by circularity.

\par\medskip

The only interpretations that define probability at the same time claim that after the Stern-Gerlach experiment, the wave function is indeed equal to \par
$\di\frac1{\sqrt2}\Bigl([\uparrow\rangle_z, labo_{up}]+[\downarrow\rangle_z,labo_{down}]\Bigr)$, whereas an observer in the lab only perceives one of the two terms: \par
\begin{itemize}
\item According to the Everett interpretation [D.W.2], when passing through the magnetic field,  each term of the sum $\uparrow\rangle_x=\di\frac1{\sqrt2}\Bigl(\uparrow\rangle_z+\downarrow\rangle_z\Bigr)$ acts differently on the laboratory and even on the universe as a whole, so that the universe itself is then in the superposed state $\di\frac1{\sqrt2}\Bigl([\uparrow\rangle_z, univ_{up}]+[\downarrow\rangle_z,univ_{down}]\Bigr)$. This state describes a coexistence of two different universes. Thus, measuring the spin of the atom in the vertical direction induces the emergence of two universes. As such measurements occur at every moment and in every place of the universe, that Everett interpretation describes a universe from which emerge at every moment a multitude of new universes from which other universes emerge, according to a poorly understood emergence process [D.W.2, page 58]. In this context,
 David Wallace proposes to define probabilities based on rational betting decisions, but this leads him into mathematical complications whose consistency is not certain [A.M]. 
The Everett interpretation, however, is gaining more and more physicists to the point of sometimes being considered as the new dogma.

\item
According to the existential interpretation [W.Z.2], each of the two terms of the sum
$\di\frac1{\sqrt2}\Bigl([\uparrow\rangle_z, labo_{up}]+[\downarrow\rangle_z,labo_{down}]\Bigr)$ contains an observer (at least) aware of the outcome of the experiment. It is the same initial observer who differs in each term by his consciousness, which contains the outcome of the experiment. The collapse of the wave function is thus an illusion perceived by the observer's consciousness. The notion of observer ignorance allows Zureck [W.Z.1]  [W.Z.2] to define probabilities and establish the Born rule. However, the use of a conscious observer who is not part of the physical model used lacks rigour  [W.Z.2, page 17]. 
\end{itemize}
\par\medskip

\subsubsection{Breaking the Dead End}
I propose here to get out of this dead end by actually including a model of consciousness in quantum mechanics.  Strongly inspired by the two previous interpretations, I will then define the notion of probability and establish the Born rule.\par
The price to pay
is the acceptance of physical assumptions about the structure of  human's consciousness. 
 I will put forward some arguments to justify them but in the future these assumptions should be checked through appropriate experiments.\par
In a second part, I will use these assumptions to build a theory of human consciousness and higher cognitive functions compatible with current neuroscience data.

\subsection{Ontology of the Wave Function}

I restate the non-relativistic quantum mechanical  formalism and at the same time provide a coherent interpretation.\par\medskip

\bf{First Formal Principle (FP1):} 
In quantum mechanics, modeling an object denoted by \bf{obj} consists of giving a complex vector space ${\cal H}_{obj}$. 
It is further assumed that ${\cal H}_{obj}$ is a separable Hilbert space; if it is finite-dimensional, it only means that it is given with a Hermitian  product denoted by  $\langle .\ |\ .\rangle $. If it is infinite-dimensional, then ${\cal H}_{obj}$ must be complete according to the norm induced by $\langle .\  |\ .\rangle $ and contain a  countable dense set.\par
The possible states of \bf{obj}, also called the wave functions of \bf{obj}, are all unit vectors of ${\cal H}_{obj}$, that is, vectors of norm 1. By default, such a state will be denoted by $obj$.\par\medskip

Physicists do not agree on the interpretation that should be given to the  wave function even though it constitutes the basis of the quantum mechanical  formalism; some consider it epistemic when others consider it ontological, as I will detail.
\par\medskip
In general, the wave function $obj$ of the object \textbf{obj} is not directly accessible to the observer. The latter, by making measurements on this object, acquires only partial information concerning $obj$, and most often each measurement modifies $obj$ by collapsing the wave function. Such elusiveness, among other arguments, 
led some physicists
to grant $obj$ only a status of subjective information and observer partial knowledge about the object.
This is the epistemic interpretation, which was defended since the birth of quantum mechanics by the Copenhagen school as well as by Einstein [A.E] and which still has many defenders, such as Fuchs [C.A.F, page 9] and Spekkens [R.W.S]. More references can be found in [M.S.L, page 72]
 and [M.F.P., page 1].\par \medskip
However, a majority of scientists believe that wave function directly models object as it exists in the real world, independently of any observer. Here are some arguments in favour of this ontological interpretation.

\begin{itemize}
\item The results of  the Young's double-slit experiment [J.D.C Ch 4], [R.R, pages 4-9] are very difficult to explain without conceiving the wave function of the particle as a very real entity that splits into two waves each passing through a slit and then interferes onto the screen. Other intuitive arguments can be found in [M.S.L, pages 78-81].
\item During the nineties, the concept of protective measures was developed. It is a way to access 
to the value of a parameter of the $obj$ vector by modifying $obj$ by as little as you want. Several protective measures thus make it possible to determine the wave function completely without disturbing it. It then becomes much less elusive. This is a strong indication in favour of its ontological character
[S.G.1], [L.V], but it remains controversial [J.C],[M.S.2] and debated [S.G.3].
\item In 2011, Pusey, Barett and Rudolph [M.F.P] proved a theorem stating that if the wave function is only informative about  real objects and if two real objects prepared independently of each other have independent properties, then quantum mechanics is contradicted. This new argument for an ontological wave function is detailed and improved in [M.S.L] but challenged in [O.C]. 
\end{itemize}
\par\medskip
If the  wave function is epistemic, then its subjectivity returns us to the previous dead end. So, 
unless we conclude this article here, we must admit the ontology of the wave function:\par

\bf{First Interpretative Principle (IP1):} 
The wave function $obj$ of an object \textbf{obj} mathematically models the state of the object as it exists in the real world. It reflects the existence of \textbf{obj} and characterizes its state: 
 two different real states have different wave functions. \par
Conversely, if $obj_1$ and $obj_2$ are two linearly independent unit vectors
of ${\cal H}_{obj}$, these are two wave functions associated with different real states of $\bf{obj}$. \par
When there is a phase $\theta\in\Re$, different from 0 modulo $2\pi$, such that $obj_1=e^{i\theta}obj_2$, these two wave functions are still associated with real states of $\bf{obj}$ a priori different, however, we will see at IP3 how and when we can equate these two real states.

\subsection{Universality of Quantum Mechanics}

According to the Copenhagen interpretation, quantum mechanics should be reserved for the microscopic world and macroscopic objects should conform to classical physics. 
Such a limitation would prevent the observation of purely quantum phenomena for objects whose size exceeds a certain microscopic length $L$. In particular,  interference patterns would not appear in the Young's experiment with a particle size greater than $L$.\par
However, the evolution of experimental techniques has made it possible to highlight such interference for larger and larger objects [M.S.1, pages 258-270 and 282]. In 2013, Sandra Eibenberger and her collaborators did it with 
 macromolecules of more than 800 atoms [S.E]. In 2010  
[R.W], [A.D.O] A.D.O'Connell and his team made it with a metal vibrating blade 0.06 millimeters long. \par
These experiments require $L$ to be greater than macroscopic values.\par
Moreover quantum mechanics explains very well [M.S.1] how interaction with the environment destroys interference. This phenomenon called decoherence is all the more difficult to avoid if the object handled is large. 
In this context, classical mechanics appears only as an approximated theory of quantum mechanics for macroscopic objects. If we refuse any approximation though, as we must on the ontological level, it is still quantum mechanics which applies, even on the macroscopic scale.\par\medskip

\bf{Second Interpretative Principle (IP2):} 
Quantum mechanics is universal. It applies to everything, not just microscopic objects.\par\medskip

It applies in particular to the universe as a whole, which is therefore itself characterized by its wave function, as Hugh Everett proposed as early as 1957 [E.H., page 9].

\subsection{Schr\"{o}dinger Equation\label{schro}}

\bf{Second Formal Principle (FP2):} Suppose the object \textbf{obj} exists in the real universe between the moments $t_0$ and $t$\footnote{This assumption is not always true. We will see examples of objects that exist at the moment $t_0$  whose existence is no longer ensured at later moments.}. Let $obj$ be its state at the moment $t_0$ and $U_{t_0\rightarrow t}(obj)$ its state  at the moment $t$.
Then $U_{t_0\rightarrow t}$ is a linear function \footnote{It implies that the state $obj$ of \bf{obj} at the moment $t_0$ is variable. The theory is not only about the universe as it exists, it must also apply to the universe as it could exist.}.\par\medskip

For any $obj\in{\cal H}_{obj}$ such that $\norm{obj}=1$, we have $\norm{U_{t_0\rightarrow t}(obj)}=1=\norm{obj}$, so $U_{t_0\rightarrow t}$ is a unitary operator of ${\cal H}_{obj}$:  its adjoint $U_{t_0\rightarrow t}^*$ is equal to its inverse $U_{t_0\rightarrow t}^{-1}$. \par\medskip
 
If one derives with respect to $t$ the relationship 
$obj(t)=U_{t_0\rightarrow t}(obj(t_0))$, 
one gets \par
 $i\hbar\di\frac{\partial [obj(t)]}{\partial t}=i\hbar \frac{\partial U_{t_0\rightarrow t}}{\partial t}
U_{t_0\rightarrow t}^{-1}(obj(t))$. This is the Schr\"{o}dinger equation 
$$i\hbar\di\frac{\partial[obj(t)]}{\partial t}=\widehat{H(t)}[obj(t)],$$
where the Hamiltonian $\di\widehat{H(t)}=i\hbar \frac{\partial U_{t_0\rightarrow t}}{\partial t}
U_{t_0\rightarrow t}^{-1}$ is a Hermitian operator. Indeed,\par
$\di \bigl[\widehat{H(t)}\bigr]^*=-i\hbar U_{t_0\rightarrow t} \frac{\partial U_{t_0\rightarrow t}^{-1}}{\partial t}
=\widehat{H(t)}$
because $\di\frac{\partial \bigl[U_{t_0\rightarrow t} U_{t_0\rightarrow t}^{-1}\bigr]}{\partial t}=0$.
\par\medskip

We say that $\bf{obj}$ is completely isolated when its dynamics does not depend on the rest of the universe, that is, when it evolves as if it were the entire universe. Such isolation in the strict sense is concretely illusory, nevertheless it separates the intrinsic evolution of the object from its interaction with the environment.\par
In this case, $\widehat{H}$ does not depend on time because physical laws are immutable
 and the Schr\"{o}dinger equation is solved as
$obj(t)=e^{-\frac i{\hbar}t\widehat H}obj(0)$.
 So $U_{t_0\rightarrow t}=e^{-\frac i{\hbar}(t-t_0)\widehat H}$.\par\medskip

\subsection{Tensor Product\label{tens}}

\bf{Third Formal Principle (FP3):} Two objects of the universe, $\bf{obj}^1$ and $\bf{obj}^2$, are said to be disjoint when it is possible, at least in theory, to act on one without modifying the other.  The Hilbert space of the union of these two objects is the tensor product ${\cal H}_{obj^1}\otimes{\cal H}_{obj^2}$. As a consequence, the union is denoted by $\bf{obj}^1\otimes \bf{obj}^2$. \par
More generally, if $(\bf{obj}^k)_{1\leq k\leq n}$ is a collection of $n$ disjoint objects of the universe, their union is denoted  $\bf{obj}^1\otimes\cdots\otimes \bf{obj}^n$. So 
${\cal H}_{obj^1\otimes\cdots\otimes obj^n}={\cal H}_{obj^1}\otimes\cdots\otimes{\cal H}_{obj^n}$.\par\medskip

\bf{Third Interpretative Principle (IP3):} 
If $\bf{obj}^1\otimes \bf{obj}^2$ is formally in the state \par
$obj^1\otimes obj^2$, the associated interpretation is that the object $\bf{obj}^1$ is in the state $obj^1$ and that at the same time the object $\bf{obj}^2$ is in the state $obj^2$.\par
As for every $\theta\in\Re$, $obj^1\otimes obj^2=[e^{i\theta}obj^1]\otimes[e^{-i\theta}obj^2]$, it is equivalent to saying that the object $\bf{obj}^1$ is in the state $e^{i\theta}obj^1$ and that at the same time the object $\bf{obj}^2$ is in the state $e^{-i\theta}obj^2$. Thus, in this situation, the phases of the states $\bf{obj}^1$ and $\bf{obj}^2$ are not uniquely defined.
\par\medskip

The state $obj^1\otimes obj^2$ is said to be separable. Any state of ${\cal H}_{obj_1}\otimes{\cal H}_{obj_2}$ is a linear combination of separable states, which is generally not itself separable; 
such a state is written $\alpha obj^1_a\otimes obj^2_a+\beta obj^1_b\otimes obj^2_b$. It is said to be entangled. 

\par\medskip
A qubit is an object such that $dim({\cal H}_{qubit})=2$. \par
Quantum computing performs calculations with sets of qubits.
Consider $n$ qubits $\bf{qb}_1,\ldots,\bf{qb}_n$, where $n\geq 2$, whose set is $\bf{E}=\di\bigotimes_{1\leq i\leq n}\bf{qb}_i$.\par
When they are isolated, let $U$ be the unitary operator of  ${\cal H}_E$ that describes the dynamics of \bf{E} for a  given duration $\tau$. Thus, for any state $E(t)$ of \bf{E}
at a moment $t$, we have $E(t+\tau)=U(E(t))$.\par
Suppose we also know how to act on \bf{E} to swap states of $\bf{qb}_1$ and $\bf{qb}_2$.
 Let us denote by $T$ this operation.
In these conditions, Deutsch, Barenco and Ekert [D.D] have proved that, with few exceptions, $U$ is a universal quantum gate, that is to say that any unitary operator of ${\cal H}_{E}$ can be approached with arbitrary precision by successive applications of $U$ and $T$ in a well chosen order. \par
We can then control the dynamics of \bf{E} so that the final state is, with arbitrary precision, the image of the initial state by any unitary operator of ${\cal H}_E$.\par
The proof [D.D pages 3-5 and 9] adapts to the more general case where ${\cal H}_E$ is finite-dimensional and where $E$ decomposes into a tensor product of several parts of which we know how to swap two of them without modifying their states.\par
This makes plausible the following statement, which we admit: \par
for any object $\bf{obj}$ that we encounter in this article, it is possible to control its dynamics so that the final state is, with arbitrary precision, the image of the initial state by any unitary operator of ${\cal H}_{obj}$.

\newpage

\subsection{Coexistence}

\subsubsection{Linearity and Coexistence\label{linearite}}
Let us fix an initial moment $t_0\in\Re$.\par
We choose from ${\cal H}_{obj}$ two possible states of \bf{obj}, denoted by $obj_1(t_0)$ and $obj_2(t_0)$. \par
We fix $\alpha,\beta\in \Ce\setminus\{0\}$ such that $\norm{\alpha obj_1(t_0)+\beta obj_2(t_0)}=1$;
according to FP1, we can assume that at the moment $t_0$, \bf{obj} is in the
 state  
$obj(t_0)=\alpha obj_1(t_0)+\beta obj_2(t_0)$.\par
Let $t$ be such that $t>t_0$. 
Let $obj_1(t)=U_{t_0\rightarrow t}(obj_1(t_0))$: it would be the state of \bf{obj} at the moment $t$ if we had chosen $obj_1(t_0)$ as the state of \bf{obj} at the moment $t_0$. \par
Similarly, let $obj_2(t)=U_{t_0\rightarrow t}(obj_2(t_0))$: it would be the state of \bf{obj} at the moment $t$ if $obj_2(t_0)$ had been selected as the state of \bf{obj} at the moment $t_0$.\par

Based on the linearity of $U_{t_0\rightarrow t}$, the state of \bf{obj} at the moment $t$ is\par
$obj(t)=U_{t_0\rightarrow t}(obj(t_0))=\alpha U_{t_0\rightarrow t}(obj_{1}(t_0))+\beta U_{t_0\rightarrow t}(obj_{2}(t_0))
=\alpha obj_1(t)+\beta obj_2(t)$. \par

Thus $obj(t)$ is obtained directly from $obj(t_0)$ by applying $U_{t_0\rightarrow t}$, while we can also obtain $obj(t)$ by reading
$obj(t_0)$ as the superposed state $\alpha obj_1(t_0)+\beta obj_2(t_0)$ built from the states $obj_1(t_0)$ and 
$obj_2(t_0)$; one evaluates the future of the latter, independent of each other, by applying $U_{t_0\rightarrow t}$  to them, and one obtains $obj_1(t)$
and $obj_2(t)$. Then the state $obj(t)$ is also the superposition $\alpha obj_1(t)+\beta obj_2(t)$.\par

The linearity principle FP2 is therefore compatible with dynamically interpreting\par
 $t\ass \alpha obj_{1}(t)+\beta obj_{2}(t)$ as the simultaneous coexistence of $t\ass obj_1(t)$ and\par
 $t\ass obj_2(t)$, so that the object \bf{obj} is at all times $t$ ``in the state $obj_1(t)$ and at the same time in the state $obj_2(t)$''. If it is true at the moment $t_0$, it  stays true indefinitely, with the same coefficients $\alpha$ and $\beta$.\par\medskip
More generally, if $(obj_i)_{1\leq i\leq k}$ is a family of $k$ possible states of $\bf{obj}$ and if 
$(\alpha_i)_{1\leq i\leq k}$ is a family of non-zero complex numbers such that 
$\norm{\di\sum_{i=1}^k\alpha_i obj_i}=1$, we can consider interpreting the sum 
$obj=\di\sum_{i=1}^k\alpha_i obj_i$ as a coexistence state of states $(obj_i)_{1\leq i\leq k}$.\par 

\subsubsection{Orthogonality and Coexistence\label{orthoco}}
The purpose of this paragraph is to show that such an interpretation is acceptable if and only if the family $(obj_i)_{1\leq i\leq k}$ is orthonormal.\par\medskip

A first argument consists in asserting that the reality of such coexistence is acquired if and only if it concerns states $obj_i$ whose existence can be detected by a well chosen device, denoted by \bf{dev}. It can verify the existence of the object \bf{obj} in the state $obj_i$ if and only if, starting from an initial state $dev^0$ independent of $i$, 
after an interaction phase with \bf{obj}, the state of \bf{dev} gets a value $dev_i$ such that $dev_i\not=dev_j$ when $i\not=j$. This last condition means that the device can distinguish $obj_i$ from $obj_j$, so it can verify that \bf{obj} is in the state $obj_i$. After the interaction, we require that the state of \bf{obj} is still equal to the initial state $obj_i$ : the device must be able to analyze the state of \bf{obj} without modifying its reality.  It thus testifies to a reality not only past but also  present. The device must be theoretically constructible but its actual existence is not necessary. 
\par\medskip
We thus are in the usual conditions of a ``repeatable measurement'' in quantum mechanics [P.M, page 27]. If we denote by\, $U$ the unitary operator of ${\cal H}_{dev\otimes obj}$ that performs this interaction, then for all 
$i\in\{1,\ldots,k\}$, 
 $U(obj_i\otimes dev^0)=obj_i\otimes dev_i$. \par\medskip

\bf{Theorem:} The previous situation is possible if and only if the family $(obj_i)_{1\leq i\leq k}$ is orthonormal. 
\par\medskip
\bf{\it{Proof:}} \label{14}
Let us assume firstly that the previous situation is possible. \par
Let $i,j$ be in $\{1,\ldots,k\}$ with $i\not=j$.\par
Then
$\langle obj_i\ |\ obj_j\rangle =\langle obj_i\ |\ obj_j\rangle .\langle dev^0\ |\ dev^0\rangle =\langle obj_i\otimes dev^0\ |\ obj_j\otimes dev^0\rangle $.\par
  As $U$ is unitary it preserves the Hermitian product. Thus,\par
$\langle obj_i\ |\ obj_j\rangle =\langle U(obj_i\otimes dev^0)\ |\ U(obj_j\otimes dev^0)\rangle 
=\langle obj_i\otimes dev_i\ |\ obj_j\otimes dev_j\rangle $,\par
so $\langle obj_i\ |\ obj_j\rangle (1-\langle dev_i\ |\ dev_j\rangle )=0$.\par
Suppose $\langle dev_i\ |\ dev_j\rangle =$1. Then $| \langle dev_i\ |\ dev_j\rangle |=\norm{dev_i}\norm{dev_j}$, so according to the Cauchy-Schwarz
inequality, there is $\lambda\in \Ce$ such that $dev_i=\lambda dev_j$.\par
 Then\footnote{
In this paper, it is agreed that the Hermitian product $\langle x | y\rangle $ is linear as a function of $y$ and semilinear as a function of $x$.}
$1=\langle dev_i\ |\ dev_j\rangle =\bar{\lambda}$, so $dev_i=dev_j$. This is false,\par
so $\langle dev_i\ |\ dev_j\rangle \not=1$ and $\langle obj_i\ |\ obj_j\rangle =0$.\par

To establish the converse, assume that $(obj_i)_{1\leq i\leq k}$ is orthonormal. \par
The families 
$(obj_i\otimes dev^0)_{1\leq i\leq k}$ and $(obj_i\otimes dev_i)_{1\leq i\leq k}$ being both orthonormal, there is a unitary operator $U$ which transforms the first into the second. This completes the proof
according to the end of paragraph  \ref{tens}.\par
We can provide a more constructive proof\label{constructible}; for any $i\in\{1,\ldots,k\}$, let $p_i$ be the orthogonal projection onto the line whose direction vector is $obj_i$ and let $\lambda_i$ be a real,
such that
$\lambda_i\not=\lambda_j$ when $i\not=j$. \par
Let
$\di \widehat{H_{obj}}=\di \sum_{i=1}^k \lambda_ip_i$. It is a Hermitian operator on
${\cal H}_{obj}$.\par
Let\label{13} $\widehat H=\widehat{H_{obj}} \otimes \widehat{H_{dev}}$ where $\widehat{H_{dev}}$ is a Hermitian operator on  ${\cal  H}_{dev} $.\par
  It is a usual form of interaction Hamiltonian [M.S.1, page 78]. We admit
we know how to build a device such that the interaction Hamiltonian  between $\bf{obj}$ and $\bf{dev}$ is given by $\widehat H$. 

Before interaction, we start from the state $obj_j\otimes dev^0$. 
Then, after an interaction of a duration T, we are in the state
\footnote{In this article, $i$ refers to both an integer and a complex number such as $i^2=-1$, as long as the context allows to remove this ambiguity.} $E=
e^{-\frac {iT}{\hbar}\widehat{H_{obj}}\otimes \widehat{H_{dev}}}
(obj_j\otimes dev^0)$.\par
So, $E=\di\sum_{n=0}^{+\infty}\frac1{n!}\bigl(-\frac {iT}{\hbar}\bigr)^n 
\widehat{H_{obj}}^n(obj_j)\otimes \widehat{H_{dev}}^n(dev^0)$.\par
The family $(obj_i)_{1\leq i\leq k}$ is orthonormal, so when $i\not=i'$, $p_ip_{i'}=0$.\par
Thus 
$\widehat{H_{obj}}^n(obj_j)=\di\sum_{i=1}^k\lambda_i^np_i(obj_j)=\lambda_j^n obj_j$.\par
This implies that 
$E=obj_j\otimes dev_j$ \par
where $dev_j=\di\sum_{n=0}^{+\infty}\frac1{n!}\bigl(-\frac {iT}{\hbar}\bigr)^n \lambda_j^n\widehat{H_{dev}}^n(dev^0)=e^{-\frac {iT}{\hbar}\lambda_j\widehat{H_{dev}}}(dev^0)$.
Q.E.D.

Under these conditions, when \bf{obj} is in the state $obj=\di\sum_{i=1}^k\alpha_i obj_i$,  according to the linearity principle, interaction with \bf{dev} can be summarized in the form \label{vonneumann}
$$obj\otimes dev^0\longrightarrow \di\sum_{i=1}^k\alpha_i obj_i\otimes dev_i.$$
Regardless of any interpretation of the wave function collapse and the notion of probability, everyone agrees that under these conditions, the experimenter observes all the states $obj_i\otimes dev_i$, with a relative frequency of about $|\alpha_i|^2$. We can say under these conditions that the device $\bf{dev}$ can indeed  detect the presence of the states $(obj_i)_{1\leq i\leq k}$ when $\bf{obj}$ is in the state $\di\sum_{i=1}^k\alpha_i obj_i$.
We have therefore shown that, if the reality of a coexistence is equivalent to its detectability, then the interpretation of 
$obj=\di\sum_{i=1}^k\alpha_i obj_i$ as a coexistence of the states $obj_i$ is valid if and only if the family $(obj_i)_{1\leq i\leq k}$ is orthonormal. \par\medskip

A second argument consists in accepting to interpret $obj=\di\sum_{i=1}^k\alpha_i obj_i$ as a coexistence of the states $obj_i$ only if they have a certain independence from each other, that is  if it is possible 
 to modify the phase of one without modifying that of the others. For example in the Mach-Zender interferometer [P.M, page 71], when a photon meets the first semireflecting mirror, its wave function has the form 
$pho=\di\frac1{\sqrt2}(pho_1+pho_2)$, where $pho_1$ (resp. $pho_2$) represents the photon state if it is (resp. is not)
reflected by the mirror. A translucent material that slows down the photon can be placed on the path corresponding to the reflection. 
The photon wave function becomes\par
 $pho=\di\frac1{\sqrt2}(e^{i\delta}pho_1+pho_2)$. The same could be done in the Young experiment by placing the translucent material in front of one of the two slits.\par
More generally, depending on whether the phase of $obj_1$ is changed or not, there are two dynamics defined by two unitary operators $V$ and $V'$ such that for every $i\geq $2, $V(obj_i)=V'(obj_i)$ and $V(obj_1)=e^{i\delta}V'(obj_1)$, with $\delta$ different from 0 modulo $2\pi$.\par
Let $U=V^{-1}V'$. Then by preserving the Hermitian product,
 for any $i\geq 2$, \par
$\langle   obj_i|obj_1\rangle  =\langle   U(obj_i)|U(obj_1)\rangle  =e^{-i\delta}\langle   obj_i|obj_1\rangle  $, so $obj_i$ and $obj_1$ are orthogonal. \par\medskip

\bf{Fourth Interpretative Principle (IP4):} 
For an object \bf{obj} in the universe, let $(obj_i)_{1\leq i\leq k}$ be a family of states of ${\cal H}_{obj}$ and 
$(\alpha_i)_{1\leq i\leq k}$ a family of non-zero complex numbers. 
Then the equality
$obj=\di\sum_{i=1}^k\alpha_i obj_i$ is interpreted as a coexistence of the states $obj_i$ if and only if
 $(obj_i)_{1\leq i\leq k}$ is orthonormal and if $\di\sum_{i=1}^k|\alpha_i|^2=1$.
\par\medskip

When ${\cal H}_{obj}$ has a finite dimension equal to $n$, 
by completing $(obj_i)_{1\leq i\leq k}$ in an orthonormal basis of ${\cal H}_{obj}$,
the previous relationship is the decomposition of $obj$  in the basis $(obj_i)_{1\leq i\leq n}$.\par
Thus IP4, which is already an idea that is difficult to accept, hides another equally eccentric idea, equally unacceptable if it was not dictated by quantum formalism and by the impossibility of finding another simple and coherent interpretation.
 Indeed,
if $(obj_i)_{1\leq i\leq n}$ and $(obj'_i)_{1\leq i\leq n}$ are two orthonormal basis of ${\cal H}_{obj}$, \par
writing  
$obj=\di\sum_{i=1}^n\alpha_i obj_i=\di\sum_{i=1}^n\alpha'_i obj'_i$ models a coexistence of certain states $obj_i$ and meanwhile of certain states $obj'_i$. We must admit that the object is in a state of a simultaneous coexistence of  the states $obj_i$, and  that this first reading of the state of \bf{obj} does not prevent other readings: it is also in the state of a simultaneous coexistence of the states $obj'_i$. These two readings are two consistent ways to read the state of existence of $\bf{obj}$.
\par
For example, the following relationships can be interpreted in this way with respect to 
 the spin of an electron: 
  $\uparrow\rangle_z=\di\frac1{\sqrt2}(\uparrow\rangle_x+\downarrow\rangle_x)
=\di\frac1{\sqrt2}(\uparrow\rangle_y+\downarrow\rangle_y$)  [J.H, page 99].

\subsubsection{Interference and Interaction}

Let $(obj_i)_{1\leq i\leq k}$ be an orthonormal family of ${\cal H}_{obj}$ and $(\alpha_i)_{1\leq i\leq k}$ a family of non-zero complex numbers
 such that $\di\sum_{i=1}^k|\alpha_i|^2=1$. Suppose $obj=\di\sum_{i=1}^k\alpha_i obj_i$.
We have seen that a device that can detect the states $obj_i$ provides only one outcome among them in each trial. It is therefore tempting to argue that when $obj=\di\sum_{i=1}^k\alpha_i obj_i$, the object is in one of the $k$ states $obj_i$, without anyone knowing exactly which one. However, the Young's double-slit experiment in particular disqualifies such an interpretation; if at each trial the particle passed only through one of the two slits, the set of the  particle positions on the screen after multiple trials would not reveal an interference pattern. 
I will explain this theory on page \pageref{inter} and we will see that more generally, we can observe the presence of interference between the $k$ terms  $obj_1, \dots obj_k$.

\par\medskip
Thus, the linearity on the formal side and the presence of interference on the experimental  side, lead us to accept the idea that a given object can coexist at the same time in several states, constituting an orthonormal family. The story
$t\ass obj(t)=\di\sum_{i=1}^k\alpha_i obj_i(t)$ of the object $\bf{obj}$ can be read as the coexistence of $k$ stories 
$t\ass obj_i(t)$.\par
The linearity principle ensures that the different states that coexist do not interact with each other. 
Indeed, if we use the notations of  paragraph \ref{linearite}, the evolution of $obj_1$, which corresponds to $t\ass U_{t_0\ass t}(obj_1(t_0))$ is absolutely not influenced by the evolution of $obj_2$, given by $t\ass U_{t_0\ass t}(obj_2(t_0))$.\par
Interferences are experimental evidence of the coexistence of several terms, so it is useless to make them disappear to interpret these terms as several realities which coexist without interacting between them. This is though what is done in the modern version of the Everett interpretation [D.W.2, page 84-85] using decoherence. Is it due to confusion between the notions of interference and interaction?
\par\medskip

The interaction between $obj_1$ and $obj_2$, though, corresponds to a very different situation, of which we already gave an example page \pageref{constructible},  during the constructive part of the proof; $obj_1$ and $obj_2$ are then states of two different objects \bf{obj} and \bf{obj'}, 
they interact when the Hamiltonian of $\bf{obj}\otimes\bf{obj'}$ is decomposed into the form \par
$\widehat{H}=\widehat{H_{obj}}\otimes Id_{{\cal H}_{obj'}}+Id_{{\cal H}_{obj}}\otimes\widehat{H_{obj'}}
+\widehat{H_{int}}$,
with [M.S page 78]  $\widehat{H_{int}}$ of the form \par
$\widehat{H_{int}}=\di\sum_{\alpha}E_{\alpha}\otimes E'_{\alpha}$, where $E_{\alpha}$ (resp :
$E'_{\alpha}$) is a linear operator of ${\cal H}_{obj}$ (resp. ${\cal H}_{obj'}$)  different from identity. 
$\widehat{H_{obj}}$ and $\widehat{H_{obj'}}$ describe the intrinsic dynamics of \bf{obj} and \bf{obj'}.
When $\widehat{H_{int}}\not=0$, if $\bf{obj}\otimes\bf{obj'}$ starts from the initial state 
$obj_0\otimes obj'_0$, the final state of $\bf{obj}\otimes\bf{obj'}$ is usually an entangled state 
$\di\sum_{i=1}^k\alpha_i obj_i\otimes obj'_i$,  where $obj_i$ may depend on $obj'_0$ and $obj'_i$ may depend on $obj_0$.  When that happens, we say that there is interaction between \bf{obj} and \bf{obj'}. 

So an interaction involves two different objects in the real world, whereas an interference involves only one object, which coexists in several states.
The coexistence $t\ass \di\sum_{i=1}^k\alpha_i obj_i(t)$ tells $k$ stories that run parallel with the same object in each story,  in different states though, and not a single story with several objects likely to interact between them.  The $k$ stories are independent of each other, what happens in one has no impact on what happens in the other. 

\subsubsection{Entangled States}

Consider two objects $\bf{obj}^1$ and $\bf{obj}^2$ being in the entangled state
$E=\di\sum_{i=1}^k\alpha_i obj^1_i\otimes obj_i^2$. We assume that the family $(obj^1_i\otimes obj_i^2)_{1\leq i\leq k}$ is orthonormal and always that the $\alpha_i$ are non-zero, with $\di\sum_{i=1}^k|\alpha_i|^2=1$.
Then $E$ is interpreted as the coexistence for $\bf{obj}^1\otimes \bf{obj}^2$ of $k$ states 
$state_i=obj^1_i\otimes obj_i^2$, where the  $state_i$ corresponds to the fact that $\bf{obj}^1$ is in the state $e^{i\theta_i}obj^1_i$ and that $\bf{obj}^2$ is in the state $e^{-i\theta_i}obj^2_i$, $\theta_i$ being any real number. 
 We are thus in the presence of a coexistence of situations in which the state of $\bf{obj}^1$ is conditioned by that of $\bf{obj}^2$, and vice versa.\par
In this situation, the object $\bf{obj}^1\otimes \bf{obj}^2$ exists in the superposed state $E$, whereas neither $\bf{obj}^1$, nor $\bf{obj}^2$ exists, because they cannot be associated with a single state.
One can generalize and interpret in the same way any relationship of the form \par
$E=\di\sum_{i=1}^k\alpha_i\bigotimes_{j=1}^p obj^j_i$. 
\par\medskip

\label{intr}
So when a set of parts exists in an entangled state, its parts do not exist! However the whole exists in a state of coexistence of several separable states, in each of which the part $\bf{obj}^j$ does exist in a perfectly defined state.
So we must relativize the non-existence of $\bf{obj}^j$ which only means that we cannot attribute a single state to it. Indeed, if an object exists in the universe, it must possess a state. The contrapositive implies non-existence. However, in the case of an entangled state, this non-existence corresponds rather to a poorly specified existence. 
\par\medskip
In the EPR experiment [P.M, pages 100-102], $\bf{obj}^1$ and $\bf{obj}^2$ are the spin vectors of two 
spin $\frac12$ particles. Let us prepare $\bf{obj}^1\otimes\bf{obj}^2$ in the entangled state \par
$E=\di\frac1{\sqrt2}\bigl(\uparrow\rangle_z\otimes\downarrow\rangle_z-
\downarrow\rangle_z\otimes\uparrow\rangle_z\bigr)$.
It is a singlet state, which is invariant by rotation, so the previous equality is valid for any direction $Oz$.\par
Then we move the two particles away from each other by a distance $d$.\par
The interaction of the first particle with a device that can measure its spin in the $Oz$ direction is given by
$$E\otimes dev^0\longrightarrow
\di\frac1{\sqrt2}\bigl(dev_{up}\otimes\uparrow\rangle_z\otimes\downarrow\rangle_z-
dev_{down}\otimes\downarrow\rangle_z\otimes\uparrow\rangle_z\bigr)=H.$$

Even if we do not have a valid explanation for it for the moment, we know that after this measurement, the experimenter located in the neighbourhood of the first particle observe that its spin in the $Oz$ direction is $\uparrow\rangle_z$
about one-half the time. Then in this case, he can affirm that the spin state of the second particle is $\downarrow\rangle_z$, meaning he has  instantly information about an object located at a distance $d$ from him, as large as one wants. 

Would this information have travelled faster than light, which is inconsistent with the special theory of relativity?\par
In fact $H$ represents a coexistence of two stories. If we include the observer in the device, we just described the first story, in which $\bf{obj}^1$ is in the state $\uparrow\rangle_z$ and $\bf{obj}^2$ in the state $\downarrow\rangle_z$.  There is no information transmission, because for each story, the information is already present in each of the two particles. This information was already present in the state $E$. Measuring the spin of the first particle includes the experimenter into each of the two stories (what remains to be explained). So we do not contradict the special theory of relativity, but we must accept the non-local nature of the entangled state $E$ of $\bf{obj}^1\otimes\bf{obj}^2$, in the sense that its description as a coexistence, whatever the direction $Oz$ chosen, involves two particles distant from $d$. The violation of Bell's inequalities  shows that it is impossible to interpret $E$ by attributing to 
$\bf{obj}^1$ (and to $\bf{obj}^2$) 
the status of an existing object, 
 whose outcomes of measurements would only be stochastic.
More generally, the state $E=\di\sum_{i=1}^k\alpha_i\bigotimes_{j=1}^p obj^j_i$ is a priori non-local.

\subsubsection{Uniform Coexistences}
The interpretation of $\di obj=\sum_{i=1}^k\alpha_i obj_i$ in terms of coexistence does not for the moment involve the coefficients $\alpha_i$. However, when $\alpha_1=0$ with $\alpha_i\not=0$ for every $i\geq 2$, it is only a coexistence of $obj_2,\ldots,obj_k$, and as soon as $\alpha_1$ is not zero, however small is its modulus, we are in the presence of $obj_1,\ldots,obj_k$. Moreover in this case, a device that can detect the states $obj_1,\ldots,obj_k$  provides $obj_i$ as a result with a relative frequency close to 
$|\alpha_i|^2$. This is the indication that in the coexistence $\di obj=\sum_{i=1}^k\alpha_i obj_i$, each state $obj_i$ exists in a way that depends on $\alpha_i$. 
This way of existing is called the presence of $obj_i$ in the coexistence.\par\medskip
The quantum mechanical formalism  largely imposes its own interpretation, however we must not forget that, as a last resort, any interpretation is based on primitive notions that reflect the reality of the universe in which we live and for which quantum mechanics does not provide a \it{definition}. This is the case with  the notions of \it{coexistence} and \it{presence}.\par
Nevertheless, I  justify below that when all the coefficients $\alpha_i$ are equal, each $obj_i$ has the same presence in the coexistence, which I  then call a uniform coexistence.  When the coexistence is non-uniform, I show below that the quantity $|\alpha_i|^2$ corresponds to the proportion of the number of states $obj_i$ present in the coexistence, once reduced to a uniform coexistence.

\par\medskip
Let us start by showing that the presence of $obj_i$ in the coexistence\par
$\di obj=\sum_{j=1}^k\alpha_j obj_j$ depends neither on the other coefficients $\alpha_j$, nor on the vector $obj_i$. \par
\diamant  We can write
 $obj=\alpha_i obj_i+\beta\di\sum_{j\not=i}\frac{\alpha_j}{\beta} obj_j$ whatever the complex  number
$\beta$ such that $|\alpha_i|^2+|\beta|^2=1$. As $obj'=\di\sum_{j\not=i}\frac{\alpha_j}{\beta} obj_j$ is a unit vector of ${\cal H}_{obj}$, it is a state of \bf{obj}. 
$obj=\alpha_i obj_i+\beta obj'$, so $obj$ is both a coexistence of $obj_1,\ldots,obj_k$ according to the coefficients $\alpha_1,\ldots,\alpha_k$ and a coexistence of $obj_i$ and $obj'$ according to the coefficients 
$\alpha_i$ and $\beta$. For these two readings of the state $obj$, the way the state $obj_i$ is present in this coexistence is the same. It therefore does not depend on
 $\alpha_j$ when $j\not=i$.\par

\diamant Let $obj'_i$ be a unit vector of ${\cal H}_{obj}$.
When $j\not=i$, we suppose that $obj'_i$ and $obj_j$ are orthogonal. 
There exists a unitary operator $U$ such that $U(obj_i)=obj'_i$ and for all $j\not=i$, $U(obj_j)=obj_j$.  We admitted at the end of paragraph \ref{tens} that we can control the dynamics of \bf{obj}
so that between two moments $t_0$ and $t_1$ well chosen, we have \footnote{To ensure an exact equality, we have sometimes to  let $t_1$ tend to $+\infty$.}
 $obj(t_1)=U(obj(t_0))$. \par
If $obj(t_0)=\alpha_i obj_i + \di\sum_{j\not=i} \alpha_j obj_j$, then 
$obj(t_1)=\alpha_i obj'_i+\di\sum_{j\not=i} \alpha_j obj_j$.\par
$obj(t_0)$ and $obj(t_1)$ both represent, at different times, the same coexistence of states subjected to a particular dynamic; for any $t\in[t_0,t_1]$, we can write \par
$obj(t)=\alpha_i obj_i(t) + \di\sum_{j\not=i} \alpha_j obj_j(t)$, where for all $k$, 
$obj_k(t)=U_{t_0\rightarrow t}(obj_k)$. The linearity principle ensures that each scenario $t\ass obj_i(t)$ coexists with the other scenarios without interaction. So the presence of $obj_i$ in  $obj=\alpha_i obj_i + \di\sum_{j\not=i} \alpha_j obj_j$ 
and that of $obj'_i$ in $\alpha_i obj'_i + \di\sum_{j\not=i} \alpha_j obj_j$ correspond to the presence of $obj_i(t)$ in the coexistence $obj(t)$, for $t=t_0$ and $t=t_1$.
This shows that the presence of $obj_i$ in the coexistence $\alpha_i obj_i + \di\sum_{j\not=i} \alpha_j obj_j$ is equal to the presence of $obj'_i$ in the coexistence $\alpha_i obj'_i + \di\sum_{j\not=i} \alpha_j obj_j$, so it does not depend on $obj_i$. 
\par\bigskip
So the presence of $obj_i$ in the coexistence $\di\sum_j\alpha_j obj_j$ depends only on $\alpha_i$.\par

In particular, when for all $i,j$, $\alpha_i=\alpha_j$, each $obj_i$ is present in the coexistence in the same way, each $obj_i$ has the same presence.\par\medskip

Let  $i$ and $j$ be in $\{1,\ldots,k\}$ with $i\not=j$.
There exists a unitary operator $U$ such as 
$U(obj_i)=obj_j$, $U(obj_j)=obj_i$ and for any $h$ other than $i$ and $j$, $U(obj_h)=obj_h$.\par
Again, we can control the dynamics of \bf{obj}
so that between two moments $t_0$ and $t_1$ well chosen, $obj(t_0)=obj$ and $obj(t_1)=U(obj)$.\par
 If 
$obj=\alpha\di\sum_{i=1}^k obj_i$, we have $U(obj)=obj$.
Thus the coexistence of $obj_1,\ldots,obj_k$, according to equal coefficients $\alpha_1,\ldots,\alpha_k$, is not modified if the states $obj_i$ and $obj_j$ are swapped. Such symmetry is an additional argument to justify that in this case the different states $obj_i$ have the same presence. 
\par\medskip

\bf{Fifth Interpretative Principle (IP5): }\par
Let $(obj_i)_{1\leq i\leq k}$ be an orthonormal family of $k$ possible states of the same object $\bf{obj}$.
Then the state $\di\frac{1}{\sqrt k}\sum_{i=1}^k obj_i$ is called 
 the uniform coexistence state of these $k$ states. 
It is a coexistence in which each  state $obj_i$ exists in the same way as the other. 
\par\medskip
\diamant 
Let us assume now
 that $obj=\di\sum_{i=1}^k\alpha_i obj_i$ and that, for every $i\in\{1,\ldots,k\}$, $|\alpha_i|^2$ is a non-zero rational number.
There are then $S, q_1,\ldots,q_k\in\Ne^*$ and $\theta_1,\ldots,\theta_k\in\Re$ such that for all $i$, 
$\alpha_i=\di\sqrt{\frac{q_i}S}e^{i\theta_i}$.
 So $obj=\di\frac1{\sqrt S}\sum_{i=1}^k\sqrt{q_i}e^{i\theta_i}obj_i$.\par
Let $\bf{env}$ be an object disjoint from $\bf{obj}$ such that the dimension of ${\cal H}_{env}$ is as large as necessary; that is the only reason  why $\bf{env}$ is called an environment.\par
We assume that the system made up of the object and the environment is in the state $obj\otimes env$, that is to say that the superposition $obj=\di\sum_{i=1}^k\alpha_i obj_i$ is not entangled with the environment, at the moment considered. It also means $\bf{obj}$ and $\bf{env}$ do not interact.\par
For every $i$, there is
 an orthonormal family $(env_{i,j})_{1\leq j\leq q_i}$ of vectors in ${\cal H}_{env}$  such that  $env=\di\frac{e^{-i\theta_i}}{\sqrt{q_i}}\sum_{j=1}^{q_i}env_{i,j}$\footnote{Indeed, if
$(e_{j})_{1\leq j\leq q_i}$ is an orthonormal family of ${\cal H}_{env}$, let
$e=\di\frac{e^{-i\theta_i}}{\sqrt{q_i}}\sum_{j=1}^{q_i}e_{j}$. There is a unitary operartor $U$ such that 
$env=U(e)$, so $env=\di\frac{e^{-i\theta_i}}{\sqrt{q_i}}\sum_{j=1}^{q_i}U(e_{j})$.}.\par
So\footnote{This argument is borrowed from Zurek [W.Z.1, page 7], [W.Z.3].}
 $obj\otimes env=\di\frac1{\sqrt S}\sum_{i=1}^k obj_i\otimes \sqrt{q_i}e^{i\theta_i}env
=\di\frac1{\sqrt S}\di\sum_{i=1}^k\sum_{j=1}^{q_i}obj_i\otimes env_{i,j}$. \par
Let us fix $(i_1,j_1)$ and $(i_2,j_2)$ \label{ortho}
 such that  $i_1,i_2\in\{1,\ldots,k\}$, $1\leq j_1\leq q_{i_1}$ and $1\leq j_2\leq q_{i_2}$. Suppose $(i_1,j_1)\not=(i_2,j_2)$. Then $i_1\not=i_2$ in which case $obj_{i_1}\bot obj_{i_2}$, or $i_1=i_2$ and $j_1\not=j_2$, in which case 
$env_{i_1,j_1}=env_{i_2,j_1}\bot env_{i_2,j_2}$. So the family $(obj_i\otimes env_{i,j})_{1\leq i\leq k\atop 1\leq j\leq q_i}$ is orthonormal.\par
According to IP5, $obj\otimes env$ is therefore a coexistence of the states $obj_i\otimes env_{i,j}$ where each of these states has the same presence, which gives meaning to counting them;  if we fix $i_0$, among the $S$ states $obj_i\otimes env_{i,j}$, $q_{i_0}$ are exactly such that \textbf{obj} is in the state $obj_{i_0}$.  Thus $obj\otimes env$ is a uniform coexistence in which the proportion of states where the object is in the state $obj_i$ is equal to $\di\frac{q_i}S=|\alpha_i|^2$. This is true regardless of the choice of $env$, so we can say that in the coexistence $\di\sum_{i=1}^k\alpha_i obj_i$, out of a total of $S$ states,  the state $obj_i$ appears $S|\alpha_i|^2$ times, using an environment to distinguish between them the different occurrences of $obj_i$. 
As we can replace $S$ and $q_i$ with $hS$ and $hq_i$, for any $h\in\Ne^*$, the total number of states inside the coexistence $obj$ when reduced to a uniform coexistence is not precisely defined. However the quotient of the number of occurrences of the state $obj_i$ by the total number of occurrences of $obj_j$ for $j\in\{1,\ldots,k\}$ remains constant, equal to $|\alpha_i|^2$. 
Thus, we can say that in the coexistence $\di\sum_{i=1}^k\alpha_i obj_i$, 
$|\alpha_i|^2$ is the proportion of states equal to $obj_i$ among all available states.

\par\medskip
\diamant
Let $i$ be in $\{1,\ldots,k\}$ and assume that $|\alpha_i|^2$ is a rational number still denoted by $\di\frac{q_i}S$, without assuming though that \label{19}
 $|\alpha_j|^2$ is rational when $j\not=i$. Let us rewrite 
$obj$ in the form $obj=\alpha_i obj_i+\beta obj'$,where $obj'=\di\sum_{j\not=i}\frac{\alpha_j}{\beta} obj_j$ is a unit vector. 
So $|\beta|^2=\di\frac{S-q_i}S$, and based on the previous point with $k=2$, 
$obj\otimes env$ is a uniform coexistence of $q_i$ states for which $\bf{obj}$ is in the state $obj_i$ and 
of $S-q_i$ states for which $\bf{obj}$ is in the state $obj'$ which is a coexistence of the other states $obj_j$ for $j\not=i$.
Thus, here again, the proportion of the states $obj_i$ in the coexistence $obj$ is equal to $\di\frac{q_i}S=|\alpha_i|^2$.
\par\medskip

\diamant
Let $i$ be again in $\{1,\ldots,k\}$ and now assume that $|\alpha_i|^2$ is an irrational number. Let us consider an additional object, denoted by \bf{int}, through which we are going to return to rational numbers. We only assume that the dimension of ${\cal H}_{int}$ is larger than 2 and that \bf{int} does not interact with $\bf{obj}$ or $\bf{env}$.\par
Let $\beta_1$ be a complex number such that $|\beta_1|<1$  and $|\beta_1 \alpha_i|^2$ is a non-zero rational number. Let
 $\beta_2$ be a second complex number such that $|\beta_1|^2+|\beta_2|^2=1$.  We can split the state of \bf{int} into the form 
$int=\beta_1 int_1+\beta_2 int_2$, where $(int_1,int_2)$ is an orthonormal family of ${\cal H}_{int}$.
So $obj\otimes int=\beta_1\alpha_i .obj_i\otimes int_1+\gamma. vect$, where 
$\gamma$ is a complex number such that  $|\alpha_i\beta_1|^2+|\gamma|^2=1$ and where 
$vect=\di\frac1{\gamma}(\beta_2\alpha_i obj_i\otimes int_2+\sum_{j\not=i}\alpha_j obj_j\otimes int)$ is a unit vector.

\par
There are $p,q$ in $\Ne\setminus\{0\}$ such that $|\beta_1\alpha_i|^2=\di\frac pq$, so $obj\otimes int\otimes env$ is a uniform coexistence containing $p$ states, out of a total of $q$ states, for which \bf{obj}
is in the state $obj_i$ and  $\bf{int}$ is in the state $int_1$. We can therefore claim that the proportion of states for which \bf{obj} is equal to $obj_i$ is greater than $\di\frac pq$, for all complex number $\beta_1$ such that 
$|\beta_1|<1$ and $|\beta_1\alpha_i|^2=\di\frac pq$, so this proportion, denoted by $p_i$, is greater than
$\di\sup_{\beta_1\in\Ce\hbox{\tiny\ such that }|\beta_1|<1\atop\hbox{\tiny and }|\beta_1\alpha_i|^2\in\Qe}|\beta_1\alpha_i|^2=|\alpha_i|^2$. \par
The expression $\di\sum_{i=1}^k p_i$ represents the proportion of the states $obj_1,\ldots,obj_k$ in the coexistence 
$obj=\di\sum_{i=1}^k \alpha_i obj_i$, so it equals 1. Thus
$\di\sum_{i=1}^k(p_i-|\alpha_i|^2)=0$ and for all $i$, $p_i-|\alpha_i|^2\geq 0$. Necessarily, for any $i\in\{1,\ldots,k\}$, 
$p_i=|\alpha_i|^2$. 

 \par\medskip

\bf{Sixth Interpretative Principle (IP6): }\par
For an object \bf{obj} in the universe, let $(obj_i)_{1\leq i\leq k}$ be an orthonormal family of states of ${\cal H}_{obj}$ and 
$(\alpha_i)_{1\leq i\leq k}$ a family of non-zero complex numbers.\par
Then the equality
$obj=\di\sum_{i=1}^k\alpha_i obj_i$ is interpreted as a uniform coexistence where each state $obj_i$ is present according to a certain number of occurrences, whose proportion is $|\alpha_i|^2$, based on the total number of occurrences of $obj_1,\ldots,obj_k$. 

\par\medskip
\bf{Some comments :}
\diamant 
This total number is necessarily infinite when $|\alpha_i|^2$ is an irrational number. However, in that case, we have a finite total number of occurrences if we accept to equate $|\alpha_i|^2$ with a rational number of the form $\beta_1|\alpha_i|^2$ where $\beta_1$ is close to, but less than 1, and if we simply estimate the proportion of occurrences of the state $obj_i\otimes int_1$ in the coexistence 
$obj\otimes (\beta_1 int_1+\beta_2 int_2)$.
\par\medskip

\diamant 
We are closer to the Born rule. All the same, we still have to explain how measuring with a suitable device gives only one state $obj_i$ with a probability equal to $|\alpha_i|^2$. \par
Above all, it remains to define this notion of probability.

\section{Consciousness and Quantum Mechanics}
\subsection{Consciousness and Multiplicities}
Under the previous conditions, equality $obj=\di\sum_{i=1}^k\alpha_i obj_i$ means a coexistence of several states and such expression for the same state $obj$ is not unique.
These two levels of multiplicity shock our rationality a priori. This is though consistent with our conscious experience.
 Our free will allows us to decide between two possibilities, in an arbitrary way, that is to say, we believe, independently of the physical laws that govern our universe. If we choose $A$ freely, we might as well have chosen $B$. Meanwhile, what does this arbitrary mean, what does the conditional ``we might have chosen'' mean? The solution, which we will develop in chapter \ref{soi} of the second part, is to consider that the universe, denoted by $\bf{univ}$, after a decision made by an individual is in a state of coexistence: 
$univ=\alpha univ_A+\beta univ_B$, where $univ_A$ (resp. $univ_B$) is a state of the universe in which the individual has chosen $A$
(resp. $B$). So we have the power to arbitrarily choose between $A$ and $B$ because in fact we choose both. \par
After several decisions, the universe can be written $univ=\di\sum_i\alpha_i univ_i$ where $univ_i$ is a universe in which 
the individual's choices have certain values and where more generally his consciousness has a certain well-defined content.\par
According to the second level of multiplicity, this expression of $univ$ is not unique. It is welcome because, for a second human being, we have $univ=\di\sum_i\alpha'_i univ'_i$ where $univ'_i$ is a universe in which the choices of the latter have certain values and where his consciousness has a certain content. Thus to each conscious individual corresponds a certain decomposition of the universe into the multiplicity of his states of consciousness. This helps to explain how several billion individuals each have a consciousness that corresponds to a unique experience. \par
In sum, from the multiplicity of the expressions $\di\sum_i\alpha_i univ_i$ or $\di\sum_i\alpha'_i  univ'_i$, we only access the one we are and, within this expression, from the multiplicity of the terms $univ_i$, we only access the term corresponding to our consciousness as we experience it. 

\subsection{Consciousness and Unconsciousness}
What is consciousness?  We experience it continuously and intimately, however, we know almost nothing about it. It is even hard to define it.\par
Among the usual definitions of consciousness, I exclude those that assimilate it to the higher functions of the brain; the faculties of analyzing its sensations, 
to identify objects, to think, to know, to decide, to act, to make moral judgments are certainly related to consciousness, however,  they do not constitute its essence as I understand it.\par
Here are some definitions gathered from the Internet which come close to the notion of consciousness that is relevant to me in this article: 
\begin{itemize}
 \item Man's perception of his own existence and the world around him.
\item Immediate knowledge, intuitive or  reflexive, that everyone has of his existence and  the outside world.
\item Knowledge, intuition or feeling that a subject possesses about himself, his states and his actions.
\item Organization of the psyche which, by providing knowledge of its states, its acts and their moral value, enables to feel to exist, to be present to oneself.
\item Intuition by which man acquires at any moment an immediate and direct knowledge, more or less complete and clear, of his existence, his states and his acts.
\end{itemize}

These different definitions disagree about the nature of consciousness. Is it a psychic representation, a perception, an immediate knowledge, an intuition, a feeling? \par
However, all these definitions agree on the content of consciousness, which is twofold; consciousness contains information about the universe as well as a strong sense of self existence and of the rest of the world. \par\medskip

Let us arbitrarily split the universe into two parts, which we call  also arbitrarily the mind and the environment, and suppose that the state of the universe at the moment $t$ can be written as 
$\di\sum_i\alpha_i mind_i\otimes env_i$, the terms being pairwise orthogonal, i.e. the universe is a coexistence of different situations, where the $i$th situation involves a mind in the state $mind_i$ with an environment in the state $env_i$. If we accept to gather certain terms, we can assume that if $i\not=j$, 
 the families $(min_i,mind_j)$ and $(env_i,env_j)$ are linearly independent. 
Then $mind_i$ is the only state of mind that is associated with $env_i$, so that $mind_i$
 contains information 
about $env_i$. Thus \bf{mind} in the state $mind_i$ has information about the state of the environment in which he lives.
Moreover when the universe is in the state 
$mind_i\otimes env_i$, the mind does exist in the state $mind_i$,  according to IP1 and IP3.\par

We thus find a situation that combines existence and information. It could be a first definition of consciousness. 
It describes a mind that 
 has an ``immediate consciousness'' of its environment. 
However, if this mind has no memory, this ``immediate consciousness'' is at once erased by a change of state, as Bergson explains [H.B]: 
``A consciousness that would preserve nothing of its past, that would constantly forget itself, should perish or be reborn at every moment: how else to define unconsciousness? When Leibniz said that matter is
\it{an instant mind}, 
did  he not declare it insensitive, willy-nilly? All consciousness is therefore memory - preservation and accumulation of the past in the present''.

\par\medskip
This ``immediate consciousness'' that any part of the universe actually has in interaction with its environment is therefore synonymous with unconsciousness. An object could not experience a consciousness similar to ours without being first a memory. \par
We must therefore  define first of all the notion of memory
 within the framework of quantum formalism.

\subsection{Memory}

To communicate with each other, ants use pheromones; they are odorant molecules
which ants release on their way to inform their congeners, for example of the presence of food.
Let $\bf{mem}$ be the set of pheromones released by an ant in a certain location. The wave function of this set of particles is very complex, it evolves with time and gets entangled with the environment. However, as long as enough of these pheromones remain close to the original location, \bf{mem} constitutes a memory that can be read by other ants.  Formally,
this reading is possible as long as $mem\in M$, where $M$ is the sub-vector space of ${\cal H}_{mem}$ containing the states of \bf{mem} that correspond to the fact that enough pheromones of \bf{mem} remain close to the original location. \par
In this first example, the object \bf{mem} only serves as memory once.\par
This is not the case with the abacus. The arrangement of the beads on a 13-wire abacus allows to store a natural number $n$ less than $N=10^{13}-1$.\par
 Thus, for every $n\in\{0,\ldots, N-1\}$, there corresponds a
sub-vector space $M_n$ of ${\cal H}_{abacus}$ which contains all the possible states of the abacus whose  beads are displayed according to an arrangement that encodes $n$.
For the abacus to retain the memory of  $n$, it is necessary and sufficient that its state remains in $M_n$. This state is however very variable because the many particles of the abacus vibrate and collide, because the abacus is in constant interaction with the atmosphere which surrounds it etc.\par\medskip

Thus, on the formal level, a memory is an object \bf{mem} whose Hilbert space ${\cal H}_{mem}$ has a finite number of sub-vector spaces $M_1,\ldots,M_n$ such that, if during a recording phase, by interaction between \bf{env} and \bf{mem}, $\bf{univ}=\bf{mem}\otimes\bf{env}$ is set in a state that belongs to the sub-vector space $M_i\otimes{\cal H}_{env}$, then the state of $\bf{univ}$ remains in that same sub-vector space for a while, ideally as long as $\bf{mem}$ is not modified by a new recording. In this case, We say that \bf{mem} has memorized the event $M_i$, also denoted by
``$mem\in M_i$''. \par
Of course, during a period when the event $mem\in M_i$ is memorized, the state of \bf{mem} varies in $M_i$ and it gets entangled with the state of the environment. Thus, if at the moment $t$, \bf{mem} keeps in memory the event $M_i$, the state of the universe is of the form $\di\sum_k\alpha_k mem_k\otimes env_k$, where for all $k$, 
$mem_k\in M_i$.
A priori, neither the terms of this sum, nor the sub-vector spaces $M_i$ are pairwise orthogonal.\par
The reading of the recorded event ``$mem\in M_i$'' corresponds to another type of interaction that transforms any state of $M_i\otimes {\cal H}_{env}$ into a state 
of $M_i\otimes G_i$: the recorded event is not erased, the environment takes into account the event 
$mem\in M_i$ by setting itself in $G_i$.
In order to allow the recorded event to be read at any time, the spaces $M_i$ must remain constant with time. This point will become important thereafter. 
\par\medskip

Many memories are made of disjoint  smaller memories, which I call registers: 
$$\bf{mem}=\di\bigotimes_{1\leq k\leq N}\bf{reg}^k.$$ 
For example, the previous abacus is the tensor product of its 13 wires, each fitted with its beads. The different types of  computer memories  also have such a structure. This is still the case with the
 DNA molecule that we will discuss later. \par
For each $k\in\{1,\ldots,N\}$,  the ``memory'' spaces of $\bf{reg}^k$ are denoted by
$(M_i^k)_{0\leq i\leq I_k-1}$.\par
Thus $mem$ can record events of the form 
$mem\in \di\bigotimes_{1\leq k\leq N}M_{i_k}^k$. \par
The advantage of such a structure is that one register can be modified without changing the content of the others.

\subsection{Human Consciousness}
In this chapter, I formulate and justify three founding hypotheses on the notion of conscious observer, in conformity with our condition as human beings.\par\medskip

\bf{Materialistic Hypothesis (H1) : } 
A conscious observer is an object of the universe.
The heart of his consciousness is a part of this object, which I call his mind and denote by \bf{mind}. 
The consciousness of the observer at the moment $t$ is equal to the wave function of his mind at the same instant. 
\par\medskip
We thus reject any dualism, which assumes that the mind is made up of an immaterial substance, unable to interact with matter because of its very immateriality. This absence of interaction between mind and matter requires any dualistic quantum theory to follow the principle of psychophysical parallelism [J.N, pages 418-420]. The existing many-minds interpretations are all dualistic, to my knowledge [HD.Z],[B.M],[L.F], apart from Everett's original interpretation [H.E.1, paragraph 5], [H.E.2, part IV]. The interpretation presented in this article draws much inspiration from it. It is a  materialistic many-minds interpretation. 
\par\medskip

Our conviction that we exist, this unique and deep feeling coiled at the heart of our consciousness, allows us to link H1 and the ontological principle IP1. Indeed, if H1 is true, then we directly experience the ontological character of our mind's wave function. If the laws of physics are identical everywhere and always, we deduce that the wave function of any object in the universe also possesses the ontological property IP1. Conversely, IP1$\Longrightarrow$H1, in the sense that IP1 gives any object in the universe a  status of ``being'',
noble enough to suit the mind of a human being.
\par\medskip

All experimental data from neurosciences indicate that the content of our consciousness depends closely on the activity of some of our neurons. This leads us to a second hypothesis that specifies the first.\par\medskip
\bf{Cerebral Hypothesis (H2):} $\bf{mind}$ is an object of the universe disseminated in some neurons of the observer's brain, which I call c-neurons \footnote{``c'' for ``conscious''.}.\par
\par\medskip

Despite the feeling of unity that is part of our conscious experience, which I will explain later by the construction of self-consciousness, a little introspection shows us  that the content of our consciousness can be split  into several parts; a conscious experience contains images and sounds in particular, each image is divided into its colors and shapes, it also contains feelings about our emotional state and many other components, which in turn are made up of sub-components. \par
According to IP3, our mind is therefore a  memory which is separable into a tensor product. Without loss of generality, we can assume that its registers are \it{bits}, that is, by reusing the notations of the previous paragraph, that for any $k\in\{1,\ldots,N\}$, $I_k=2$. To remind us that these bits are the units of our consciousness, I call them c-bits. So, $\bf{mind}=\di\bigotimes_{1\leq k\leq N}\bf{cb}_k$
and each $\bf{cb}_k$ can store two different events $cb_k\in M_0^k$ and $cb_k\in M_1^k$. \par\medskip

The idea of equating the observer to a memory to interpret quantum mechanics was used by Everett [H.E.1], [H.E.2] as well as by Zurek [W.Z.1], but without relying on a precise modelling of the notions of memories and observers.\par\medskip

When the c-bit $\bf{cb}_k$ stores one of the two events $M_i^k$, we saw that its state remains constantly in $M_i^k$, but it can vary over time in $M_i^k$ and become entangled with the environment.  Bergson's criticism then remains valid: \bf{mind} is still an ``immediate consciousness'' that forgets itself unceasingly, that perishes or is reborn at every moment. 
It's still an unconscious memory.

There is one noteworthy exception to this situation, however, when the $M_i^k$ are all one-dimensional. 
Then, if you choose in each $M_i^k$ a unit vector $f_i^k$, 
the ``memory'' states of $\bf{mind}$ has the form $\di\bigotimes_{1\leq k\leq N}f_{c(k)}^k$, where $c$ is a function from $\{1,\ldots,N\}$ to $\{0,1\}$, because any phase $e^{i\theta}$  can be moved  into the environment state.  This state remains constant between two recording phases. This implies a \it{discontinuity} of consciousness, which we justify in the next paragraph.\par
In these conditions, the mind has a stable consciousness which can be further split into several memorized attributes whose recordings are more or less old, thus ensuring the 
``preservation and accumulation of the past in the present''. \par
I insist on the fact that the constant character of $M_i^k$ and therefore of $f_i^k$ over time is a necessary condition to allow the reading of these states at any time. Then the conscious state of our mind can continually influence its environment. It is a form of action of the mind on the rest of the universe. 
\par\medskip

In summary, we have just seen that the human mind satisfies a necessary condition: 
$\bf{mind}=\di\bigotimes_{1\leq k\leq N}\bf{cb}_k$ and for all $k$, $\bf{cb}_k$ 
behaves like a memory, based on two different \it{one-dimensional} sub-vector spaces $M_0^k$ and $M_1^k$. \par\medskip
Under these conditions, we necessarily have, for every $k\in\{1,\ldots,N\}$,
$M_0^k\bot M_1^k$. Indeed, the reading of this memory must not change its state, 
otherwise it is not a memory and it is essential to allow repeated readings, 
so it conforms to the following dynamics: 
$f_i^k\otimes env\longrightarrow f_i^k\otimes env_i$, where $env_0\not=env_1$ because the environment can make the distinction between the two states memorized by the $k$-th c-bit. We can then adapt the proof on page \pageref{14} and show that 
$\langle f_0^k | f_1^k \rangle=0$. \par\medskip
Let $\cal C$ be the set of all  functions from $\{1,\ldots,N\}$ to $\{0,1\}$ and for all $c\in\cal C$, let 
$mind_c$ be the state $\di\bigotimes_{1\leq k\leq N}f_{c(k)}^k$. 
When he is awake, the state of \bf{mind}\ belongs to the orthonormal 
 family $(mind_c)_{c\in {\cal C}}$; it is the family of conscious states.  \par\medskip

The third chapter will show that by using only observers that satisfy this necessary condition, we can define the notion of probability and rebuild the usual probabilistic quantum mechanics.\par

I do not think it is possible to extend this theory of quantum measurement to any memory whose spaces $M_i^k$ would have a dimension larger than 2. Anyway, the
observation of  wave function collapse concerns only human beings; they are the only observers who  can testify to their perceptions. \par
Even if our notion of coexistence corresponds to Everett's original view, we depart from his approach by at least two points: probabilities will not be based on  relative frequencies and we replace the use of decoherence, added to Everett's theory by many authors [D.W.2, page 75], [W.Z.2], [F.B, pages 12-18], by the orthogonality of the different conscious states (cf page \pageref{50}). 
\par\medskip

During a recording phase,  the mind moves from a state 
$mind_c$ to another state $mind_d$, where $c,d\in{\cal C}$. Such a transition 
cannot be immediate. 
Indeed, Schr\"{o}dinger equation implicitly assumes that the state of an object is a differentiable, therefore continuous function of time. Thus, this transition is achieved by going through variable states
located in\footnote{$\hbox{Vect}(mind_c)_{c\in {\cal C}}$ is the sub-vector space generated by the family $(mind_c)_{c\in {\cal C}}$.}
 ${\cal H}_{mind}\setminus \hbox{Vect}(mind_c)_{c\in {\cal C}}$; they are
unconscious.  Thus a recording phase takes place between two phases of consciousness. This is a brief period of unconsciousness, 
at least for the c-bits modified during the recording. As we have already said, this approach implies a discontinuous behavior of consciousness which we justify in the next paragraph. 
\par\medskip

I summarize these arguments in the form of a third hypothesis about the modelling of an observer's consciousness. The whole article can be viewed as  a defence of this hypothesis. 
\par\medskip

\bf{Fundamental Hypothesis (H3)} : $\bf{mind}=\di\bigotimes_{k=1}^N \bf{cb}_k$,  
where for every $k\in\{1,\ldots,N\}$, $\bf{cb}_k$ is a memory with exactly two 
\it{one-dimensional}
memory spaces $M_0^k$ and $M_1^k$, which are necessarily orthogonal.
\par
Let $f_0 ^k$ and $f_1^k$ be unit vectors of $M_0^k$ and $M_1^k$ respectively. 
\par
\begin{itemize}
\item
During the waking period,  every c-bit $\bf{cb}_k$ 
 memorizes certain characteristics of the past and present activity of the c-neuron that shelters it;
for a duration denoted by $T$ of about 50 milliseconds, the state of $\bf{cb}_k$ is a constant vector, equal to $f_0^k$ or $f_1^k$. \par
Just before the next period of the same duration $T$, the state of $\bf{cb}_k$ is possibly updated. In this case, the permutation between the two states $f_0^k$ and $f_1^k$ occurs over a duration denoted by $\tau$ of about one millisecond, during which $cb_k(t)$
comes out of $\hbox{Vect}(f_0^k,f_1^k)$ and, for any $\delta t>0$, $cb_k(t)$ and $cb_k(t+\delta t)$ are linearly independent. 
	\item 
	During periods of unconsciousness, the state of \bf{mind} at the moment $t$ is a coexistence of  separable states of the form $\di\bigotimes_{k=1}^N cb_k(t)$  a priori entangled with the environment. 
The linearity principle FP2 allows us to take into account only one of these states.\par
 We can therefore assume that $mind(t)=\di\bigotimes_{k=1}^N cb_k(t)$.
We impose that for every $k$, $cb_k(t)$ takes its values
  in the orthogonal complement of $\hbox{Vect}(f_0^k,f_1^k)$ and that, for any $\delta t>$0,
$cb_k(t+\delta t)$ and $cb_k(t)$ are linearly independent. So no c-bit in \bf{mind} is in a conscious state.\par
The entanglement with the environment further complicates this dynamic.  
\end{itemize}
\par\medskip

This hypothesis deserves some remarks: \par\medskip

\diamant
The time indications  $T=50ms$ and $\tau=1ms$ are hypothetical, Their orders of magnitude only are relevant.
The value of $T$ may vary depending on the observer's emotional state. \par\medskip

\diamant
During the waking period, H3 imposes severe constraints on \bf{mind}.
We propose further a possibility of biological implementation.\par
In the second part of this article, we will see that  hypothesis H3 is a good foundation for explaining self-consciousness, as we experience it daily. 
 \par\medskip

\diamant The brain architecture  developped in the second part allows us to estimate 
that one in 100 neurons is a c-neuron. If we further assume that each c-neuron houses about twenty c-bits to code its activity, then $N$ is in the order of $2.10^{10}$. Thus a conscious state of mind contains $20$ gigabits of information, updated every 50 milliseconds.\par\medskip

\diamant
According to our definition of consciousness, for any part  $I$ of $\{1,\ldots,N\}$, 
$\di\bigotimes_{k\in I}\bf{cb}_k$ is also conscious. This seems to deny the sense of unity we have of our own consciousness, 
which Giulio Tononi calls the \it{integration axiom}  [G.T].  We will see on page \pageref{90}  that the unity of our consciousness is appropriately achieved through self-consciousness.

\subsection{Discontinuity of Consciousness}

H3 assumes a discontinuous behaviour of consciousness; the conscious mind would adopt a fixed state for a certain duration and then move to a second state for the same duration and then continue periodically. \par\medskip

Several scientific observations [S.P, page 1], [R.C.1], [D.P.1], now give favor to theories according to which consciousness evolves in this way.  Indeed, several optical illusions are difficult to explain without this hypothesis: the phi phenomenon (especially in color) [H.MH], the flash-lag effect [R.C.1], the continuous wagon wheel illusion [D.P.1] etc. \par
Moreover, some neurons are only sensitive to movement (direction and speed) for one of our senses [D.P.2, pages 278-279]. These neurons are located in specific areas of the brain. The rare testimonies of individuals accidentally deprived of the cervical area sensitive to movements in the visual domain plead for the discontinuous character of consciousness [J.Z]. They describe a visual conscious experience in which the perception of movement is replaced by the perception of successive static paintings, without anything between two paintings. The patient studied in [J.Z] lost visual perception of movement following a stroke, although she retained her other mental and visual faculties. When she pours tea, she sees only a static image similar to a tea glacier between the teapot and the cup. She reports her discomfort when she is waiting in a room where other people are moving because 
individuals seem to move immediately from one place to another, as if by magic.
\par\medskip
For certain conscious experiences, such discontinuity is natural: 
a sensation of fear is stable during the twentieth of a second  and can vary discontinuously, the same is true for the color of an object, for an imagined drawing, etc.\par
Yet why does the information seem to scroll continuously in our minds? How to explain the continuity and fluidity of the music, the movements of the body, the swinging of a shrub under a light wind? \par
A solution compatible with H3 is to accept that these sensations of continuity and fluidity are  coded under a static format within each period of consciousness, which specifies for example the average speed and the average acceleration of certain zones of the moving objects [H.MH].  \par
Television  has the same discontinuity, with 25 or 30   frames per second.
Nethertheless we have the illusion of a continuity between these images, probably built by the brain to increase the raw reality using these additional encodings. 

\par\medskip
H3 requires a synchronized update of the values of the different c-bits. Such synchronization is also a hypothesis frequently put forward to explain the binding problem [AK.S].  It seems necessary but not sufficient. According to H2, this assumes synchronization between the different c-neurons. 
It is suspected that this type of neuronal harmonization is essential for brain function [PJ.U].  One known way to achieve such a mechanism is to connect the c-neurons to a network of synchronized pacemaker neurons. The link between a c-neuron and a pacemaker neuron must be local in order to neglect the propagation time. Synchronization of the pacemaker neural network is ensured by electrical synapses [D.P.2, pg 95]. These synapses, minority in the brain, ensure a faster transmission of the signal than the chemical synapses and seem especially used by the brain for synchronization operations [MV.LB], [S.O].

\par\medskip
The frequency of renewal of conscious frames, in the range of 
 20Hz, should not be confused with the dominant global frequency of neuronal activations of the brain, which differs according to whether the individual concerned is awake, drugged, anaesthetised, or immersed in one of the different phases of sleep. According to the second part of this article, it is likely that there is no more than one c-neuron per 100 neurons. Thus, the frequency induced by H3 participates little in the global frequency, especially since the update of $\bf{mind}$ within a c-neuron does not imply that it delivers an action potential; pacemaker neurons can transmit their tempo to each c-neuron\label{26} using a dedicated synapse, which does not necessarily have a decisive influence on the  overall electrical activity of the c-neuron.\par
 According to [S.A.1] and [S.A.2], if we look at the brain at a sufficiently coarse spatial scale, neural signals are assimilable to waves propagating in a particular network, which corresponds to the overall organization of the brain. The latter therefore has its own resonance modes, each vibrating according to its own frequency. The  selection by the brain of the fundamental frequency as well as the main harmonics would essentially depend on the mean of the excitatory or inhibitory character of the synapses, which differs precisely according to whether the observer is awake, drugged or asleep.

\subsection{The Mind Experiments Coexistences\label{coex}}
Let us consider an individual whose mind is denoted by \bf{mind}. Let \bf{env} be the rest of the universe. At the moment $t$,  the state of the universe has the form \label{27}
$$univ(t)=\di\sum_{c\in C}\alpha_c mind_c\otimes env_c
+\sum_{i\in I}\beta_i mind_i\otimes env_i+univ'(t),$$
 where $mind_c$ with $c\in C$ stands for conscious states of \bf{mind} and where $mind_i$ with $i\in I$ stands for unconscious states, even states where the individual is dead, and where $univ'(t)$ represents a coexistence of universes in which 
$\bf{mind}$ does not exist because at least one of its particles does not exist.
The linearity principle allows us to isolate one of the terms, for example that of index $c_0\in C$, and to study its evolution independently of the others, that is, to pretend that the universe is exactly in the state\par
 $univ(t)=mind_{c_0}\otimes env_{c_0}$. \par\medskip
Suppose $env_{c_0}$ is a  separable state of the environment, so that each particle is given by a specific wave function: $env_{c_0}=\di\bigotimes_{p\in {\cal P}}e_p$, where $\cal P$ is the set
of the particles of the environment and where 
$e_p$ represents the state of the particle $p$ at the moment $t$. It is therefore temporarily assumed  that there is neither creation nor annihilation of particles. 
It is further assumed that the moment $t$ corresponds to the creation of the conscious state $mind_{c_0}$, which therefore last until the moment $t+T$. 
\par
Suppose our individual works in an office. It contains some $10^{27}$ molecules, that is to say one billion of billions of billions of molecules mostly equal to $N_2$ or $O_2$, which I call air molecules. According to [A.A.1], each air molecule, which moves at an average speed of about $500m.s^{-1}$, collides with about 50 million other air molecules during $T=50ms$.
Informally, according to Heisenberg's uncertainty principle, the wave function of each air molecule presents a little indeterminacy of its position and speed, which according to [A.A.1] leads to  uncertainty about the occurrence of  a collision with  another given air molecule. Thus, at the moment $t+T$, several scenarios of
 collisions exist for a single molecule, which get entangled with the scenarios of the other molecules, because they are interdependent. Let 
 $\cal Q$ be the set of  the air molecules in the office. 
Each molecule $q\in\cal Q$ interacts with millions of other molecules of $\cal Q$  during the duration $T$, setting the office atmosphere at the moment $t+T$ in an entangled state $\di\sum_{s\in\cal S}\beta_s[\bigotimes_{q\in\cal Q}e_{q,s}]$,
	where $\cal S$ is the set of all possible scenarios between $t$ and $t+T$. Moreover each term of the sum gets entangled with the rest of the universe. 
So, 
$univ(t+T)$ is in the form $univ(t+T)=\di\sum_{s\in\cal S}\beta_s[\bigotimes_{q\in\cal Q}e_{q,s}]\otimes env'_s\otimes mind_s$, where $env'_s$ is the state of the set of particles  ${\cal P}\setminus{\cal Q}$. 
Even though the cardinality of $\cal S$  is difficult to estimate,  it is likely to be larger than $2^{(10^{27})}$. 

\par\medskip
The many-worlds interpretation is often explained by stating that each term \par
$\di[\bigotimes_{q\in\cal Q}e_{q,s}]\otimes env'_s\otimes mind_s$ is a world in which the scenario $s$ takes place, experienced by the individual in the state $mind_s$. However, we must take into account not only the molecules of the office but also all the particles of the universe.
We then obtain  $univ(t+T)=\di\sum_{a\in\cal A}\gamma_a mind_a\otimes env_a$,
 where $\cal A$ is the set of all the scenarios of all the particles,  much larger than the set $\cal S$, and where, for every $a\in\cal A$, $env_a$ is a separable state of the environment of the form
$env_a=\di\bigotimes_{p\in\cal P}e_{a,p}$, for which the state of each particle is known. 

 Thus, at each moment, each world would split into a gigantic number of new worlds each containing an occurrence of our individual, which themselves would immediately split in the same way.
\par\medskip
In this article I propose a different description of our reality; 
at the moment \par
$t+T$, according to H3, our individual updates his state of mind, which is assumed conscious for the moment, to simplify. 
The new state $mind(t+T)$
is one of the $2^N$ states $mind_c$, where $c\in \cal C$, with $N$ in the order of $2. 10^{10}$. From $mind_{c_0}$\par
 to $mind(t+T)$, only some c-bits of the c-neurons whose activity has changed are modified, so $mind(t+T)$ belongs to the set of the  $mind_c$  where $c$ differs only slightly from $c_0$. Let $\{mind^1,\ldots, mind^K\}$ be this set. Then $K\ll 2^{2. 10^{10}}$. It is a very modest set compared to  $\cal A$, whose cardinality is much larger than
$2^{(10^{27})}$. Indeed,\par
 $\di\frac{2^{(10^{27})}}{2^{2. 10^{10}}}
=2^{(10^{27}-2. 10^{10})}$, but $10^{27}-2. 10^{10}\simeq 10^{27}$. 
Thus, we will find the same state $mind^i$ for a huge number of scenarios $a\in\cal A$.
In practical terms, this means that $mind(t+T)$ does not depend on the position of any particular air molecule in the office, as long as the mixture between oxygen and nitrogen is relatively homogeneous.\par
 Then, we can write
$univ(t+T)=\di\sum_{i=1}^K mind^i\otimes\Bigl(\sum_{a\in {\cal A}\hbox{\tiny\ such that}\atop 
mind_a=mind^i}\gamma_a env_a\Bigr)$.\par

Thus, from our individual's point of view, at the moment $t$ he is in the state of mind $mind_{c_0}$, then at the moment $t+T$ he will be in the state $mind^i$, which depends only on some information relating to the neuronal activity of his brain. This activity certainly depends on the evolution of the environment, however,  many environments induce the same state $mind^i$. \par
And from an external point of view to the individual,  the universe is in a coexistence of states, in each of which the individual experiences a certain conscious state, that corresponds to an environment whose state is
$\delta_i env^i=\di \sum_{a\in {\cal A}\hbox{\tiny\ such that}\atop 
mind_a=mind^i}\gamma_aenv_a$, where $\norm{env^i}=1$.
$env^i$ is the coexistence of the environments resulting from $env_{c_0}$ that induce in the observer the state of consciousness $mind^i$.
\par
If we also take into account the unconscious states of the individual, we have \footnote{without taking into account the states of the universe where one of the particles constituting $\bf{mind}$ disintegrated.}\par
$univ(t+T)=\di\sum_{i=1}^K \delta_i mind^i\otimes env^i+\di\sum_{j\in I}\delta_j mind^j\otimes env^j$,  where for all $j\in I$, $env^j$ is the state of the environment that leads the mind to a state $mind^j$ of unconsciousness or death.
\par\medskip
To study the evolution of $univ_i(t+T)=mind^i\otimes env^i
=\di \sum_{a\in {\cal A}\hbox{\tiny\ such that}\atop 
mind_a=mind^i}\frac{\gamma_a}{\delta_i} mind^i\otimes env_a$, we can again invoke the linearity principle and study the evolution of each $mind^i\otimes env_a$. As $env_a=\di\bigotimes_{p\in\cal P}e_{a,p}$,  we are back to the situation we were studying at the moment $t$.\par
It is however simpler to write that if at the moment $t+T$, we limit ourselves to the universe experienced by \bf{mind} in the state $mind^i$, we have $univ=mind^i\otimes env^i$. Between $t+T$ and $t+2T$, the environment evolves in an extremely complex way, which induces some changes in the state of $\bf{mind}$, but in a less complex way. Let $mind'_1,\ldots,mind'_M$ be the different states of mind accessible by \bf{mind} from the situation $mind^i\otimes env^i$ at the moment $t+T$.  Then at the moment $t+2T$, we have simply\par
 $univ(t+2T)=\di\sum_{j=1}^M \alpha'_j mind'_j\otimes env'_j$, where $env'_j$ denotes the coexistence of  separable states  of the environment that bring \bf{mind} to the state $mind'_j$ at the moment $t+2T$.  If we now accept to consider particle creation and annihilation, the state $env'_j$ is no longer a coexistence of separable states at the particle level, meanwhile it remains the global state of the environment that leads \bf{mind} to the state $mind'_j$.
\par\medskip
Thus this a priori baroque notion of coexistence of states is in fact an omnipresent banality in the world as we experience it daily. The observer that we are systematically experiences the coexistence of the multitude of microscopic states of the universe that are compatible with his own conscious experience.\par
The atmosphere of the office in which I work does not have the incredible complexity of billions of billions of molecules which vibrate, rotate and collide but rather the simplicity of a coexistence of such overdetermined states, which constitutes a sort of sub-determined average. 
\par\medskip

Most events that occur several miles from us usually have no impact on our neural activity, at least as encoded by each  c-bit within each c-neuron. If, in a laboratory of which I have no knowledge, in another country, physicists repeat the Stern-Gerlach experiment while placing a (
Schr\"{o}dinger's) cat  in a cage provided with a pernicious mechanism which kills the cat if the silver atom moves upwards and which leaves the cat alive if the atom moves downwards, then for my mind, the cat will be in a state of coexistence of these two stories, and this as long as the result of this experiment has no consequence on the state of my mind.
 \par\medskip
I conclude this paragraph with two remarks: \par \medskip
\diamant
Descriptions using the language of classical physics, notably those that follow in this article, are always narratives from an internal point of view to an observer, real or fictional; I say that an object $\bf{obj}$ is in a classical state $F$ for an observer $\bf{mind}$ when, from the point of view of $\bf{mind}$, the state of the universe is \par
 $univ=mind_F\otimes\di\sum_j\alpha_j obj_j\otimes env_j$, where for all $j$, $obj_j\in F$, $F$ being a sub-vector space of ${\cal H}_{obj}$.\par
This implies that the observer's brain can detect the local event 
$obj\in F$ and then memorize it in his mind. 
So, $mind_F$ is one of the conscious states $mind_c$ where $c\in\cal C$.

\par\medskip

\diamant
If we now consider two individuals\label{30}  whose minds are denoted by $\bf{mind}^1$ and $\bf{mind}^2$, limiting ourselves to conscious states to simplify, we can write at the moment $t$: \par
$univ=\di\sum_{i,j}\alpha_{i,j}mind^1_i\otimes mind^2_j\otimes env_{i,j}$. So the universe is a coexistence of states in each of which $\bf{mind}^1$ and $\bf{mind}^2$ have some conscious experience. $env_{i,j}$ is the state of the environment for which the minds $\bf{mind}^1$ and $\bf{mind}^2$ are in the conscious states $mind^1_i$ and $mind^2_j$. One can write 
$univ=\di\sum_i mind^1_i\otimes\Bigl(\sum_j \alpha_{i,j}mind^2_j\otimes env_{i,j}\Bigr)$. 
Thus, $univ$ is a coexistence of states in each of which $\bf{mind}^1$
 is consciously experiencing  the state $mind^1_i$ at the moment $t$, in which case its environment is 
$\di\sum_j \alpha_{i,j}mind^2_j\otimes env_{i,j}$. This means that, in general, others present themselves to us in the form of a coexistence of several states of mind, which makes them particularly unpredictable. This social relationship is symmetrical because $univ=\di\sum_j mind^2_j\otimes\Bigl(\sum_i \alpha_{i,j}mind^1_i\otimes env_{i,j}\Bigr)$.

\subsection{Biological Mind}
\subsubsection{Eukaryotic Cell}
Each c-neuron is an eukaryotic cell, like all cells in our body.
It is a solution of large molecules called proteins, often containing several million atoms, separated from the external environment by a membrane, itself made of proteins. 
Each cell, with a diameter of less than one tenth of a millimeter, has several subcells also separated by membranes. The nucleus is one of these subcells. It contains 23 pairs of chromosomes, in the case of human cells [D.R], [P.VG], which together form DNA. Each of these chromosomes is a macromolecule of about ten billion atoms, consisting of a very long helicoidal stack of subunits, made up of 4 possible bases: adenine which systematically matches thymine and cytosine which matches guanine. They are denoted by A, T, C, and G and each have a few dozen atoms. Each subunit is one of the base pairs AT, TA, GC or CG, so DNA is a set of 46 words written with this four-letter alphabet that constitutes the so-called \it{genetic information}. The cell
can read it to make and regulate its proteins.  In particular a gene is a sequence of consecutive letters directly involved in the design of a protein; after a complex initialization process involving other parts of the DNA, one protein separates the two strands of the DNA at the gene location, then another protein makes a copy of the gene, replacing though the letter T with uracil, denoted by U. This copy is arn that migrates into the cell, possibly outside the nucleus, to meet a particular protein assembly called ribosome, 
which can translate arn into protein according to a precise code. Each protein is a word whose letters are chosen among about twenty amino acids, which are molecules of a few tens of atoms.
It fulfils a precise function within the cell, for its nutrition, its transmembrane exchanges, the regulation of other proteins, its movements, its duplication etc.
\par\medskip

The DNA is therefore a memory which the cell can read to plan its operation. Writing in this memory is very scarce. These are mutations which natural selection then  eliminate or perpetuate.\par

Except for a few mutations, genetic information is identical in every cell of an animal throughout its life. 
We now know that DNA carries other information, qualified as epigenetic, which differs from cell to cell and changes over time. 
This additional information is only partially transmissible during cell replication and the gametogenesis process. It allows the cell to adapt to its environment. It explains how cell differentiation takes place within the same animal.  
This information is essentially built using two processes which often combine [K.I]:
\begin{itemize}
\item A protein, by binding to a segment of DNA, suppresses or enhances the transcription of a certain gene [CW.G]. 
\item A protein adds a methyl group on a cytosine occurrence in DNA or removes it [PF.C]. This modifies the affinity of other proteins for the part of the DNA containing this base which also suppresses or enhances
 the transcription of a certain gene.
\end{itemize}
These processes thus make it possible to code at molecular scale certain characteristics of the environment detected by the cell. For example, an increasing outside temperature may be detected by a transmembrane protein whose outer part is heat-sensitive, triggering a series of events within the cell which induce either the binding of a protein to a particular location in the DNA, or the methylation or demethylation of certain cytosines in the DNA, or both. These superficial modifications of the DNA are a molecular coding of the temperature rise. They also control the expression of certain genes which allow the cell to adapt to this change in the environment. \par
This epigenetic coding therefore also behaves like a memory, which manages the cell specific information according to the three operations of recording, storing and reading.
\par\medskip

It is plausible to assume that c-neurons use these two processes to code certain characteristics of their axonal and synaptic activities, in accordance with mechanisms that are responsible for long-term synaptic plasticity of neurons [D.P.3, pages 168 and 178]. However, even if this coding is already at the molecular level, here is why in my opinion it is not 
 compatible with assumptions H2 and H3.\par

The object $\bf{mind}$ must remain the same during each waking phase of the individual, under penalty of disturbances of the consciousness. Moreover, on the scale of the whole life, to ensure a continuity of the 
self-consciousness, it can only change very weakly during the unconscious periods of sleep. These constraints are compatible with the remarkable fact that neurons are the rare cells whose longevity is equal to that of the organism, 
but they impose the same longevity on c-bits of c-neurons. Additionally it must be possible to maintain each c-bit in the same quantum state for a duration $T$. \par

Let us firstly study the process of 
methylation/demethylation. Particles added to DNA by methylation are removed during demethylation so
  they cannot take part in the physical definition of a c-bit.
 The cytosine atoms involved have the expected longevity because they are part of the DNA, so they are rarely renewed by the maintenance mechanisms of the DNA. Depending on whether the cytosine is methylated or demethylated, the wave functions of the electrons and nuclei of its atoms are different, but in a way too complex to be controlled;  for example, the electron orbitals depend on the entire molecule [C.L, chapter 11], which is here one of the chromosomes in its entirety. There is therefore little hope of finding a part of this system whose quantum state would be compatible with H3. \par

About the process of binding a protein to DNA, the same problems are encountered; the lifetime of the protein should  be firstly of the same order as the longevity of the organism, which is rare
 [BH.T]. When the protein gets the agreed signal, it changes shape, possibly detaching from the DNA,
and these changes affect the expression of certain genes.
Secondly the protein is a very large molecule and the wave functions of its particles are no more controllable than in the previous process.  There might be a solution, in relation to the protein folding problem; 
if the protein has
 two stable  folding structures, each associated with the exact value of an angle between  two atomic bonds, this angle can then constitute a c-bit. One can use [L.L.] as an inspiration to dig this  track. 

\par\medskip
However, it seems more likely to me that nature has set up a specific process to translate this molecular coding of the c-neuron activity  into a quantum coding compatible with H3.\par

\subsubsection{C-Proteins and C-Bits}

To be absolutely sure of H3, 
 it would be enough to study the quantum state of the particles which make up neurons and to verify that certain neurons contain certain particles, of a longevity equal to that of human beings, whose quantum state depends on the axonal or synaptic activity of the neuron and remains constant over a duration of about fifty milliseconds.  However such cellular exploration does not seem within the reach of current technologies. We are reduced to imagining a process compatible with H3, as precisely as possible, to show its plausibility.
\par\medskip

Let us specify the problem to be solved; after a certain period of inactivity, a c-neuron emits
a sequence of action potentials along its axon. How to code this event in the mind?  Let us assume that it is encoded at the molecular level in the following way; the burst of action potentials is detected by proteins denoted by $p_1$ at the beginning of the axon. They send proteins denoted by $p_2$\label{32} into the neuron nucleus, which can find in the DNA a particular succession of base pairs, that is, a particular word written with the alphabet 
$\{$AT, TA, GC, CG$\}$. This word has a unique occurrence in the DNA that defines a specific location denoted by $\ell$, to which proteins $p_2$ converge. They attach a protein $p_3$ to $\ell$. However, when the c-neuron becomes inactive again, the non-renewal of protein $p_2$ allows $p_3$ to leave the location $\ell$.
So the presence or absence of $p_3$ in $\ell$ is the molecular coding of the axonal activity of the c-neuron.\par
The study of ion channels [D.P.2, page 69] shows that certain proteins
can play with some particles, guide each of them by means of specific channels on particular sites at the heart of the protein, and modify the relative positions of these sites under the activation of such or such signal received by the protein. So this makes it possible to imagine interesting mechanisms for making quantum states [M.F, page 3]. In addition, electron spin selection has recently been discovered when electrons pass through certain helicoidal parts of certain proteins [I.C], [M.K], allowing local control of the electromagnetic field applied to the protein core. Such proteins
can thus implement ion trapping techniques sometimes envisaged for the realization of quantum computers [N.M, pages 309-319]. \par

So next to $\ell$, we will assume that such a protein is attached to the DNA, 
a protein whose core houses an object made up of a few particles, a protein
which can control the quantum state of that object. Such a protein will be called a c-protein and the quantum object is then a c-bit\label{33}, if they  really have the following behaviour:
If the c-protein detects that the protein $p_3$ is missing from $\ell$, 
it sets its c-bit in a state $f_0$ during the next recording phase whose duration is $\tau$. 
The state $f_0$ is preserved as long as $p_3$ is absent. But as soon as $p_3$ appears in $\ell$, 
the c-protein imposes on its c-bit a state $f_1$ at the next recording phase, $f_0$ and $f_1$ being orthogonal.  Besides, 
during a deep sleep phase, 
the increase in inhibitory neurotransmitter concentrations in the brain is detected by 
 the c-neuron at its synapses using proteins denoted by $p_4$, that massively send proteins $p_5$ into the nucleus, which are detected by the c-protein. It then sets the c-bit in an unconscious state, that is, a state 
never constant which belongs to the  orthogonal complement of $\hbox{Vect}(f_0,f_1)$. 
This is a description that is compatible with H3, still sketchy though. 
In chapter \ref{realiste}, by specifying further
the functioning of a c-protein,
we analyze the difficulties raised by the implementation of H3
and we propose solutions.\par\medskip

Using other c-proteins, a c-neuron can encode at the quantum level in the mind other events which are encoded at the molecular level, such as the more precise form of the axonal activity period, the history of previous activity period or the activity of certain synapses. \par\medskip

The lifetime of a c-protein must be longer than one day. The complexity of its functions undoubtedly excludes a higher longevity. The repair and renewal of c-proteins can be carried out during periods of unconscious sleep. The particles composing the c-bits must be carefully preserved during this operation.  So, from one day to the next, it is always the same set of particles which corresponds to $\bf{mind}$. However, a low variability of the object $\bf{mind}$  must be accepted to take into account learning as well as the destruction of c-proteins or c-neurons, accidental or due to aging.  \par\medskip
A c-bit is not at all protected from entanglement with the environment. On the contrary, its state becomes  entangled with that of the c-protein which in turn becomes entangled with the cellular medium. From this point of view, c-bits are fundamentally different from quantum computer qubits. I do not assume that the observer behaves like a quantum computer and maintains superpositions of quantum states which are not entangled with the environment. I thus depart from certain theories currently in vogue [S.H] and [M.F] which seem though contradicted by the calculations   of the very short time of decoherence in  biological systems  [M.T].

\section{The Notion of Probability}

\subsection{ Idealized Born rule \label{born}}
We start with an idealized situation, which we will make more realistic in the next paragraph.\par
Let us take the situation and the notations from paragraph \ref{orthoco}; we consider an object $\bf{obj}$ and an orthonormal family $(obj_i)_{1\leq i\leq n}$ of ${\cal H}_{obj}$. It is assumed that an observer is
provided with a device $\bf{dev}$ such that:\par
\begin{itemize}
	\item Before any interaction of \bf{dev} with \bf{obj}, the observer systematically sets the device in a reference state $dev^0$.
\item	If \bf{obj} is in the state $obj_i$ before interaction with \bf{dev}, then after interaction, \bf{obj} is still in the state $obj_i$, meanwhile \bf{dev} has evolved to the state $dev_i$. We can summarize this dynamic by writing: $obj_i\otimes dev^0 \longrightarrow obj_i\otimes dev_i$.
	\item The observer's senses can differentiate a state $dev_i$ from another state $dev_j$ when $i\not=j$. This means that the device has a macroscopic component that clearly indicates the state $dev_i$. 
\end{itemize}

Ideally, during a quantum measurement, the universe is initially in a separable form:
$univ(t_0)=mind\otimes obj\otimes dev^0\otimes env$, where $mind$ is the state, assumed conscious, of the observer's mind and where $env$ is the state of the rest of the universe.\par
We assume that $obj=\di\sum_{i=1}^n\alpha_i obj_i$. During a first phase,  called the premeasurement, the object becomes entangled with the device and inevitably with the environment, but not with the
observer's mind, because this first phase takes place within a conscious period 
whose duration is $T$, during which the state of the mind remains constant, 
or because the observer does not observe for the moment at all the measuring device. At the end of this first phase, the state of the universe is\par
$univ(t_1)=mind\otimes\di\sum_{i=1}^n \alpha_i obj_i\otimes dev_i(t_1)\otimes env_i(t_1)$. \par
Then, during the measurement phase, the observer becomes aware of the result provided by the device. Each state $dev_i$ of the device modifies its mind into a state of consciousness $mind_i$. It is assumed that if
$i\not=j$, $mind_i\not=mind_j$, i.e. the observer can actually read the indication provided by the device. 
 The universe is then in the state 
$univ(t_2)=\di\sum_{i=1}^n \alpha_i mind_i\otimes obj_i\otimes dev_i(t_2)\otimes env_i(t_2)$.

Initially, $mind\otimes \Bigl(\di\sum_{i=1}^n\alpha_i obj_i\Bigr)\otimes dev^0\otimes env$ represents a state of the universe in which different parts exist independently of each other. The object is in a state of coexistence a priori non-uniform of the various states $obj_i$. As we can also write  \par
$univ(t_0)=\di\sum_{i=1}^n
\alpha_i mind\otimes obj_i\otimes dev^0\otimes env$,  initially, the universe is a coexistence of $n$ states that will each evolve from the state $mind\otimes obj_i\otimes dev^0\otimes env$
at the moment $t_0$ to the state $mind\otimes obj_i\otimes dev_i(t_1)\otimes env_i(t_1)$ at the moment $t_1$ then to the state $mind_i\otimes obj_i\otimes dev_i(t_2)\otimes env_i(t_2)\otimes env_i(t_2)$.\par
After the premeasurement, $mind\otimes\di\sum_{i=1}^n \alpha_i obj_i\otimes dev_i(t_1)\otimes env_i(t_1)$ represents a state where the observer remains independent from the rest of the universe, which is now described as a non-uniform coexistence of different scenarios, the $i$th describing an object in the state $obj_i$, consistent with a device in the state $dev_i$ and an environment in the state $env_i$.\par

We can also write 
$univ(t_1)= 
 \di\sum_{i=1}^n\alpha_i mind\otimes obj_i\otimes dev_i(t_1)\otimes env_i(t_1)$;
the universe is a non-uniform coexistence of different scenarios, the $i$th describing a universe in which the object is in the state $obj_i$, the device in the state $dev_i$ and the environment in the state $env_i$, and where the observer, in the state $mind$ independent of $i$, has no awareness of the state of $\bf{obj}$.\par
At the end of the measurement, $univ(t_2)=\di\sum_{i=1}^n \alpha_i mind_i\otimes obj_i\otimes dev_i(t_2)\otimes env_i(t_2)$ is interpreted in the same way.\par\medskip
Let us now study this measurement process from the point of view of the observer whose consciousness \it{is} rigorously his state of mind; this state \it{is} also all the information available to the observer. Before the measurement, the observer is supposed to know that he is about to measure the state of \bf{obj} according to the family $(obj_i)_{1\leq i\leq n}$. After reading the device, the observer knows the result of the measurement because it is integrated in the value $mind_i$ of his state. \par
Which of these $n$ observers is he among $mind_1,\ldots,mind_n$? The question assumes that the observer is provided with some sort of individuality, identity, soul or other thing that ensures his uniqueness within the sum for $i$ ranging from 1 to $n$.  In this article, I suppose that is not the case; The identity of the observer, in that he is aware of being, is modified by the result obtained, differently in each term. There is a coexistence of different scenarios, and in each of these scenarios, there is always a single observer, whose identity and knowledge are specific. \par

In this context, the probability that the observer experiences the $i$th result can be defined in accordance with our intuition; it is the number of observers experiencing the $i$th result relative to the total number of observers. \par\medskip
We have seen though that such a count only makes sense in the context of uniform coexistences, in which each term is present in the same way. \par
As $univ(t_0)=\di\sum_{i=1}^n \alpha_i mind\otimes obj_i\otimes dev^0\otimes env$,  according to 
IP6,  $univ(t_0)$ is a uniform coexistence in which
the ratio of states $mind\otimes obj_i\otimes dev^0\otimes env$ to all states is equal to $|\alpha_i|^2$. These states become at the moment $t_1$
$mind\otimes obj_i\otimes dev_i(t_1)\otimes env_i(t_1)$. In particular, at the moment $t_1$, all states correspond to an observer whose mind is in the state
$mind$, so we are in the presence of a unique conscious experience of the observer. Then between $t_1$ and $t_2$, the states\par 
$mind\otimes obj_i\otimes dev_i(t_1)\otimes env_i(t_1)$ become 
$mind_i\otimes obj_i\otimes dev_i(t_2)\otimes env_i(t_2)$. Thus, in a proportion of $|\alpha_i|^2$, at the moment $t_2$, the observer experiences the state $mind_i$, i.e. he measures the state $obj_i$ of the object thanks to the device \bf{dev}.  The situation being reduced to uniform coexistences, we thus have at the moment $t_0$ or $t_1$ a coexistence of states in which all the observers are in the same state $mind$, and among these a proportion of $|\alpha_i|^2$ becomes $mind_i$ at the moment $t_2$.\par

We \it{define} the probability, at the moment $t_0$ or $t_1$, that the object will be at the moment $t_2$ in the state $obj_i$, as the number of observers who will experience this reality at the moment $t_2$ divided by the total number of observers considered. 
Thus, in our idealized situation, this probability is equal to $|\alpha_i|^2$. 
 We have established Born rule. \par
This is the exact idea of Brian Greene in his popularization work ``the hidden reality'' [B.G, page 314]. It is made coherent thanks to a modelling of the conscious observer, inside the formalism of quantum mechanics, which we will consolidate in the second part.  It is indeed the definition of the notion of probability that is important here, and not the derivation of Born rule, because there are already several convincing proofs: Gleason's theorem [A.M.G], [R.C.2], the proofs of Zurek [W.Z.1] and Wallace [D.W.2], [F.B]. \par
So $|\alpha_i|^2$ is the probability at the moment $t_0$ or $t_1$, that the observer $\bf{mind}$ experiences at the moment $t_2$ a state equal to $mind_i$. As defined, the notion of probability has a subjective aspect, because it concerns the observer's state of mind. It is however objective because its value does not depend on the observer but only on the state \par
$obj=\di\sum_{i=1}^n \alpha_i obj_i$. As defined, this probability has a clear physical interpretation. It exists without the need for anyone to evaluate it.
\par\medskip
The proof of IP6, adapted to the current situation, uses objects $\bf{int}$ and $\bf{env'}$
 that do not interact with each other nor with 
$\bf{mind}\otimes\bf{obj}\otimes\bf{dev}\otimes\bf{env}$, that is with the entire universe. In addition,
 the dimension of ${\cal H}_{env'}$ must be as large as necessary. 
It means assuming that there are other universes, independent of ours, or to put it another way, that we can split our universe into several parts that have never interacted with each other.

\subsection{Generalized Born Rule} 

Let us recall the three key moments of a measure, in the previous idealized situation: \par
\begin{itemize}
\item Initialization: 
$univ(t_0)=\di\sum_{i=1}^n\alpha_i mind\otimes obj_i\otimes dev^0\otimes env$.
\item Premeasure: 
$univ(t_1)=\di\sum_{i=1}^n \alpha_i mind\otimes obj_i\otimes dev_i(t_1)\otimes env_i(t_1)$. 
\item Measure: $univ(t_2)=\di\sum_{i=1}^n \alpha_i mind_i\otimes obj_i\otimes dev_i(t_2)\otimes env_i(t_2)$.
\end{itemize}

Between the moments $t_0$ and $t_2$, $\bf{dev}$ and $\bf{env}$ interact, so the state of 
$\bf{dev}\otimes\bf{env}$  a priori does not remain separable and it is appropriate to replace
 $dev_i(t_1)\otimes env_i(t_1)$ by
$\di\sum_{j}\beta_{i,j} dev_{i,j}(t_1)\otimes env_{i,j}(t_1)$, which will be denoted by
$[dev,env]_i(t_1)$.\par
Between $t_0$ and $t_1$, $\bf{obj}$ interacts with $\bf{dev}\otimes\bf{env}$.
The fact that $\bf{dev}$ has correctly measured \bf{obj} in the state $obj_i$ translates into the fact that, for every $j$, $dev_{i,j}(t_1)\in F_i$, where $F_i$ is a sub-vector space of  ${\cal H}_{dev}$ characteristic of $obj_i$, 
that is to say into the fact that $[dev,env]_i\in F_i\otimes{\cal H}_{env}$. 
Thus $\bf{dev}\otimes\bf{env}$ memorizes that $obj=obj_i$ in each term by preserving the physical property 
``$dev\in F_i$''.
\par
The observer's brain is the union of $\bf{mind}$ and part of $\bf{env}$. 
It is assumed that he can read this property and then set his mind accordingly in the state $mind_i$.\par
Thus, we  replace thereafter the equalities at moments $t_1$ and $t_2$ by : \par
$univ(t_1)=\di\sum_{i=1}^n \alpha_i mind\otimes obj_i\otimes [dev,env]_i(t_1)$ and\par
$univ(t_2)=\di\sum_{i=1}^n \alpha_i mind_i\otimes obj_i\otimes [dev,env]_i(t_2)$.\par

\subsubsection{Variability of the Object State\label{var}}

A totally constant state $obj_i$ during an interaction is illusory, if a whole mechanism is not put in place to achieve it, as is the case for the observer's mind during a phase of consciousness. For all $k\in\{0,1,2\}$, it is therefore appropriate to replace 
$obj_i$ by $obj_i(t_k)$ in the expression $univ(t_k)$.\par

For all
$i\in\{1,\ldots,n\}$, by writing $obj_i(t_0)=obj_i$, \par
$mind\otimes obj_i(t_1)\otimes  [dev,env]_i(t_1)
=U_{t_0\rightarrow t_1}(mind\otimes obj_i\otimes dev^0\otimes env)$ and \par
$mind_i\otimes obj_i(t_2)\otimes  [dev,env]_i(t_2)
=U_{t_0\rightarrow t_2}(mind\otimes obj_i(t_0)\otimes dev^0\otimes env)$,\par
so the orthogonality of the family $(obj_i)_{1\leq i\leq n}$ is transmitted to
the families\par
$(mind\otimes obj_i(t_1)\otimes  [dev,env]_i(t_1))_{1\leq i\leq n}$
 and $(mind_i\otimes obj_i(t_2)\otimes  [dev,env]_i(t_2))_{1\leq i\leq n}$. This still allows the expressions $univ(t_k)$ to be interpreted as coexistences and that extends the validity of the previous arguments to this more general situation. \par

$|\alpha_i|^2$ is still interpreted as the probability at the moment $t_0$ or $t_1$, that at the moment $t_2$ the observer will subjectively experience the state  $mind_i$. However, in the idealized situation, experiencing this state of consciousness goes hand in hand with the fact that $\bf{obj}$ is in the state $obj_i$. This is no longer the case.  We can no longer say that $|\alpha_i|^2$ is the probability that at the moment $t_2$, the observer is in the presence of the object in the state $obj_i$. This is now the probability for the observer to be at the moment $t_2$ in the presence of the object in the state $obj_i(t_2)$, exactly as if at the moment $t_0$, the observer had already been in the presence of the object in the state $obj_i(t_0)=obj_i$, which would then have evolved to the state $obj_i(t_2)$.\par

If $obj_i(t_2)$ is completely different from $obj_i$, then $obj_i(t_2)=\di\sum_j\beta_j obj_j$, with 
possibly $|\beta_i|$ far less than 1, so if at the moment $t_2$, the observer makes a second measurement using the same device, he probably will not find the same result; it is not a repeatable measurement.  For this reason,  we prefer to measure objects according to basis $(obj_i)_i$ such that, if \bf{obj} is in the state $obj_i$ at the moment $t_0$, then it will evolve very little during interaction with the device and with the environment. In this case, $|\beta_i|^2$ remains very close to 1 and the observer can consider with a good confidence that the state of the object at the moment $t_2$ is still $obj_i$. 
This is important when the observer makes these measurements 
 in order to adapt to his environment.\par\medskip

In the most general case, between $t_1$ and $t_2$, the state of $\bf{obj}$ is  again entangled with 
$\bf{dev}\otimes \bf{env}$ so that we should write
$univ(t_2)=\di\sum_{i=1}^n \alpha_i mind_i\otimes[obj,dev,env]_i(t_2)$. Even if the following arguments generalize to this situation, for the sake of clarity, we will remain in the idealized case where $obj_i$ is constant during the measurement process.

\subsubsection{Partial Measure\label{partielle}}
For any $\theta\in\Re$, $obj_i\otimes env_i=(e^{i\theta}obj_i)\otimes (e^{-i\theta }env_i)$, so after the previous idealized measure, the observer can only assert that the state of $\bf{obj}$ belongs to the one-dimensional space
$\hbox{Vect}(obj_i)$.
We now turn to more general measures that simply specify
a subspace of ${\cal H}_{obj}$ in which the state of the object is located; rather than starting from an orthonormal family 
$(obj_i)_i$, a family of subspaces $E_1,\ldots,E_n$ is given into ${\cal H}_{obj}$, pairwise orthogonal and  such that 
${\cal H}_{obj}=\di\bigoplus_{1\leq i\leq n}^{\bot}E_i$. \par
Between $t_0$ and $t_1$, the measuring device conforms to the following dynamics: \par
for all $i\in\{1,\ldots,n\}$, 
for any $x \in E_i$,
$x\otimes dev^0\otimes env\longrightarrow x\otimes [dev,env]_{i}$, 
where $[dev,env]_i$ is a state of $\bf{dev}\otimes\bf{env}$ that memorizes an event of the form 
``$dev\in F_i$'' which then sets 
$\bf{mind}$ in the state $mind_i$ at the time $t_2$.\par
If we denote by
$p_{E_i}$ the orthogonal projector on $E_i$ and if we define
   $obj_i$ as $p_{E_i}(obj)$, then
$univ(t_0)=mind\otimes obj\otimes dev^0\otimes env$, where
$obj=\di\sum_{i=1}^n obj_i$. \par

In general $\norm{obj_i}\not=1$, so to put $obj$ in the form of a coexistence, one must write $obj=\di\sum_{1\leq i\leq n\atop\hbox{\tiny\ such that }  obj_i\not=0}\alpha_i . \frac{obj_i}{\norm{obj_i}}$, where $\alpha_i$ denotes $\norm{obj_i}$. 
\par
The premeasure leads to $univ(t_1)=mind\otimes\di\sum_{i=1}^n obj_i\otimes  [dev,env]_{i}$,\par

then $univ(t_2)=\di\sum_{i=1}^n  mind_i\otimes obj_i\otimes  [dev,env]_{i}(t_2)$. 
\par
One can write
$univ(t_1)=\di\sum_{1\leq i\leq n\atop \hbox{\tiny\ such that }   obj_i\not=0}  \alpha_i mind\otimes \frac{obj_i}{\norm{obj_i}}\otimes  [dev,env]_{i}$\par
and
$univ(t_2)=\di\sum_{1\leq i\leq n\atop \hbox{\tiny\ such that }   obj_i\not=0}   \alpha_i mind_i\otimes  \frac{obj_i}{\norm{obj_i}}\otimes  [dev,env]_{i}(t_2)$. \par
The orthogonality of the  family $(obj_i)_i$ makes it possible to reproduce the arguments seen in the idealized case. Thus, the probability at moment $t_0$ or $t_1$ that the observer is at the moment $t_2$ in a state $mind_i$ and that the object is in a state belonging to $E_i$ is equal to 
$|\alpha_i|^2=\norm{obj_i}^2=\norm{p_{E_i}(obj)}^2$.   This is the first generalization of the Born rule.\par\medskip

This kind of measurement, associated with the projector family $(p_{E_i})_{1\leq i\leq n}$,  is called a PVM (projection-valued measurement). There are even more general measures, called POVMs. These can be
though interpreted as PVMs applied to an auxiliary object that previously interacted with \textbf{obj} [C.A.F, page 15].
We will therefore limit ourselves in this article to PVMs.

\subsubsection{Initial Entanglement of the Device with the Environment}
It is illusory in practice to assume that one can set a macroscopic device in a precise initial quantum state $dev^0$. So let us assume that\par
$univ(t_0)=mind\otimes obj\otimes\di\sum_k\beta_k dev^0_k \otimes env_k$, where the family $(dev^0_k)_k$ is 
orthonormal. We assume again that \bf{dev} can measure the projector family 
$(p_{E_i})_i$,\par
 where ${\cal H}_{obj}=\di\bigoplus_{1\leq i\leq n}^{\bot}E_i$.
With the notations from the previous paragraph,\par
$univ(t_0)=\di\sum_{i,k}\beta_k mind\otimes obj_i\otimes dev_k^0\otimes env_k$.
 \par
After the premeasure, $univ(t_1)=\di\sum_{i,k}\beta_k mind\otimes obj_i\otimes  [dev,env]_{i,k}(t_1)$, \par
then
$univ(t_2)=\di\sum_{i,k}\beta_k mind_i\otimes obj_i\otimes  [dev,env]_{i,k}(t_2)$.
\par
Again, using the unitary operators $U_{t_0\rightarrow t_1}$ and $U_{t_0\rightarrow t_2}$ ensures that in these sums, the terms are pairwise orthogonal.   So the probability  that at the moment $t_2$, the observer experiences $mind_i$, therefore that he is in the presence of an object whose state belongs to
$E_i$ is equal to 
$\di\sum_{k}|\beta_k|^2\norm{obj_i}^2=\norm{obj_i}^2$.\par
Thus without loss of generality, we can keep assuming that the initial state of $\bf{dev}\otimes \bf{env}$ is separable and equal to 
$dev^0\otimes env$.

\subsubsection{A Richer Consciousness\label{riche}}

Between $t_0$ and $t_2$, the transition from the  state  $mind$ to a state $mind_i$ that depends only on the object belonging to $E_i$ can seem very poor in front of the richness of our conscious experiences. \par

Starting again from the initial state 
$univ(t_0)=mind\otimes obj\otimes dev^0\otimes env$,
if we accept that our observer's mind performs between $t_0$ and $t_1$
measures on other objects, then $univ(t_1)$ has the form 
$univ(t_1)=\di\sum_{i,k}\beta_k mind_k\otimes obj_i\otimes[dev,env]_{i,k}$; for every $k$, the mind is in the state $mind_k$ because he has measured some information in the environment and on the device.\par
The family $(mind_k\otimes obj_i)_{i,k}$ being orthonormal, \par
we have 
$1=\norm{univ(t_1)}^2=\di\sum_{i,k}|\beta_k|^2\norm{obj_i}^2$,
therefore 
$\di\sum_{k}|\beta_k|^2=1$.\par
$univ(t_1)$ is then the coexistence $\di\sum_{i,k\atop \hbox{\tiny such that } obj_i\not=0}\beta_k\norm{obj_i} univ_{i,k}(t_1)$,\par
where $univ_{i,k}(t_1)=mind_k\otimes \di\frac{obj_i}{\norm{obj_i}}\otimes[dev,env]_{i, k}$.\par
If we fix $i$ and $k$,  if the universe is in the state $univ_{i,k}(t_1)$ at the time $t_1$ and if other observer's measurements between $t_1$ and $t_2$ are taken into account, then the state of the universe at the time $t_2$ is
\par 
$univ_{i,k}(t_2)=\di\sum_{h}\gamma_{i,k,h} mind_{i,k,h}\otimes \di\frac{obj_i}{\norm{obj_i}}\otimes [dev,env]_{i,k,h}(t_2)$.  Again, the family \par
$(mind_{i,k,h})_h$ being orthonormal, we have 
$1=\norm{univ_{i,k}(t_2)}^2=\di\sum_h|\gamma_{i,k,h}|^2$.\par
Globally,  
$univ(t_2)=\di\sum_{i,k,h}\beta_k\gamma_{i,k,h} mind_{i,k,h}\otimes obj_i\otimes 
[dev,env]_{i,k,h}(t_2)$.\par
At the moment $t_0$, the observer experiences the unique state  $mind$. The probability that at the moment $t_2$, he experiences a state of mind according to which $obj\in E_i$ is always by definition the proportion of states of $univ(t_2)$ for which the state of $\bf{mind}$ is of the form $mind_{i,k,h}$, therefore this probability is equal to 
$\di\sum_{k,h}|\beta_k\gamma_{i,k,h}|^2\norm{obj_i}^2=\norm{obj_i}^2$.\par
From the observer's point of view, at the moment $t_1$, he is experiencing one of the state $mind_k$, so for him, the universe is just $univ(t_1)=mind_k\otimes \di\sum_iobj_i\otimes 
[dev,env]_{i,k}$  and then
$univ(t_2)=\di\sum_{i,h}\gamma_{i,k,h} mind_{i,k,h}\otimes obj_i\otimes 
[dev,env]_{i,k,h}(t_2)$. 
 Thus, from the observer's point of view at the moment $t_1$, the probability
that he will  experience at the moment $t_2$ a state of mind according to which $obj\in E_i$ is still equal to 
$\norm{p_{E_i}(obj)}^2$. 
\par
We can verify that the result remains the same if we combine a rich consciousness and an initial state of the  device entangled with the environment. \par\medskip
In sum, to study probabilities in quantum mechanics, we can assume  
 that $obj_i$ is constant during the measurement, that the initial state of $\bf{dev}\otimes\bf{env}$ is separable and that the mind simply moves from the single state $mind$ at the moment $t_0$ to 
a state $mind_i$ at the moment $t_2$ that depends only on the event $obj\in E_i$.\par
We make these assumptions  most often afterwards.

\subsubsection{The Problem of the Preferred Basis}
The  measurement problem in quantum mechanics contains as a sub-problem
that of the basis choice [M.S.1, page 53]; why do we observe \bf{obj} according to an orthonormal basis $(obj_i)_i$ of ${\cal H}_{obj}$ rather than another, or according to the sum 
${\cal H}_{obj}=\di\bigoplus_i^{\bot}E_i$ rather than another? \par
This problem is solved here by the fact that there is not a single way of writing the universe in the form of a coexistence, so there is in fact no choice of any basis from a global point of view, external to any observer.
\par\medskip
For a given individual, provided with the orthonormal family of his conscious states 
$(mind_c)_{c\in\cal C}$, an object is measured according to the sum ${\cal H}_{obj}=\di\bigoplus_i^{\bot}E_i$ and not another, because \bf{dev} is sensitive to this sum and because the brain can translate the state of the device into a state $mind_c$ of the mind. 
\par\medskip

Let ${\cal V}$ be the sub-vector space $\hbox{Vect}(mind_c)_{c\in \cal C}$. We can however wonder about the relevance of working in the particular basis $e=(mind_c)_{c\in\cal C}$ of $\cal V$. 
Consider a second orthonormal basis $f=(f_d)_{d\in \cal C}$ of $\cal V$. 
Let us denote by $P=(p_{d,c})_{d,c\in\cal C}$ the transition matrix from $f$ to $e$. \par
So for all $c\in\cal C$, $mind_c=\di\sum_{d\in\cal C}p_{d,c}f_d$, 
and for any $d\in C$,
 $f_d=\di\sum_{c\in\cal C}\bar{p_{d,c}}mind_c$.\par
Thus, during a T period of consciousness,
the conscious state $mind_c$ can be written as a coexistence of the states $f_d$. Moreover, since each vector $f_d$ is a linear combination of vectors $mind_c$ with constant coefficients, if the state of $\bf{mind}$ is  $f_d$ at the beginning of a T conscious period, it remains constantly equal to $f_d$ during that period. Then, the universe from our observer's point of view, during a T conscious period, can also be written as $univ=\di\sum_{d\in\cal C}\gamma_d f_d\otimes env_d$. It is a coexistence of universes in which each universe contains the observer in the state $f_d$, constant over the duration $T$.
So from that point of view, there would be one observer for each orthonormal basis of $\cal V$. \par
It is likely that an observer according to a basis $f$ other than $e$ does not satisfy the properties of a conscious memory required by H3 so that such an observer is actually unconscious. It is likely, but it is better to prove it.
\par\medskip
For this purpose, it would be too restrictive to assume that when measuring $\bf{obj}$, the mind is set in a state $mind_i$ which depends only on the event $obj\in E_i$. 
So we keep using the context of the previous paragraph. \par

For any $i\in\{1,\ldots,n\}$, let ${\cal C}_i$ be  the set of $c\in \cal C$ such that $mind_c$ is a possible state of mind when he experiences the event $obj\in E_i$.\par
If there was $c\in {\cal C}_i\cap{\cal C}_j$ with $i\not=j$, then the mind in the state $mind_c$ would not know whether the object belongs to $E_i$ or $E_j$, which is false if we assume that the observer can measure \bf{obj} according to the sum
${\cal H}_{obj}=\di\bigoplus_i^{\bot}E_i$.\par
The family $({\cal C}_1,\ldots, {\cal C}_n)$ is therefore  pairwise disjoint.
\par\medskip

With these notations, after the measurement, the universe takes the form of \par
$univ(t_2)=\di\sum_{i=1}^n obj_i\otimes \di\sum_{c\in {\cal C}_i}
\beta_c mind_c\otimes[dev,env]_{i,c}$.\par
For every $i\in\{1,\ldots,n\}$, there exists $c\in {\cal C}_i$ such that the state of \bf{mind} is 
$mind_c$, so that the mind, according to the basis $e$, actually experiences the event $obj\in E_i$. The choice of $c$ among all states of ${\cal C}_i$ depends on the other measures made by the observer between $t_0$ and $t_2$, therefore it depends on the different possibilities of evolution of $\bf{dev}\otimes\bf{env}$. \par\medskip
In the basis $f$, we can write \par
$\eqalign{univ(t_2)&=\di\sum_{i=1}^n obj_i\otimes \di\sum_{c\in {\cal C}_i}
\beta_c \di\sum_{d\in{\cal C}}p_{d,c}f_d\otimes[dev,env]_{i,c}\hfill\cr
&=\di\sum_{d\in{\cal C}}f_d\otimes
\di\sum_{i=1}^n obj_i\otimes \di\sum_{c\in {\cal C}_i}
\beta_c p_{d,c}[dev,env]_{i,c}.\hfill\cr}$\par
Thus, for the ``observer according to the basis $f$'', the universe at time $t_2$ is in one of the states 
$univ(t_2)=\di f_d\otimes
\di\sum_{i=1}^n obj_i\otimes \di\sum_{c\in {\cal C}_i}
\beta_c p_{d,c}[dev,env]_{i,c}$.\par
The observer in the state $f_d$ is then entangled with a state of 
$\bf{obj}\otimes\bf{dev}\otimes\bf{env}$ which is generally non-separable. The information retrieved by the mind is no longer just about the object $\bf{obj}$, but globally about
$\bf{obj}\otimes\bf{dev}\otimes\bf{env}$. He has not measured the object according to the sum $\di {\cal H}_{obj}=\bigoplus_i^{\bot}E_i$, nor any other writing of ${\cal H}_{obj}$ alone. 
Thus, the observer according to the basis $f$ does not memorize the state of the object. Worse, he only has access to global information about the entire universe. So the observer according to the basis $f$ does not have the properties of a consciousness. \par\medskip

To avoid this non-separable situation, for any $d\in\cal C$, into the expression\par
$univ(t_2)=\di f_d\otimes
\di\sum_{i=1}^n obj_i\otimes \di\sum_{c\in {\cal C}_i}
\beta_c p_{d,c}[dev,env]_{i,c}$, the first sum must be reduced to a single non-zero term, therefore there must exist $i_d\in\{1,\ldots,n\}$ such that for all $c\notin C_{i_d}$, $\beta_c p_{d,c}=0$. 
We have then $univ(t_2)=\di f_d\otimes
obj_{i_d}\otimes \di\sum_{c\in {\cal C}_{i_d}}
\beta_c p_{d,c}[dev,env]_{i_d,c}$ and the observer according to the basis $f$ indeed measures 
\bf{obj} according to the sum
${\cal H}_{obj}=\di\bigoplus_i^{\bot}E_i$.\par
$\beta_c$ depends on the interaction between the observer and $\bf{dev}\otimes\bf{env}$, so it is reasonable to assume that $\beta_c$ is non-zero for many initial conditions of $\bf{dev}\otimes\bf{env}$.\par
Thus, for the observer  for the basis $f$ to be conscious, it is necessary that for every $d\in\cal C$, there exists $i_d\in\{1,\ldots,n\}$ such that for every $c\notin C_{i_d}$, $p_{d,c}=0$. Moreover this must happen for any observed object, so for a large number of families 
$({\cal C}_i)_{1\leq i\leq n}$ of parts of $\cal C$ which are pairwise disjoint. 
 \par\medskip

Let us suppose that there exists $d_0\in\cal C$ and 
$c_1,c_2\in\cal C$ with $c_1\not=c_2$, such that $p_{d_0,c_1}\not=0$ and $p_{d_0,c_2}\not=0$.\par
It is likely that there exists an object $\bf{obj}$ and a relationship of the form ${\cal H}_{obj}=\di\bigoplus_{1\leq i\leq n}^{\bot}E_i$
that defines a family $({\cal C}_1,\ldots,{\cal C}_n)$ of parts of $\cal C$ which is pairwise disjoint and such that $c_1\in {\cal C}_1$ and $c_2\in {\cal C}_2$.\par
Then, with $d=d_0$, the measure of $\bf{obj}$ sets the universe, from the  point of view of the observer according to the basis  $f$, into the state\par
 $univ(t_2)=\di f_{d_0}\otimes
\di\sum_{i=1}^n obj_i\otimes \di\sum_{c\in {\cal C}_i}
\beta_c p_{d_0,c}[dev,env]_{i,c}$.  The first sum contains at least 2 non-zero terms, for $i=1$ and $i=2$. Then the observer according to $f$ does not memorize the state of the object.\par
Thus, for an observer according to $f$ to be conscious, it is necessary that for every $d\in\cal C$, there is a single $c(d)\in\cal C$ such that $p_{d,c(d)}\not=0$. $P$ being unitary, there is $\varphi_d\in\Re$ such that $p_{d,c(d)}=e^{i\varphi_d}$. So 
$f_d=\di\sum_c\bar{p_{d,c}}mind_c=e^{-i\varphi_d}mind_{c(d)}$.\par
In conclusion, regardless of the phase $\varphi_d$ and  the order of vectors, $e$ is the only basis according to which the observer is conscious. 

\subsection{Conditional Probabilities}

\subsubsection{Non Contextuality\label{noncon}}

Let $(E_1,\ldots,E_n)$ and $(F_1,\ldots,F_k)$ be  two families of subspaces of ${\cal H}_{obj}$ such that\par
 ${\cal H}_{obj}
=\di\bigoplus^{\bot}_{1\leq i\leq n}E_i=\bigoplus^{\bot}_{1\leq j\leq k}F_j$. 
If we have two devices, $\bf{dev}_1$ and $\bf{dev}_2$ that can  premeasure respectively 
$(E_1,\ldots,E_n)$ and $(F_1,\ldots,F_k)$, an observer can start with the object in the state $obj$ and measure it according to either family. The probability that after a measurement using $\bf{dev}_1$, he experiences that $obj\in E_i$ is 
$Pr_{dev_1}(obj\in E_i)=\norm{p_{E_i}(obj)}^2$ and the probability that he experiences, after a measure using $\bf{dev}_2$, that $obj\in F_j$ is equal to
$Pr_{dev_2}(obj\in F_j)=\norm{p_{F_j}(obj)}^2$.\par
In particular, if there is $i_0,j_0$ such that $E_{i_0}=F_{j_0}$, then these two probabilities are equal. So, 
$Pr_{dev_1}(obj\in E_i)$ does not depend on the device used, nor on the family $(E_1,\ldots,E_n)$, but only on $obj$ and $E_i$. This is the noncontextuality property of quantum measurement 
[C.A.F., page 14].\par
We can thus denote this probability  by $Pr(obj\in E_i)$. 

\subsubsection{The Law of Total Probability\label{condi}}
The notations of the previous paragraph are used again.\par
We denote by
$Pr(obj\in F_j | obj\in E_i)$ the probability of the event $obj\in F_j$ given that $obj\in E_i$.
To define it in a natural way, we assume that the observer performs two successive measurements, firstly with the device $\bf{dev}_1$ then with $\bf{dev}_2$. Then $Pr(obj\in F_j | obj\in E_i)$ is the probability that the observer experiences the event $obj\in F_j$ when the universe before the second measure is restricted to the terms in which he experiences the event $obj\in E_i$.
 \par\medskip
More precisely, by denoting $t_1$ and $t_2$ the moments of these two successive measures, we have
$univ(t_0)=mind\otimes obj\otimes dev_1^0\otimes dev_2^0\otimes env$, 
then if we define
   $obj_i=p_{E_i}(obj)$,
\par
$univ(t_1)=\di\sum_{1\leq i\leq n\atop  \hbox{\tiny\ such that } obj_i\not=0} \norm{obj_i}. mind_i\otimes \frac{obj_i}{\norm{obj_i}}
\otimes [dev_{1},env]_i\otimes dev_2^0$,\par
and finally, if we define, when $obj_i\not=0$,
$obj_{i,j}=\di p_{F_j}\Bigl(\frac{obj_i}{\norm{obj_i}}\Bigr) $, \par
$univ(t_2)=\di\sum_{1\leq i\leq n\atop  \hbox{\tiny\ such that } obj_i\not=0}\!\!\!\! \!\!\!\! \norm{obj_i}.\!\!\!\!  \!\!\!\!  \di\sum_{1\leq j\leq k\atop \hbox{\tiny\ such that } obj_{i,j}\not=0}\!\!\!\!  \!\!\!\!  \norm{obj_{i,j}} mind_{i,j}\otimes \frac{obj_{i,j}}{\norm{obj_{i,j}}}
\otimes  [dev_{1},dev_{2},env]_{i,j}$.\par\medskip

We assume that the states $mind_i$ and $mind_{i,j}$ are conscious states and that $mind_{i,j}$ keeps the memory of the first measure.\par
We know that the states $mind_i$ are pairwise orthogonal.
This is fundamental\label{50} because 
it allows us to consider $univ(t_1)$ not only as a coexistence of the states\par
  $\di mind_i\otimes \frac{obj_i}{\norm{obj_i}}
\otimes [dev_{1},env]_i\otimes dev_2^0$, but also, by substituting
 $\di \frac{obj_i}{\norm{obj_i}}$\par
 with $\di\sum_{1\leq j\leq k\atop  \hbox{\tiny\ such that }  obj_{i,j}\not=0}\norm{obj_{i,j}} .   \frac{obj_{i,j}}{\norm{obj_{i,j}}}$, as  a coexistence of the states \par
$\di mind_{i}\otimes \frac{obj_{i,j}}{\norm{obj_{i,j}}}
\otimes [dev_{1},env]_i\otimes dev_2^0$. 
 Each of these states becomes at the moment $t_2$\par
 $\di mind_{i,j}\otimes \frac{obj_{i,j}}{\norm{obj_{i,j}}}
\otimes [dev_{1},dev_{2},env]_{i,j}$ and once $univ(t_2)$ is interpreted as a uniform coexistence, each of these states is present in a proportion equal to $ \norm{obj_i}^2\norm{obj_{i,j}}^2$. \par
  We deduce that the proportion at the moment $t_2$ of minds who experience the $j$-th result for the second measure, i.e. by definition 
 the probability at the moment $t_0$ that the observer experiences at the moment $t_2$ the event $obj\in F_j$ is 
$$Pr(obj\in F_j)=\di\sum_{1\leq i\leq n\atop \hbox{\tiny\ such that }  obj_i\not=0} \norm{obj_i}^2\norm{obj_{i,j}}^2.$$
Besides, if we fix $i$ in $\{1,\ldots,n\}$, given
 that after the first measurement the observer experiences the event $obj\in E_i$, then $obj_i\not=0$ and the universe reduces for this observer at the moment $t_1$ to  
$\di univ(t_1)= mind_i\otimes \frac{obj_i}{\norm{obj_i}}
\otimes [dev_{1},env]_i\otimes dev_2^0$. 
Given that $obj\in E_i$, we must consider that this is the initial state of the universe when we begin the second measurement, which therefore leads to \par
$univ(t_2)=\di\sum_{1\leq j\leq k\atop tq\ obj_{i,j}\not=0} \norm{obj_{i,j}} mind_{i,j}\otimes \frac{obj_{i,j}}{\norm{obj_{i,j}}}
\otimes [dev_{1},dev_{2},env]_{i,j}$.\par
Thus, at the moment $t_0$,
$Pr(obj\in F_j | obj\in E_i)=\norm{obj_{i,j}}^2$.\par
So, when $obj_i\not=$0 for all $i$, we have shown the law of total probability :\par
\label{tot}
$Pr(obj\in F_j)=\di\sum_{i=1}^n Pr(obj\in E_i) . Pr(obj\in F_j|obj\in E_i)$.\par
We can also write 
$Pr(obj\in F_j | obj\in E_i)
=\di\frac1{\norm{obj_i}^2}
\norm{p_{F_j}(obj_i)}^2$, so
$$Pr(obj\in F_j | obj\in E_i)=\di\frac{\norm{p_{F_j}\circ p_{E_i}(obj)}^2}{Pr(obj\in E_i)}.$$

In general, the numerator cannot be interpreted as a probability because $p_{F_j}\circ p_{E_i}$ is not an orthogonal projector. 

\subsubsection{Bayes' Theorem}

More specifically, $p_{F_j}\circ p_{E_i}$ is an orthogonal projector if and only if $p_{F_j}$ and $p_{E_i}$ commute.
In this case, $p_{F_j}\circ p_{E_i}$ is the orthogonal projector on $F_j\cap E_i$, 
and the previous formula becomes 
$Pr(obj\in F_j | obj\in E_i)=\di\frac{Pr(obj\in E_i\cap F_j)}{Pr(obj\in E_i)}$, which corresponds to the classic definition of a conditional probability.\par
In this case, if we switch the two measures, we get \par
$Pr(obj\in E_i | obj\in F_j)=\di\frac{Pr(obj\in E_i\cap F_j)}{Pr(obj\in F_j)}$. That proves the Bayes'  theorem: 
$$Pr(obj\in F_j | obj\in E_i)=\di\frac{Pr(obj\in E_i | obj\in F_j) . Pr(obj\in F_j)}{Pr(obj\in E_i)}.$$

As an example of such a situation, let us assume that $\bf{obj}=\bf{obj}_1\otimes \bf{obj}_2$.\par
Let $E_1$ be a subspace of ${\cal H}_{obj_1}$ and $E_2$ be a subspace of ${\cal H}_{obj_2}$. Then
$p_1=p_{E_1}\otimes Id_{{\cal H}_{obj_2}}$ \par
and
 $p_2= Id_{{\cal H}_{obj_1}}\otimes p_{E_2}$ are two orthogonal projectors of ${\cal H}_{obj}$. 
\par
Moreover, $p_1\circ p_2=p_{E_1}\otimes p_{E_2}=p_2\circ p_1$.\par
 So we can write 
$Pr(obj_1\in E_1 | obj_2\in E_2)=\di\frac{Pr(obj_1\in E_1\hbox{ and } obj_2\in E_2)}{Pr(obj_2\in E_2)}$, if we consider that the event $obj_1\in E_1$ corresponds more rigorously 
to
$obj\in E_1\otimes{\cal H}_{obj_2}$.

\subsection{Young's Double-Slit Experiment and Its Generalization}

Let us go back to the situation and notations presented at the beginning of paragraph \ref{noncon}. 
Young's double-slit experiment is a special case of this general situation; 
 $\bf{obj}$ is then a particle,   
$obj\in E_1$ if and only if the particle goes through the first slit and $obj\in E_2$ if it goes through the second slit, these $n=2$ events being detectable by the device $\bf{dev}_1$ when it is activated.\par
In this case, when the particles are sent one by one, it is observed that they go through one or the other slit, unpredictably and with the same frequency. 
The link between frequency and probability will be established in paragraph \ref{conf}, however we could present it now because it only uses previous knowledge. 
So, when a particle is sent towards the slots, it is in the state $obj=obj_1+obj_2$ where $obj_1=p_{E_1}(obj)$ and $obj_2=p_{E_2}(obj)$ have the same norm $\di\frac{1}{\sqrt 2}$. \par
The second device can detect the impact position of the particle on the screen located after the slits. We divide the screen into $k$ small squares called pixels. The event $obj\in F_j$ corresponds to the case where the particle hits the screen in the $j$-th pixel. 

\subsubsection{Interference\label{inter}}
In this paragraph, only $\bf{dev}_2$ is activated: we measure only the object according to the second family 
$(F_1,\ldots,F_k)$.\par
In the particular case of Young's double-slit experiment, we do not observe through which slit the particle goes but only the position of its impact on the screen. \par
For all $i\in\{1,\ldots,n\}$ and $j\in \{1,\ldots,k\}$, we define
$obj_i=p_{E_i}(obj)$ and\par
$obj_{i,j}=p_{F_j}(obj_i)$. Then, 
$obj=\di\sum_{i=1}^n obj_i=\di\sum_{j=1}^k\sum_{i=1}^n obj_{i,j}$.\par
We  assume that $obj_i\not=$0 for all $i$.
\par
$p_{F_j}(obj)=\di\sum_{i=1}^n obj_{i,j}$,  so when we measure \bf{obj} with $\bf{dev}_2$,
 we get \par
$Pr(obj\in F_j)=\norm{p_{F_j}(obj)}^2=\di\sum_{i=1}^n\norm{obj_{i,j}}^2+\sum_{i_1\not= i_2}
\langle  obj_{i_1,j} | obj_{i_2,j}\rangle  $.\par

As $\norm{obj_{i,j}}^2=\di\norm{p_{F_j}\circ p_{E_i}(obj)}^2
=Pr(obj\in E_i) . Pr(obj\in F_j|obj\in E_i)$, we see that\par
$Pr(obj\in F_j)\not=\di\sum_{i=1}^n Pr(obj\in E_i) . Pr(obj\in F_j|obj\in E_i)$.\par
Thus, when $\bf{dev}_1$ is not used, the law of total probability is wrong.\par
The experimental verification of this inequality forbids us to interpret  the relationship \par
$obj=\di\sum_{i=1}^n obj_i=\di\sum_{i=1}^n \norm{obj_i} . \frac{obj_i}{\norm{obj_i}}$ as
``at the moment $t_0$, 
the object $\bf{obj}$ is in only one of the states $\di\frac{obj_i}{\norm{obj_i}}$, with a probability (whatever its definition) equal to $\norm{obj_i}^2$'':
the existence of the additional terms 
$\langle  obj_{i_1,j} | obj_{i_2,j}\rangle  $,  with $i_1\not= i_2$, forces us to accept that
the two components $obj_{i_1}$ and $obj_{i_2}$ are both present.
Thus these terms of \textit{interference} indicate that the equality $obj=\di\sum_{i=1}^n obj_i$ must be interpreted as the simultaneous coexistence of all states $\di\frac{obj_i}{\norm{obj_i}}$.\par
In the case of Young's double-slit experiment, this means that the particle $obj$ is in a state of coexistence of two scenarios: $obj$ goes through the first slit for the first scenario and through the second slit for the second scenario. Both scenarios really exist because they both take part in the calculation of the probability that the particle hits the screen on the $j$-th pixel. \par\medskip

However, the interference between the different states $obj_i$ does not mean that they interact with each other. Rather, they coexist independently. Indeed, the initial situation is 
$univ(t_0)=\di\sum_{i=1}^n mind\otimes obj_i\otimes dev_2^0\otimes env$, if we include $\bf{dev}_1$ in the environment.  It is a coexistence of the $n$ states $mind\otimes \di\frac{obj_i}{\norm{obj_i}}\otimes dev_2^0\otimes env$.\par
We also have 
$univ(t_0)=\di\sum_{i=1}^n mind\otimes\Bigl(\sum_{j=1}^k obj_{i,j}\Bigr)\otimes dev_2^0\otimes env$, so that at the moment $t_2$ when the position on the screen is measured, \par
$univ(t_2)=\di\sum_{i=1}^n \sum_{j=1}^k mind_j\otimes obj_{i,j}\otimes [dev_{2},env]_j$. \par
Here, the state of $\bf{dev}_2\otimes \bf{env}$ depends only on $j$, 
because the fact that $obj_{i,j}\in E_i$ is not pre-measured by any device, by any part of the environment. Only the information ``$obj_{i,j}\in F_j$'' gets entangled with $\bf{dev}_2$ and the environment. \par

Thus, within the initial coexistence of states, the $i$-th state moved from the state\par
$mind\otimes \di\frac{obj_i}{\norm{obj_i}}\otimes dev_2^0\otimes env
=\di\frac{1}{\norm{obj_i}}\sum_{j=1}^k mind\otimes obj_{i,j}\otimes dev_{2}^0\otimes env$
at the moment $t_0$, to the state 
$\di\frac{1}{\norm{obj_i}}\sum_{j=1}^kmind_j\otimes obj_{i,j}\otimes [dev_{2},env]_j$ at the moment $t_2$, independently of the evolution of the $i'$-th state for $i'\not=i$.\par

In the case of Young's double-slit experiment, the interference occurs between the two scenarios mentioned above,
however, we cannot describe them until the end of the experiment, unless we integrate the observer, the devices and the environment into each scenario. Then the first scenario, at the end of the experiment, becomes a coexistence of universes in which the observer measures the particle in position $j$ on the screen while the particle has gone through the first slit, this last information being neither entangled with $\bf{dev}_2\otimes \bf{env}$, nor with the mind. 
 It interferes with the second scenario because this coexistence of $n=2$ scenarios can be written as another coexistence: \par
$\eqalign{univ(t_2)&=\di\sum_{j=1}^k\sum_{i=1}^n mind_j\otimes obj_{i,j}\otimes [dev_{2},env]_j\hfill\cr
&=\!\!\!\!  \!\!\!\! \di\sum_{1\leq j\leq k\atop  \hbox{\tiny\ such that } \alpha_j\not=0}\!\!\!\!  \!\!\!\! \alpha_j. mind_j\otimes \Bigl(\frac1{\alpha_j}\sum_{i=1}^n obj_{i,j}\Bigr)\otimes [dev_{2},env]_j,
\hbox{ where }
\alpha_j=\di\norm{\sum_{i=1}^n obj_{i,j}}.\hfill\cr}$\par
Thus, the non-unicity of the way the universe can be interpreted as a coexistence plays a fundamental role in the phenomenon of interference. 

\par\medskip
 Mathematically, the presence of interference comes from the non-nullity of the quantity $\di\sum_{i_1\not= i_2}
\langle  obj_{i_1,j} | obj_{i_2,j}\rangle  $, which in the case of Young's double-slit experiment, comes down to 
$\langle  obj_{2,j},obj_{1,j}\rangle  =\langle  p_{F_j}(obj_2),p_{F_j}(obj_1)\rangle  $.  As $obj_1\in E_1$ and $obj_2\in E_2$ are orthogonal,  the interference phenomenon is based on the non conservation of the orthogonality by the function $u_j\ : \ x\ass \di\frac{p_{F_j}(x)}{\norm{p_{F_j}(x)}}$. \par
 Besides, for the observer having experienced that the particle is in position $j$ on the screen (that $obj\in F_j$ in the general case), the universe is reduced at the moment $t_2$ to\par
$univ(t_2)=\di mind_j\otimes \Bigl(\frac1{\alpha_j}\sum_{i=1}^n obj_{i,j}\Bigr)\otimes [dev_{2},env]_j
=mind_j\otimes u_j(obj) \otimes [dev_{2},env]_j$, whereas at the moment $t_0$, for the same observer, 
$univ(t_0)=mind\otimes obj\otimes dev_2^0\otimes env$.  So $u_j$ is the operator that transforms $obj$ between moments $t_0$ and $t_2$, from the observer's point of view\footnote{When $p_{F_j}(obj)=0$, the observer does not experience $mind_j$ in any term of the coexistence, so it is not necessary to define $u_j(obj)$.}.  It is not unitary. 

\subsubsection{Second Generalization of the Born Rule\label{dec}}

In practice, it is difficult to prevent the environment from becoming entangled with
the different scenarios ``$obj\in E_i$''. In particular, in the case of Young's double-slit experiment, the environment is modified, even very slightly, according to 
 the slit through which the particle goes. 
To simplify, we idealize the situation by limiting entanglement with the environment to entanglement between $\bf{obj}$ and $\bf{dev}_1$, which we now assume activated. Under these conditions, at the beginning of the experiment, keeping general notations,
$univ(t_0)=mind\otimes  obj\otimes dev_1^0\otimes dev_2^0\otimes env$, then when the particle goes through the slits, 
$univ(t_1)=mind\otimes \Bigl(\di\sum_{i=1}^n obj_i\otimes dev_{1,i}\Bigr)\otimes dev_2^0\otimes env$.
Thus, $\bf{dev}_1$ performs a premeasurement of  the family $(E_1,\ldots,E_n)$. However, the observer does not proceed with the measurement; he does not observe the result given by $\bf{dev}_1$.
After the measure according to the family $(F_1,\ldots,F_k)$, we have \par
$univ(t_2)=\di\sum_{j=1}^k mind_j\otimes \Bigl(\sum_{i=1}^n obj_{i,j}\otimes dev_{1,i}\Bigr)\otimes
 [dev_{2},env]_j$.\par 

In this new context, the probability at the moment $t_0$ or $t_1$, that the observer experiences $mind_j$ at the moment $t_2$, that is, that he experiences the event $obj\in F_j$, is equal to
$Pr(obj\in F_j)
=\norm{\di\sum_{i=1}^n obj_{i,j}\otimes dev_{1,i}}^2$.
The calculation of this probability is always based on its definition in terms of observer proportions.
 This is the second generalization of Born rule.
We can also write: \par
$Pr(obj\in F_j)=\di\sum_{i=1}^n\norm{obj_{i,j}}^2+\sum_{i_1\not=i_2}
\langle  obj_{i_1,j} | obj_{i_2,j}\rangle    .  \langle  dev_{1,i_1} |  dev_{1,i_2} \rangle  $.
So the interference terms are now modulated by the quantities $\langle  dev_{1,i_1} |  dev_{1,i_2} \rangle  $.

\subsubsection{Decoherence}

In the idealized situation of the previous paragraph, between $t_0$ and $t_1$, the system $\bf{obj}\otimes\bf{dev}_1$ is isolated from the rest of the universe, then between $t_1$ and $t_2$, $\bf{dev}_1$ remains isolated while $\bf{obj}$ interacts with $\bf{dev}_2\otimes\bf{env}$. \par
We assume that between $t_0$ and $t_1$, the interaction dynamics  between $\bf{obj}$ and $\bf{dev}_1$ is defined by a Hamiltonian of the form $\widehat H =\di\Bigl(\sum_{i=1}^n \lambda_ip_{E_i}\Bigr)\otimes 
\widehat{H_{dev_1}}$, where $\lambda_i\not=\lambda_j$ if $i\not=j$. 
  The calculation developed page \pageref{13} adapts and shows that, if we start from an initial state equal to $obj_i\otimes dev_1^0$, where $obj_i\in E_i$, then after an interaction duration $t$, the state of 
$\bf{obj}\otimes \bf{dev}_1$ becomes $obj_i\otimes dev_{1,i}$, with 
$dev_{1,i}=e^{-\frac {it}{\hbar}\lambda_i\widehat{H_{dev_1}}}(dev_1^0)$.\par
The decoherence theory [M.S.1] shows that when $\bf{dev}_1$ is macroscopic, for realistic values of $\widehat{H_{dev_1}}$, 
for all $i_1,i_2\in \{1,\ldots,n\}$ with $i_1\not=i_2$, $\langle dev_{1,i_1} | dev_{1,i_2} \rangle $ tends toward 0 when $t$ tends toward infinity. Moreover, this convergence is extremely rapid, to the point that an interaction of less than $10^{-12}$ second is generally enough to make the quantity $|\langle dev_{1,i_1} | dev_{1,i_2} \rangle |$ indistinguishable from zero, in the context of an experimental estimate of
 $Pr(obj\in F_j)$ by relative frequencies. This phenomenon is called decoherence. It is omnipresent at macroscopic scale, even when the Hamiltonian of $\bf{obj}\otimes \bf{dev}_1$ also takes into account the intrinsic dynamics of $\bf{obj}$ and $\bf{dev}_1$.
It is likely that the decoherence phenomenon 
is even more pronounced in realistic non-idealized situations.
\par
The subspaces $E_i$ according to which decoherence occurs always consist of states $obj_i$ that change very little during an interaction with the environment [M.S.1, page 73]. Thus, the decoherence concerns precisely the families of subspaces $(E_1,\ldots,E_n)$ according to which human beings tend to measure an object (cf paragraph \ref{var}).\par\medskip

In case of decoherence due to the device $\bf{dev}_1$, we can write: \par
$Pr(obj\in F_j)\simeq\di\sum_{i=1}^n\norm{obj_{i,j}}^2
=\di\sum_{i=1}^n Pr(obj\in E_i) \times Pr(obj\in F_j|obj\in E_i)$.  For this reason, in the presence of decoherence, which is practically impossible to avoid when $\bf{obj}$ and $\bf{dev}_1$ are both macroscopic,  the equality $obj=\di\sum_{i=1}^n obj_i$ seems to have the statistical interpretation that the object $\bf{obj}$ is in one of the states $\di\frac{obj_i}{\norm{obj_i}}$ with a probability equal to $\norm{obj_i}^2$.  
\par
However, this interpretation is based on an equality that, strictly speaking, is false. The interference terms remain non-zero, and even very small they continue to forbid the previous interpretation.\par

With respect to  the calculation of the probabilities relative to a second measurement, 
the measurement according to $\bf{dev}_1$ leads to results that a human being cannot distinguish from those obtained during a premeasurement according to the same device $\bf{dev}_1$ if it is subjected to decoherence.
However, these two situations are of a different nature. Confusing both gives too much theoretical importance to decoherence in my opinion.\par
For example, in Everett interpretation as presented by David Wallace 
[D.W.2], decoherence is at the heart of the theory to explain probabilities and the emergence of  several worlds. In this article, on the contrary, the decoherence phenomenon  is not used to justify the notions of coexistence and probability.

\subsection{Density Matrix}

\subsubsection{Pure States}
Let us go back to the situation and the notations in paragraph \ref{partielle}.  \par
$E_i$ often refers to the set of states for which a certain quantity associated with the object has a well-defined value $u_i\in\Re$ ; it is displayed by the device when $obj\in E_i$. 
For a certain choice of units, this quantity can for example correspond to its energy, or to one  coordinate of its position or momentum. 
By construction, $u_i\not=u_j$ when $i\not=j$, so $\bf{dev}$
can simply give the value $u_i$ when $obj\in E_i$. The expectation of the result returned by the device when \bf{obj} is in the initial state $obj$ is, by definition, 
$(2)\ :\ E=\di\sum_{i=1}^n u_i Pr(obj\in E_i)=\di\sum_{i=1}^n u_i \norm{p_{E_i}(obj)}^2$.\par
Let $U$ be  the unique linear operator of ${\cal H}_{obj}$ such that, for all $i\in\{1,\ldots,n\}$ and $x\in E_i$, $U(x)=u_i x$. $U$ is a Hermitian operator which is called the observable of the quantity associated with $(u_i)_{1\leq i\leq n}$. \par\medskip

For any vector $x$ of ${\cal H}_{obj}$, we define its ``ket'': it is the function $|x\rangle  \ :\ \alpha\ass \alpha x$ from $\Ce$ to ${\cal H}_{obj}$. We also define its ``bra'': it is the linear operator
$\langle  x |\ : y\ass \langle  x | y\rangle   $ from
${\cal H}_{obj}$ to $\Ce$. So, if $x,y\in {\cal H}_{obj}$, 
the composite function $\langle  x| \circ |y\rangle   $ is equal to $\langle  x | y\rangle    Id_{\Ce}$, which we can safely equate with the complex $\langle x | y\rangle $, so that $\langle  x| \circ |y\rangle   =\langle  x | y\rangle   $ .
This equality justifies these notations a posteriori. \par\medskip

Let  $i$ be in $\{1,\ldots,n\}$. $p_{E_i}$ being a self-adjoint operator,\par
$\norm{p_{E_i}(x)}^2=\langle  x  | p_{E_i}(x)\rangle  
=Tr(\langle  x | \circ p_{E_i} \circ | x\rangle  )
=Tr(p_{E_i} \circ| x\rangle  \circ \langle  x |)$.\par
So, $E=Tr\Bigl(\Bigl(\di\sum_{i=1}^n u_i p_{E_i}\Bigr) | obj\rangle   \langle  obj|\Bigr)$,  then 
$$(3)\ :\ E=Tr(U\rho),$$
where $\rho=|obj\rangle  \langle  obj| $ is the orthogonal projector on the one-dimensional space\par
$\hbox{Vect}(obj)$. $\rho$ is also called the  density matrix of the state $obj$.\par\medskip
In particular, when $U=p_{E_j}$, where $j\in\{1,\ldots,n\}$, 
$Tr(p_{E_j}\rho)=\norm{p_{E_j}(obj)}^2$. Thus the relationship $(3)$ is equivalent to the first generalization of Born rule.

\subsubsection{Mixed States}
In the situation of the previous paragraph, we say that the state $obj$ is \textit{pure}, as opposed to the situation of paragraph \ref{dec} where, before the measurement by 
$\bf{dev}_2$, the object got entangled with a part of the environment, denoted  by $\bf{dev}_1$, which set 
$\bf{obj}\otimes\bf{dev}_1$ at the moment $t_1$ in the state
 $X=\di\sum_{i=1}^n obj_i\otimes dev_{1,i}$.  In this case, we saw page \pageref{intr} that $obj$ no longer exists in the universe $univ(t_1)$, but that $univ(t_1)$ is a coexistence of universes in each of which $\bf{obj}$ exists in the state $\di\frac{obj_i}{\norm{obj_i}}$. In this case, quantum mechanics textbooks say that $\bf{obj}$ is in a mixed state.\par\medskip

With the notations in paragraph \ref{dec}, let us associate to $\bf{dev}_2$ an observable $U$, defined by :
$\forall j\in\{1,\ldots,k\}$, $\forall x\in F_j$, $U(x)=u_j  x$, where $u_1,\ldots,u_k$ are pairwise distinct real numbers.
  Then the expectation of the result returned by $\bf{dev}_2$ is \par
$E=\di\sum_{j=1}^k u_j . Pr(obj\in F_j)=\di\sum_{j=1}^k u_j . 
\norm{\sum_{i=1}^n obj_{i,j}\otimes dev_{1,i}}^2$. \par
As $obj_{i,j}=p_{F_j}(obj_i)$ and $obj=\di\sum_{i=1}^n obj_i$,  we have
$\di \sum_{i=1}^n obj_{i,j}\otimes dev_{1,i}=
(p_{F_j}\otimes Id_{{\cal H}_{dev_1}})(X)$.\par

We have proved that $E=\di\sum_{j=1}^k u_j  \norm{p_{F_j}\otimes Id_{{\cal H}_{dev_1}})(X)}^2$, that is, if we compare with the relationship $(2)$, when we measure an object $\bf{obj}$ according to a family 
of subspaces $(F_1,\ldots, F_k)$ while $\bf{obj}$ was previously entangled with another object 
here denoted by $\bf{dev}_1$, we actually measure the object $\bf{obj}\otimes \bf{dev}_1$, which is the only that exists, according to the family 
 $(F_1\otimes{\cal H}_{dev_1},\ldots, F_k\otimes{\cal H}_{dev_1})$. We can also say that, 
when one measures an object $\bf{obj}$ according to an observable $U$, whereas $\bf{obj}$ is firstly entangled with another object $\bf{dev}_1$, one actually measures the object $\bf{obj}\otimes \bf{dev}_1$ according to the observable $U\otimes Id_{{\cal H}_{dev_1}}$.
\par\medskip
Onto the set of linear operators of ${\cal H}_{obj}\otimes {\cal H}_{dev_1}$, we define the notion of ``partial trace over ${\cal H}_{dev_1}$'' by the relationship: 
$Tr_{{\cal H}_{dev_1}}(A\otimes B)=Tr(B). A$.\par
According to the relationship $(3)$, $E=Tr([U\otimes Id_{{\cal H}_{dev_1}}]. [|X\rangle  \langle  X |])$, \par
so thanks to properties of partial trace, \par
$E=Tr\Bigl(Tr_{{\cal H}_{dev_1}}([U\otimes Id_{{\cal H}_{dev_1}}]. [|X\rangle  \langle  X |])\Bigr)
=Tr\Bigl(U \circ Tr_{{\cal H}_{dev_1}}(|X\rangle  \langle  X |)\Bigr)$.\par
Let $\rho=Tr_{{\cal H}_{dev_1}}(|X\rangle \langle X|)$. Then
the relationship $(3)$ is still valid.\par
$\rho=Tr_{{\cal H}_{dev_1}}(|X\rangle \langle X|)$ is called the density matrix of  the  mixed state $\bf{obj}$.\par\medskip
IP1 must not be generalized to a mixed state modelled by its density matrix;
When $\bf{obj}$ is in a mixed state, it has no existence of its own. The universe though is decomposed into a coexistence of universes in each of which $\bf{obj}$ actually exists in a pure state.  Thus, the density matrix of a mixed state is only a mathematical tool, quite practical because it makes the relationship $(3)$ also equivalent to the second generalization of the Born rule. 

\subsubsection{Density Matrix and Decoherence}
$X=\di\sum_{i=1}^n obj_i\otimes dev_{1,i}$, so \par
$\eqalign{|X\rangle  \langle  X|&=\di\sum_{1\leq i_1,i_2\leq n} 
[|obj_{i_1}\rangle  \otimes |dev_{1,i_1}\rangle  ]\circ [\langle  obj_{i_2} |\otimes \langle  dev_{1,i_2}|]
\hfill\cr
&=\di\sum_{1\leq i_1,i_2\leq n} [|obj_{i_1}\rangle  \circ \langle  obj_{i_2} |]\otimes
[|dev_{1,i_1}\rangle  \circ \langle  dev_{1,i_2}|].\hfill\cr}$\par
 So, according to the partial trace definition,\par
$\rho=\di\sum_{1\leq i_1,i_2\leq n}Tr(|dev_{1,i_1}\rangle  \circ \langle  dev_{1,i_2}|)\ .\ [|obj_{i_1}\rangle  \circ \langle  obj_{i_2} |]$,\par
 yet 
$Tr(|dev_{1,i_1}\rangle  \circ \langle  dev_{1,i_2}|)=Tr(\langle  dev_{1,i_2}|\circ |dev_{1,i_1}\rangle  )
=\langle  dev_{1,i_2}|dev_{1,i_1}\rangle  $,\par
 therefore 
$\rho=\di\sum_{i=1}^n |obj_i\rangle  \langle  obj_i|
+\sum_{i_1\not=i_2}\langle  dev_{1,i_2}|dev_{1,i_1}\rangle   . |obj_{i_1}\rangle  \langle  obj_{i_2}|$.\par
Let us study the case where $E_i$ are one-dimensional spaces. We choose for all\par
$i\in\{1,\ldots,n\}$ a unit vector $e_i$ in $E_i$ 
and we denote by $e=(e_1,\ldots,e_n)$ the orthonormal basis of ${\cal H}_{obj}$ thus constituted. For any $i$, there is $\alpha_i\in\Ce$ such that $obj_i=\alpha_i e_i$.\par
Then the matrix of $\rho$ in the basis $e$ has as $i$-th diagonal coefficient $|\alpha_i|^2$ and as\par
 $(i_1,i_2)$-th non-diagonal coefficient $\alpha_{i_1}\bar{\alpha_{i_2}}\langle  dev_{1,i_2}|dev_{1,i_1}\rangle  $.
This proves that the decoherence phenomenon occurs if and only if the non-diagonal coefficients of the density matrix are approximately zero.

\subsection{Probabilities and Frequencies}
\subsubsection{Non-Repeatable Events}
The notion of probability as defined in this article
 also applies to non-repeatable events. 
For example, for the observer  I am now at time $t_0$ in the state $mind$, the probability that it will rain tomorrow at time $t_1$ is equal to $|\alpha|^2$ if 
$univ(t_0)=mind\otimes(\alpha . env_1+\beta . env_2)$, where $env_1$ is the coexistence of all the states of the environment at the moment $t_0$ which will give at the moment $t_1$ a universe where it rains.  At time $t_1$, we have\par
$univ(t_1)=\di \alpha \sum_{i}\beta_i mind_{1,i}\otimes env_{1,i}
+\beta\sum_i\gamma_i mind_{2,i}\otimes env_{2,i}$, \par
with 
$1=\di\sum_i|\beta_i|^2=\di\sum_i|\gamma_i|^2$, where $env_{1,i}$ is a raining environment with the occurrence of some other events whose the observer is conscious. Starting from a single observer at the moment $t_0$, who  can be seen as a coexistence of as many identical observers as necessary, the proportion of those who will experience  rain is well $|\alpha|^2$. 
\par\medskip
Thus, for non-reproducible events, the calculation of probabilities is sometimes accessible thanks to quantum mechanics, however we then  have 
 no experimental way to verify this theoretical result.

\subsubsection{Weak Law of Large Numbers\label{loi}}
Fortunately, this is not the case for reproducible events; next I am proving that the probability of such an event is indeed approximated by its relative frequency of occurrence.
It has been pointed out in paragraph \ref{fre} that an experiment is never rigorously reproducible. Often the object is even changed from one experiment to another. 
To reproduce the measurement of the object $\bf{obj}$ under the conditions of paragraph \ref{partielle}, 
 we thus assume we have $N$ objects $\bf{obj}^k$.
  For any $k\in\{1,\ldots,N\}$, we assume that $\bf{obj}^k$ is premeasurable by a device $\bf{dev}$ (independent of $k$ only to simplify) according to a family  of subspaces $(E_i^k)_{1\leq i\leq n}$ such that ${\cal H}_{obj^k}=\di\bigoplus_{1\leq i\leq n}^{\bot}E_i^k$.\par
	Initially, we have
$univ(t_0)=\di mind\otimes \Bigl(\bigotimes_{1\leq k\leq N} obj^k\Bigr)\otimes dev^0\otimes env$. \par
We assume that for any $i\in\{1,\ldots,n\}$, the quantity $\norm{p_{E_i^k}(obj^k)}^2$ can be approximated by  $\norm{p_{E_i}(obj)}^2=p_i$. This is in fact the only condition required to consider $obj^k$  as a good reproduction of  $obj$.  This considerably extends the notion of reproducible event. \par

Let $obj_i^k=p_{E_i^k}(obj^k)$. So,
$\di \bigotimes_{1\leq k\leq N} obj^k=\bigotimes_{1\leq k\leq N} \sum_{i=1}^n obj_i^k$.
This tensor product of $N$ sums can be written as a sum of tensor products,  where each term is obtained by choosing in each sum $obj^k$ one of its terms
$obj_{g(k)}^k$,
 with $g(k)\in\{1,\ldots,n\}$, then forming their tensor product. So, if
${\cal F}(\{1,\ldots,N\},\{1,\ldots,n\})$ denotes the set of the functions from $\{1,\ldots,N\}$ to $\{1,\ldots,n\}$,
 we get\label{51} $(4)\ :\ \di \bigotimes_{1\leq k\leq N} obj^k
=\di\sum_{g\in {\cal F}(\{1,\ldots,N\},\{1,\ldots,n\})}\bigotimes_{1\leq k\leq N} obj_{g(k)}^k$.

Between $t_0$ and $t_1$, the $N$ objects are premeasured, at the same time or in any order, while the observer does not read the result (again for simplicity only).   \par

So, 
$univ(t_1)=mind\otimes 
\di\sum_{g\in {\cal F}(\{1,\ldots,N\},\{1,\ldots,n\})}\Bigl[\bigotimes_{1\leq k\leq N} obj_{g(k)}^k
\Bigr]\otimes [dev,env]_g$.\par
It is assumed for simplicity, that the device only provides  the observer  with the relative frequency of the event $obj^k\in E_{i_0}^k$, where $i_0$ is given in $\{1,\ldots,n\}$. \par
When the $N$ objects are in the state $\di\bigotimes_{1\leq k\leq N} obj_{g(k)}^k$, this frequency is equal to $\frac hN$, where $h$ is the cardinality of $\{k\in\{1,\ldots,N\}/g(k)=i_0\}$.\par
If $A$ is a set, its cardinality will be denoted by $\#A$. \par
Then, after measurement, we can write \label{50b}
the following relationship:\par
$(5)\ :\ univ(t_2)=\!\!\!\!  \!\!\!\! \di\sum_{f\in\{\frac hN/h\in\{0,\ldots,N\}\}}\!\!\!\!  \!\!\!\! 
mind_f\otimes \!\!\!\!  \!\!\!\! \di\sum_{g\in {\cal F}(\{1,\ldots,N\},\{1,\ldots,n\})\atop
\hbox{\tiny\ such that } h=\#\{k\in\{1,\ldots,N\}/g(k)=i_0\}}\!\!\!\!  \!\!\!\! 
\Bigl[\bigotimes_{1\leq k\leq N} obj_{g(k)}^k
\Bigr]\otimes [dev,env]_g$.\par

I fix $\epsilon > 0$. We want to calculate the probability that the relative frequency $f$ of the event ``$obj^k\in E_{i_0}^k$'' differs from $p_{i_0}$ by more than $\epsilon$ in absolute value.
It is by definition, among  all the minds initially in the same state $mind$, the proportion of those who experience a frequency $f$ such that $|f-p_{i_0}|\geq \epsilon$. This probability is well defined, even if the mind in the state $mind_f$ does not know the value of $p_{i_0}$.
\par

$\eqalign{\di Pr(|f-p_{i_0}|\geq \epsilon)
&=\di\sum_{f\in\{\frac hN/h\in\{0,\ldots,N\}\}\atop \hbox{\tiny\ such that }  |f-p_{i_0}|\geq \epsilon}
\norm{
 \di\sum_{g\in {\cal F}(\{1,\ldots,N\},\{1,\ldots,n\})\atop
\hbox{\tiny\ such that }  h=\#\{k\in\{1,\ldots,N\}/g(k)=i_0\}}
\Bigl[\bigotimes_{1\leq k\leq N} obj_{g(k)}^k
\Bigr]}^2\hfill\cr
&=\di \di\sum_{f\in\{\frac hN/h\in\{0,\ldots,N\}\}\atop \hbox{\tiny\ such that } |f-p_{i_0}|\geq \epsilon}
\di\sum_{g\in {\cal F}(\{1,\ldots,N\},\{1,\ldots,n\})\atop
\hbox{\tiny\ such that } h=\#\{k\in\{1,\ldots,N\}/g(k)=i_0\}}\prod_{k=1}^n p_{g(k)}.\hfill\cr
}$\par

To obtain once exactly all the functions $g$ in ${\cal F}(\{1,\ldots,N\},\{1,\ldots,n\}\})$  such that $ h=\#\{k\in\{1,\ldots,N\}/g(k)=i_0\}$, we start by choosing the part $A$ of 
$\{1,\ldots,N\}$ of cardinality $h$ whose elements are mapped by $g$ onto $i_0$, then we complete $g$ on $\bar A=\{1,\ldots,N\}\setminus A$ by any function from $\bar A$ to
$\{1,\ldots, n\}\setminus\{i_0\}$. So,\par
$\di Pr(|f-p_{i_0}|\geq \epsilon)
=\di\sum_{h\in\{0,\ldots,N\}\atop \hbox{\tiny\ such that }   |\frac h N-p_{i_0}|\geq \epsilon}
\sum_{A\subset\{1,\ldots,N\}\atop 
\hbox{\tiny\ such that }  \#A=h}\sum_{g\in{\cal F}(\bar A,\{1,\ldots,n\}\setminus\{i_0\})}p_{i_0}^h\prod_{k\in \bar A}p_{g(k)}$.\par

The argument used to get the relationship (4) shows that \par
$\di \sum_{g\in{\cal F}(\bar A,\{1,\ldots,n\}\setminus\{i_0\})}\ \ \prod_{k\in \bar A}p_{g(k)}
=\prod_{k\in \bar A}\ \ \sum_{i\in\{1,\ldots,n\}\setminus\{i_0\}}p_i=(1-p_{i_0})^{N-h}$, so \par
$\di Pr(|f-p_{i_0}|\geq \epsilon)
=\di\sum_{h\in\{0,\ldots,N\}\atop \hbox{\tiny\ such that }  |\frac h N-p_{i_0}|\geq \epsilon}\binomial Nh
p_{i_0}^h(1-p_{i_0})^{N-h}$.\par
 This result was expected. Indeed, we have shown that \par
$\binomial Nh
p_{i_0}^h(1-p_{i_0})^{N-h}=\norm{
 \di\sum_{g\in {\cal F}(\{1,\ldots,N\},\{1,\ldots,n\})\atop
\hbox{\tiny\ such that } h=\#\{k\in\{1,\ldots,N\}/g(k)=i_0\}}
\Bigl[\bigotimes_{1\leq k\leq N} obj_{g(k)}^k
\Bigr]}^2$, yet that last expression represents the probability that the event 
``$obj^k\in E_{i_0}^k$'' occurs exactly $h$ times, among $N$ trials that do not interact with each other, given that each of these events occurs with a probability $p_{i_0}$. We have therefore  just proved, into quantum formalism, that the number of successes of $N$ independent Bernoulli trials,  whose probability of success is $p_{i_0}$, is a random variable whose distribution is binomial with parameters  $N$ and $p_{i_0}$.  We can now write:\par
$\eqalign{\di Pr(|f-p_{i_0}|\geq \epsilon)&\leq
\di\sum_{h\in\{0,\ldots,N\}\atop \hbox{\tiny\ such that }  |\frac h N-p_{i_0}|\geq \epsilon}
\frac{(\frac hN-p_{i_0})^2}{\epsilon^2}
\binomial Nh
p_{i_0}^h(1-p_{i_0})^{N-h}\hfill\cr
&\leq
\di\sum_{h=0}^N
\frac{(\frac hN-p_{i_0})^2}{\epsilon^2}
\binomial Nh
p_{i_0}^h(1-p_{i_0})^{N-h}.\hfill\cr}$\par

If $P$ is a real polynomial, we define its Bernstein polynomial: \par
$B_N(P)(X)=\di\sum_{h=0}^N\binomial Nh P({h\over N})X^h(1-X)^{N-h}$.
So,
 $\di Pr(|f-p_{i_0}|\geq \epsilon)\leq \frac1{\epsilon^2}B_N(Q)(p_{i_0})$ with $Q=(X-p_{i_0})^2=X^2-2p_{i_0}X+p_{i_0}^2$.
 We have\par
$\eqalign{\di{X(1-X)\over N}B'_N(P)&\di=
\sum_{h=0}^N\binomial{N}{h}P({h\over N})X^h(1-X)^{N-h}
{h(1-X)-(N-h)X\over N}\hfill\cr
&\di=B_N(XP)-XB_N(P),\hfill\cr}$\par
so $\di B_N(XP)={X(1-X)\over N}B_N'(P)+XB_N(P)$.

As 
$B_N(1)=1$, 
$B_N(X)=X$, then 
$B_N(X^2)=\di X^2(1-{1\over N})+{X\over N}$. \par
Thus, 
$B_N(Q)(p_{i_0})=\di\frac1Np_{i_0}(1-p_{i_0})$. \par
So we have shown that 
$\di Pr(|f-p_{i_0}|\geq \epsilon)\leq \frac1{N\epsilon^2}$. \par
Notably,  
$Pr(|f-p_{i_0}|\geq \epsilon)$ tends toward 0 when $N$ tends toward $+\infty$
\footnote{This is even an exponential convergence, because one can show that 
$\di Pr(|f-p_{i_0}|\geq \epsilon)\leq 2 e^{-2\epsilon^2N}$  [A.A.2, page 6].}. \par
On the mathematical level the arguments used are well known. This amounts to proving, within quantum formalism, a version of the weak law of large numbers. 

\subsubsection{Confidence Interval\label{conf}}
In the context of this article, this result is interpreted as follows; in the coexistence\par
$univ(t_2)=\di\sum_{f\in\{\frac hN/h\in\{0,\ldots,N\}\}}\!\!\!\!  
mind_f\otimes\!\!\!\!  \!\!\!\!  \di\sum_{g\in {\cal F}(\{1,\ldots,N\},\{1,\ldots,n\})\atop
\hbox{\tiny\ such that } h=\#\{k\in\{1,\ldots,N\}/g(k)=i_0\}}\!\!\!\!  \!
\Bigl[\bigotimes_{1\leq k\leq N} obj_{g(k)}^k
\Bigr]\otimes [dev,env]_g$,\par
 once reduced to a uniform coexistence that allows counting, if we consider the proportion of states in which the mind experiences a frequency value $f$ that differs from $p_{i_0}$ by more than $\epsilon$ in absolute value, this proportion can be made as small as we want by setting $N$ large enough. The observer at the moment $t_2$ can then with some confidence consider that the value $f$ he has  measured is a good approximation of $p_{i_0}$. \par\medskip

To define this new notion of confidence, we consider a more general situation.
At the moment $t_0$, we reduce the universe to the term that is experienced by an observer in a state of consciousness $mind$ : $univ(t_0)=mind\otimes env$.\par
At a later moment $t_2$, the observer's mind and the environment are entangled:\par
$univ(t_2)=\di\sum_{k\in K} \alpha_k mind_k\otimes env_k$. In each term, the observer has made measurements on the environment, thus retrieving some information.\par
If $A$ is a subset of $K$, the probability that the observer experiences a state of mind $mind_k$ such that $k\in A$ is
$Pr(k\in A)=\di\sum_{k\in A}|\alpha_k|^2$.\par
It is assumed that the set $A$ is not known to the observer. This is the case in the previous situation, where $A=\{f\in\{\frac hN/h\in\{0,\ldots,N\}\}/|f-p_{i_0}|<\epsilon\}$ with a probability $p_{i_0}$ that the observer does not know and is trying to evaluate.

Let $d$ be in $[0.1]$. It is assumed that 
$Pr( k\in A)\geq d$. Then among all the observers initially in the state $mind$, the proportion of those who at the moment $t_2$ experience a value $k$ such that $k\in A$ is greater than $d$. 
If the observer knows that $Pr( k\in A)\geq d$, he is entitled to claim that he is at the moment $t_2$ in a state of mind $mind_k$ such that $k\in A$, ``with a confidence level $d$'', according to a terminology used in statistics, therefore
at the moment $t_2$ he is entitled to say that ``$k\in A$, with a confidence level $d$''. \par\medskip

If we go back on the approximation of $p_{i_0}$ from the frequency $f$, by choosing $N$ such that 
$ \di\frac{1}{N\epsilon^2}\leq 0,05$, the observer at the moment $t_2$ can claim that 
$|f-p_{i_0}|< \epsilon$ with a confidence level equal to  95\%, so that 
$p_{i_0}\in [f-\epsilon,f+\epsilon]$, with the same confidence level; since the observer knows the value of $f$, he thus obtains a confidence interval for $p_{i_0}$. 
Of course $p_{i_0}$ \it{might} not be  in that interval.
 For example, if a player rolls a 6-sided dice 50 times and gets $24$ times the result $6$, he can confidently claim that the dice is crooked and that the probability that the result is $6$ on each roll is close to $\frac12$. Then he  takes the risk of making a wrong decision because it is not impossible, with a straight dice, to get 24 times the result 6.

\subsubsection{Temptation of Frequentism}
Variations of the calculation in paragraph \ref{loi} are sometimes used to justify a frequentist \textit{definition} of probabilities in quantum mechanics. Peter Mittelstaedt [P.M, pages 48-57 and 125-129] developed it extensively and it is mentioned by Brian Greene [B.G page 465]. 
However any argument to show that the relative frequency is equal to $p_{i_0}$ when $N$ tends towards infinity comes up against the meaning of taking the limit. Since the value $N=+\infty$ is not accessible to human beings, if $N$ is the number of experiments performed, we have in fact to show that, when $N$ is large, the frequency $f$ approaches properly $p_{i_0}$. So any demonstration of this kind amounts to neglecting, when $N$ is large, the terms of the relationship (5) page \pageref{50b}
for which $f$ is too far from $p_{i_0}$, that is for which $|f-p_{i_0}|\geq \epsilon$, where $\epsilon$ is chosen as small as we want. So we would have to neglect the vector\par
$V_N=\!\!\!\!  \!\!\!\!\di\sum_{f\in\{\frac hN/h\in\{0,\ldots,N\}\}\atop \hbox{\tiny\ such that }  |f-p_{i_0}|\geq \epsilon}\!\!\!\!  \!\!\!\!
mind_f\otimes\!\!\!\!  \!\!\!\! \di\sum_{g\in {\cal F}(\{1,\ldots,N\},\{1,\ldots,n\})\atop
\hbox{\tiny\ such that } h=\#\{k\in\{1,\ldots,N\}/g(k)=i_0\}}
\Bigl[\bigotimes_{1\leq k\leq N} obj_{g(k)}^k
\Bigr]\otimes [dev,env]_g$, when $N$ is large, on the pretext that its norm tends towards 0. Nethertheless there is no reason to neglect these terms into the coexistence. They contain conscious states of  mind, for which precisely the frequency differs from the probability; the existence of these terms bans a frequentist definition.\par\medskip
Moreover, considering that certain terms of a coexistence do not exist as soon as their norm is below a certain threshold $\eta>0$ leads to a contradiction ; 
Indeed, in the coexistence $(5)$, the norm of each term 
$\di mind_f\otimes 
\Bigl[\bigotimes_{1\leq k\leq N} obj_{g(k)}^k
\Bigr]\otimes  [dev,env]_g$ is equal to  
$\di\sqrt{\prod_{1\leq k\leq N}p_{g(k)}}$, which is lower than
 $p^{N/2}$, where $p=\di\max_{1\leq i\leq n}p_i$. In general, $p<1$, so when  $N$ is large enough, the norm of each term involved in the coexistence $(5)$ is less than $\eta$. Then none of these terms would exist and the universe as a whole would not exist.

\section{Making a C-bit\label{realiste}} 
In this chapter, we further detail a possibility of biological implementation of c-bits, as described on page \pageref{33}. This   highlights some practical difficulties raised by H3 and we  see how to overcome them.

\subsection{Two Spin $\frac12$ Particles}
Let us denote by $\bf{cb}$ the c-bit we wish to build. If ${\cal H}_{cb}$ was infinite-dimensional,
the conscious states $f_0$ and $f_1$ would be part of a basis $(f_i)_{i\in I}$ with $I$ uncountable, according to Hilbert space completeness [D.M, page 133]. We cannot see how the c-protein could then control the state of \bf{cb} with the infinite precision required to select $f_0$ or $f_1$ among $(f_i)_{i\in I}$. In order to impose ${\cal H}_{cb}$ as being finite-dimensional, we are going to use spin of particles.\par\medskip
If $\bf{pa}$ is a spin $\frac12$ particle, for example an electron, we can write 
$\bf{pa}=\bf{p}\otimes \bf{s}$,  where 
 $\bf{p}$ is the  position of the particle and $\bf{s}$ is its spin.  If  $(f_0,f_1)$ denotes an orthonormal basis of ${\cal H}_s$, then the general form of the wave function of \bf{pa} is a function \par
$pa \ :\ x\ass p_0(x)f_0+p_1(x)f_1$ from $\Re^3$ to ${\cal H}_s$ such that \par
$1=\norm{pa}^2=\int_x|p_0(x)|^2dx+\int_x|p_1(x)|^2dx$. Thus $pa=p_0\otimes f_0+p_1\otimes f_1$.\par

If we take now two discernible spin $\frac12$ particles $\bf{pa}_1$ and $\bf{pa}_2$, whose 
positions are denoted by 
$\bf{p}_1$ and $\bf{p}_2$ and whose spins are denoted by $\bf{s}_1$ and $\bf{s}_2$, \par
 we have 
${\cal H}_{pa_1\otimes pa_2}=[{\cal H}_{p_1}\otimes{\cal H}_{p_2}]
\otimes [{\cal H}_{s_1}\otimes{\cal H}_{s_2}]$. \par
Yet according to Clebsch-Gordan theorem [C.L.1, page 242], [J.H.1, page 131], if for any integer or half-integer 
$j$, 
$D_j$ denotes the unitary irreducible   $(2j + 1)$-dimensional representation of the group $SU(2)$, then
$D_{j_1}\otimes D_{j_2}=\di\bigoplus^{\bot}_{|j_1-j_2|\leq  j\leq j_1+j_2}D_j$, so  we have \footnote{In the notation $D_j^k$, the exponent $k$ allows us to consider several irreducible representations of spin $j$, isomorphic  but physically different.}\par
${\cal H}_{s_1}\otimes{\cal H}_{s_2}=D_{\frac12}^1\otimes D_{\frac12}^2
=D_0\build{\oplus}_{}^{\bot} D_1$. \par
Let $(f_{0.0})$ be an orthonormal basis of $D_0$ and $(f_{1,j})_{0\leq j\leq 2}$ an orthonormal basis of $D_1$. \par
Then  the wave function of these two particles is a function from $(\Re^3)^2$ to
$D_0\build{\oplus}_{}^{\bot} D_1$ of the form : 
$\varphi\ :\ x=(x_1,x_2)\ass p_{0,0}(x)f_{0,0}+p_{1,0}(x)f_{1,0}
+p_{1,1}(x)f_{1,1}+p_{1,2}(x)f_{1,2}$\par
 such that
$1=\norm{\varphi}^2=\int_x[|p_{0,0}(x)|^2+|p_{1,0}(x)|^2+|p_{1,1}(x)|^2+|p_{1,2}(x)|^2]dx$.\par
Thus,  $\varphi=p_{0,0}\otimes f_{0,0}+p_{1,0}\otimes f_{1,0}+p_{1,1}\otimes f_{1,1}+p_{1,2}\otimes f_{1,2}$.
\par\medskip

Suppose temporarily that $\bf{cb}=\bf{s}_1\otimes\bf{s}_2$ and that the c-protein constantly imposes rotational movements on its c-bit.
 Let $R_{t_1\rightarrow t_2}$ be the rotation thus performed by the c-bit between $t_1$ and $t_2$. 
According to [J.H, page 69], 
$R_{t_1\rightarrow t_2}(f_{0,0})=f_{0,0}$ whereas, for all $j\in\{0.1.2\}$, 
$R_{t_1\rightarrow t_2}(f_{1,j})$ continuously varies over $D_1$.
So the state $f_{0.0}$, called singlet, behaves like a conscious state and the states in  $D_1$, called triplets, are compatible with unconscious states.\par

For example, these two particles could be electrons, each bound to a proton to form a hydrogen atom. When they are separated, if they are subjected to different rotations, they become discernible and the state of $\bf{cb}$ can then be any unit vector of ${\cal H}_{cb}$. On the contrary,  if the c-protein 
can combine the two hydrogen atoms into a dihydrogen molecule placed in its fundamental state, then the wave functions of $\bf{p}_1$ and $\bf{p}_2$ are equal to the only molecular orbital available [C.L.1, pages 360 and 384] and the two electrons become
 indiscernible : According to Pauli exclusion principle, the wave function of $\bf{pa}_1\otimes \bf{pa}_2$ is antisymmetric. It imposes a singlet state on the spin of $\bf{pa}_1\otimes \bf{pa}_2$, so with our notations, $cb$ remains equal to $f_{0.0}$, as long as the c-protein ensures the cohesion of the $H_2$ molecule. $cb$ would then be a conscious state. 

\subsection{Four Spin $\frac12$ Particles}

This model of c-bit has only one conscious state whereas we need
two orthogonal conscious states. Therefore our final version of c-bit will be built with four discernible spin $\frac12$ particles
$(\bf{pa}_i)_{1\leq i\leq 4}$.  Let
 $\bf{p}_{ens}=\di\bigotimes_{i=1}^4\bf{pa}_i$.\par
 By adapting the previous notations, $\bf{p}_{ens}=\Bigl[\di\bigotimes_{i=1}^4\bf{p}_i\Bigr]
\otimes\Bigl[\di\bigotimes_{i=1}^4\bf{s}_i\Bigr]$.\par
The c-bit now consists of the spins of the four particles: $\bf{cb}=\di\bigotimes_{i=1}^4 \bf{s}_i$. \par
Thus, 
${\cal H}_{cb}=(D_0^1\build{\oplus}_{}^{\bot} D_1^1)\otimes(D_0^2\build{\oplus}_{}^{\bot} D_1^2)$. 
\par
Using distributivity and Clebsch-Gordan theorem, we get\par
  ${\cal H}_{cb}=
D_0^3\build{\oplus}_{}^{\bot} D_1^3\build{\oplus}_{}^{\bot} {D}_1^4\build{\oplus}_{}^{\bot} (D_1^1\otimes D_1^2)=D_0^3\build{\oplus}_{}^{\bot} D_1^3\build{\oplus}_{}^{\bot} {D}_1^4\build{\oplus}_{}^{\bot}
 D_0^4\build{\oplus}_{}^{\bot} D_1^5\build{\oplus}_{}^{\bot} D_2$.\par

By renaming $D_0^3=F_0$, $D_0^4=F_1$, $D_1^3=F_2$, $D_1^4=F_3$, $D_1^5=F_4$ and $D_2=F_5$, we get  ${\cal H}_{cb}=\di\bigoplus^{\bot}_{0\leq i\leq 5}F_i$ where $F_0$ and $F_1$ are one-dimensional, $F_2, F_3$ and $F_4$ are 3-dimensional and $F_5$ is 5-dimensional. 
 For any $i\in\{0,\ldots,5\}$, let $(f_{i,j})_{0\leq j\leq j_i}$ be  an orthonormal  basis of $F_i$, where 
$j_i=\left\{\matrix{0&\hbox{ if }&i\in\{0,1\}\cr
2&\hbox{ if }&i\in\{2,3,4\}\cr
4&\hbox{ if }&i=5\cr}\right.$. 
Thus, the wave function of $\bf{p}_{ens}$ has the following form:
$\varphi\ :\ x=(x_k)_{1\leq k\leq 4}\ass \di\sum_{0\leq i\leq 5\atop 0\leq j\leq j_i}p_{i,j}(x)f_{i,j}$
from $(\Re^3)^4$ to ${\cal H}_{cb}$,\par
 such that
$1=\norm{\varphi}^2=\di\int_x
\sum_{0\leq i\leq 5\atop 0\leq j\leq j_i}|p_{i,j}(x)|^2
dx$. So, $\varphi=\di\sum_{0\leq i\leq 5\atop 0\leq j\leq j_i}p_{i,j}\otimes f_{i,j}$.
\par

For all $i,j$, let $\alpha_{i,j}=\sqrt{\int_x|p_{i,j}(x)|^2dx}$ and if $\alpha_{i,j}\not=0$, 
$q_{i,j}=\di\frac{1}{\alpha_{i,j}}p_{i,j}$. When $\alpha_{i,j}=0$, we let $q_{i,j}$ be any function from $(\Re^3)^4$ to $\Ce$ such that $\int_x|q_{i,j}(x)|^2dx=1$. One can then write that 
$\varphi=\di\sum_{0\leq i\leq 5\atop 0\leq j\leq j_i}\alpha_{i,j}q_{i,j}\otimes f_{i,j}$, where  $q_{i,j}$ are position wave functions of the 4 particles system. \par\medskip
Let $G=\di\bigoplus^{\bot}_{2\leq i\leq 5} F_i$, so that ${\cal H}_{cb}=F_0\build{\oplus}_{}^{\bot} F_1\build{\oplus}_{}^{\bot} G$.\par
We assume again the c-protein constantly imposes rotational movements on its c-bit. 

This time, $f_{0.0}$ and $f_{1.0}$ are invariant by the rotation $R_{t_1\rightarrow t_2}$ and for any state $g$ in $G$, 
$R_{t_1\rightarrow t_2}(g)$ continuously varies over $G$. We then have two conscious states $f_{0.0}$ and $f_{1.0}$ and all states in $G$ behave as unconscious states. \par

According to $H3$, the c-protein writing process on a c-bit takes less than
$\tau\simeq 1$ms at the beginning of a conscious period lasting $T\simeq 50$ms. It consists in setting the c-bit in one of the two states $f_{0.0}$ or $f_{1.0}$, i.e. imposing, according to a mechanism that remains to be specified, this 4 particles set to be seen by the rest of the universe as a single spin 0 particle, of a different nature according to whether it is $f_{0.0}$ or $f_{1.0}$, until the c-protein performs a new writing. 
Then, during the next $\tau$ transition period, the four particles are manipulated independently of each other by the c-protein, which allows the c-bit state to vary over the entire space
${\cal H}_{cb}$.\par

\subsection{Initialization of a C-bit on a Conscious State}
Ideally, the c-protein should be able to set the c-bit exactly in the state $f_{0.0}$ or $f_{1.0}$ at the beginning of a conscious period.  In this paragraph, we are seeing that this constraint can be released and it is enough that the c-bit state approximates $f_{0.0}$ or $f_{1.0}$.\par\medskip

Let us start by writing the general state of the universe in a form that takes into account the presence or absence of $\bf{mind}$, consisting of the spinorial part of $4N$ particles; 
$univ=\beta univ_0+\gamma univ_1$, where $univ_0$ is 
 a coexistence of states of the universe in which the 4N particles exist, where $univ_0$ and $univ_1$ are  orthogonal, and where $univ_1$ is a coexistence of states where at least one of these 4N particles does not exist, which corresponds either to universes in which our observer has never existed, or to universes in which the mind of our observer has just lost a particle by disintegration (whereas these particles are very stable) thus where undoubtedly a biological process of repair is taking place.  We have $|\beta|^2+|\gamma|^2=1$.\par

$univ_0$ evolves in a subspace of ${\cal H}_{univ}$ denoted ${\cal H}'_{univ}$ which can be decomposed into the form ${\cal H}'_{univ}=\bf{mind}\otimes \bf{pos}\otimes\bf{env}$, where $\bf{pos}$ stands for the position of the $4N$ particles of \bf{mind} and where $\bf{env}$ stands for the universe except for the $4N$ particles. 
We have $\bf{mind}\otimes\bf{pos}=\di\bigotimes_{1\leq h\leq N}\bf{p}_{ens}^h$, where $\bf{p}_{ens}^h$ represents the 4 particles system whose $h$-th c-bit is the spinorial part. 
So, we can write $univ_0$ in the form \par
$(6)\ :\ univ_0=\di\sum_b\beta_b env_b\otimes \di\bigotimes_{1\leq h\leq N}\varphi_b^h$, where $\varphi_b^h$ is the wave function of 
$\bf{p}_{ens}^h$, which depends by entanglement on the state $env_b$ of the environment.\par

With obvious notations, one can write :
$(7)\ :\ \varphi^h_b=\di\sum_{0\leq i\leq 5\atop 0\leq j\leq j_i}\alpha_{i,j,b}^h\ q_{i,j,b}^h\otimes f_{i,j}^h$.  \par
Let
$univ_b=env_b\otimes\di\bigotimes_{1\leq h\leq N}\varphi^h_b$, so that 
$univ_0=\di\sum_b\beta_b univ_b$.
\par
Let us order the couples $(i,j)$ according to the lexicographical order from the smallest, $(0,0)$, to the largest, 
$(5,4)$, and denote them in that order $r(0),\ldots,r(15)$.\par
 So 
$univ_b=env_b\otimes \di\bigotimes_{1\leq h\leq N}\di\sum_{k=0}^{15}\alpha_{r(k),b}^h\ q_{r(k),b}^h\otimes f_{r(k)}^h$. \par
 The argument used to get the relationship (4) page \pageref{51} shows that \par
$univ_b=env_b\otimes\!\!\!\!  \!\!\!\! \di\sum_{a\in{\cal F}(\{1,\ldots,N\},\{0,\ldots,15\})}\ \bigotimes_{1\leq h\leq N}
\alpha_{r(a(h)),b}^h\ q_{r(a(h)),b}^h\otimes f_{r(a(h))}^h$.
So, \par
$univ_0=\di\sum_b\!\!\sum_{a\in{\cal F}(\{1,\ldots,N\},\{0,\ldots,15\})}\!\!\!\!\!\!\beta_b
\Bigl[\prod_{h=1}^N\alpha^h_{r(a(h)),b}\Bigr].\Bigl[\bigotimes_{1\leq h\leq N}f_{r(a(h))}^h\Bigl]
\otimes\Bigl[env_b\otimes\bigotimes_{1\leq h\leq N}q_{r(a(h)),b}^h\Bigl]$.\par
Then, \par
$(8)\ :\ univ_0=\!\!\!\!\!\!\di\sum_{a\in{\cal F}(\{1,\ldots,N\},\{0,\ldots,15\})}\!\!\!\!\!\!mind_a\otimes\sum_b\beta_b
\Bigl[\prod_{h=1}^N\alpha^h_{r(a(h)),b}\Bigr]
\otimes\Bigl[env_b\otimes\bigotimes_{1\leq h\leq N}q_{r(a(h)),b}^h\Bigl]$,\par
 where
$mind_a=\di \bigotimes_{1\leq h\leq N}f_{r(a(h))}^h$. This is  the most general expression of the state of the universe, according to  $\bf{mind}$.
\par\medskip

Let us now examine how the mind moves from a conscious state to a new one.
Let $c$ be a function from $\{1,\ldots,N\}$ to $\{0,1\}$. The conscious state of mind corresponding to $c$ is 
$mind_c=\di\bigotimes_{1\leq h\leq N}f^h_{c(h),0}$. Thus, for  $\bf{mind}$ in the conscious state $mind_c$, during a conscious period from the moment $t_0$ to the moment $t_0+T$, the universe is reduced to the terms for which the $4N$ particles of its mind exist, and where for any $h\in\{1,\ldots,N\}$, 
$r(a(h))=(c(h),0)$.
 So for $t\in [t_0,t_0+T]$, the universe at the moment $t$ according to \bf{mind} in the state $mind_c$ is\par
$univ(t)=mind_c\otimes \di\sum_b\delta\beta_b\Bigl[\prod_{h=1}^N\alpha^h_{c(h),0,b}\Bigr]\otimes
\Bigl[env_b(t)\otimes\bigotimes_{1\leq h\leq N}q_{c(h),0,b}^h(t)\Bigl]=mind_c\otimes env'_c(t)$
 where 
$\delta$ is a normalizing factor ensuring that $\norm{univ(t)}=1$.\par
\medskip

$env'_c$ is a coexistence of environments that will each lead the brain to a particular pattern of neural activity, when $t>t_0+T$, and the c-bits of the mind to a particular state. The principle of linearity allows us to replace $env'_c$ by the coexistence of the only environments that will seek, by the transmission of an adequate signal to the c-proteins, to lead the mind towards a given conscious state, defined by a second function $d$ from $\{1,\ldots,N\}$ to $\{0,1\}$. This reduction resulted in
$univ=mind_c\otimes env'_d$;
at the end of the conscious period corresponding to $mind_c$, under the influence of $env'_d$, for every $h\in\{1,\ldots,N\}$, the $h$-th c-protein will attempt to set its c-bit in the state $f_{d(h),0}^h$, 
a priori only when $d(h)\not=c(h)$.
\par

The transition from $f_{c(h),0}^h$ to $f_{d(h),0}^h$ is made during the next $\tau$ transition period, while $\bf{cb}^h$ evolves into intermediate states belonging to ${\cal H}_{{cb}^h}\setminus(F_0^h\oplus F_1^h)$, which vary under the effect of the rotation imposed on the c-bit, and are therefore unconscious. 
At the atomic scale, a millisecond is a long time, during which many events can occur locally and become entangled between them. So, at the end of the transition between the two conscious states, the universe is  in a general state
$univ=\beta univ_0+\gamma univ_1$,  using the previous notations;
$univ_1$ represents the coexistence of the states of the universe in which, whereas a millisecond before, the $4N$ particles of \bf{mind} existed, at least one of them spontaneously disintegrated. The scarcity of such an event translates into the fact that $\gamma$ has a very small modulus.  Ideally, we take $\gamma=0$, and more realistically, we assume that $|\gamma|<10^{-20}$.\par

Now let us focus on $univ_0$, which at the moment $t_1=t_0+T+\tau$ is in the form $(8)$.
Ideally, each c-protein manages to set its c-bit exactly in the desired state $f_{d(h),0}^h$ at the moment $t_1$.
According to the relationships $(6)$ and $(7)$, this amounts to assuming that after the unconscious transition period, for all $b$ and for all $h\in\{1,\ldots,N\}$, $\varphi_b^h=q_{d(h),0,b}^h\otimes f_{d(h),0}^h$. So, ideally, \par
$univ(t_1)=mind_d\otimes \di\sum_b\beta_b . \Bigl[env_b(t_1)\otimes\bigotimes_{1\leq h\leq N}q_{d(h),0,b}^h(t_1)\Bigr]=mind_d\otimes env'_d(t_1)$, 
which then evolves when $t\in [t_1,t_1+T]$ in the form 
$univ(t)=mind_d\otimes env'_d(t)$, where $mind_d$ remains constant, independent of $t$.
\par\medskip

More realistically, at the moment $t_1$, the $h$-th c-protein only manages to set its c-bit in a state very close to $f_{d(h),0}^h$. According to the relationships (6) and (7), this amounts to assuming that after the unconscious transition period, for all
 $h\in\{1,\ldots,N\}$ and for all $b$,  $|\alpha_{d(h),0,b}^h|>1-\epsilon$, where $\epsilon$ is small enough.\par Specifically, we will assume that 
$\epsilon\leq 10^{-20}$.\par
Then $univ(t_1)=\beta univ_0(t_1)+\gamma univ_1$ with 
$univ_0(t_1)=univ_d+univ'_d$, where according to (8)
$univ_d=
\di\sum_b\beta_b\Bigl[\prod_{h=1}^N\alpha_{d(h),0,b}^h\Bigr]mind_d\otimes 
\Bigl[env_b\otimes\bigotimes_{1\leq h\leq N}q_{d(h),0,b}^h\Bigr]=mind_d\otimes env''_d$ and\par
$univ'_d= \di\sum_{a\in{\cal F}(\{1,\ldots,N\},\{0,\ldots,15\}\atop  \hbox{\tiny\ such that }
r\circ a\not=[h\ass (d(h),0)]}mind_a\otimes \sum_b\beta_b
\Bigl[\prod_{h=1}^N\alpha^h_{r(a(h)),b}\Bigr] . 
env_b\otimes\bigotimes_{1\leq h\leq N}q_{r(a(h)),b}^h$.\par
$univ_d$ and $univ'_d$ are orthogonal  because, for all states  $mind_a$ involved in $univ'_d$, $mind_d$ and $mind_a$ are orthogonal.\par

Since $N$ is in the order of 20 billion, the entanglement between the $4N$ particles of the mind and the environment induces a decoherence phenomenon which allows us to affirm that the family $(env_b)_b$ is approximately orthonormal, this approximation being of very high quality. 
So according to $(6)$, $\di\sum_b|\beta_b|^2\simeq 1$ and \par
$1\geq \norm{env''_d}^2\simeq\di\sum_b|\beta_b|^2\prod_{h=1}^N|\alpha_{d(h),0,b}^h|^2
\geq \di\sum_b|\beta_b|^2(1-\epsilon)^{2N}\simeq (1-\epsilon)^{2N}\simeq 1-2N\epsilon$,\par
 therefore 
$\norm{env''_d}^2\in[1-4.10^{-10},1]$. \par
Let $\eta=\norm{env''_d}$. Thus
$univ(t_1)=\beta\eta . mind_d\otimes env+\sqrt{1-(\beta\eta)^2} . univ'$,  where $univ'$ and $env$ are unit vectors and where $univ'$ and $mind_d\otimes env$  are orthogonal. \par
$univ(t_1)$ is thus a coexistence of universes among which the proportion of universes where $\bf{mind}$ is conscious and in the state of consciousness $mind_d$ is equal to $|\beta\eta|^2$, with
 $|\beta\eta|^2\geq (1-10^{-40})(1-4.10^{-10})\simeq 1-4.10^{-10}$.\par
The probability being always defined in terms of proportions, with the choices we have made for orders of magnitude, we have just shown that, for a mind in a conscious state $mind_c$ between $t_0$ and 
$t_0+T$, given that his environment should lead him at the moment $t_1=t_0+T+\tau$ to the state of consciousness $mind_d$, the probability that his c-proteins dysfunction and lead him to another state (probably unconscious) is less than $4. 10^{-10}$. If this happens, during $T=50ms$, the mind is unconscious or conscious of a fact unrelated to $mind_d$, which corresponds to a slight disorder of consciousness. 
This compares with the number $L$ of periods of consciousness that a human being experiences over a lifetime: $L$ is in the order of $4.10^{10}$. 
Thus, the average number of such disorders of consciousness over the entire existence of a human being is in the order of 
16. 

\subsection{Constancy of a Conscious State}
We have seen that  the state of the universe according to  $\bf{mind}$ in the state $mind_c$, at a moment $t_0$ that begins a period of consciousness, is of the form 
$univ(t_0)=mind_c\otimes env$.\par
$env$ is a coexistence of environments, for most of which, between $t_0$ and $t_0+T$, for all $h\in\{1,\ldots,N\}$, $\bf{p}^h_{ens}$  is actually seen as a single spin 0 particle, whose spinorial state remains constant and equal to 
$f_{c(h),0}^h$.  Meanwhile for other environments of this coexistence, in much rarer proportions, 
that is not the case.
So, at the moment $t_0+T$,
the universe is a coexistence of several universes, for most of which the state of \bf{mind} has remained constant and equal to $mind_c$ between $t_0$ and $t_0+T$, and where for some rare universes, a dysfunction has occurred for at least one of the $N$ c-bits. 
It is a second source of consciousness disorders. 
\par\medskip
In conclusion, to consider that H3 is well verified is to ignore universes, negligible in proportion, in which an error occurred when initializing the value of a c-bit at the beginning of a conscious period, or in which the conscious state did not remain constant for the required $T$ duration. \par

Born rule is valid under the assumption that 
 the observer's mind will not experience consciousness  disorders.  To guarantee this rationality condition, probability is evaluated by taking into account only universes in which, during  all periods of consciousness required for the measurement, all c-bits have remained constant and have been positioned on a correct conscious state at the beginning of the conscious phase. This is a somewhat particular form of conditional probability because it concerns the observer.  It is though consistent with the natural definition of conditional probabilities presented page
\pageref{condi}.
\newpage

\part{The Observer and Self-Consciousness}
\setcounter{section}{0}

In the first part of this paper, we have summarized the formalism of quantum mechanics without probability into three formal principles FP1, FP2 and FP3. We have provided an interpretation in six principles IP1 to IP6, based on the notion of coexistence. Within this framework, we have then modelled the conscious observer by making 
 three hypotheses H1, H2 and H3, which we have justified. 
 This allowed us to define the notion of probability and then to prove and explain the usual properties of quantum mechanics and probabilities. \par\medskip

This approach sheds a particular light on the  philosophical notions of existence, universe and mind;
the universe exists in a single pure quantum state, but there are many ways to write it in the form
$\bf{univ}=\bf{mind}\otimes \bf{env}$.\par 
 Then, the relationship
$univ=\di\sum_i mind_i\otimes env_i$ describes a coexistence of universes whose $i$-th is in the state 
$mind_i\otimes env_i$, where $\bf{mind}$ and $\bf{env}$ both exist in the states $mind_i$ and $env_i$.
When $\bf{mind}$ satisfies hypothesis H1,H2 and H3,
 we can more precisely write the state of the universe as proposed page 
\pageref{27}:\par
 $univ(t)=\di\sum_{c\in C}\alpha_c mind_c\otimes env_c
+\sum_{i\in I}\beta_i mind_i\otimes env_i+univ'(t)$. This mind goes from a period of consciousness corresponding to one of the terms $mind_c\otimes env_c$ to a new period of consciousness of the form 
$mind_d\otimes env_d$. This corresponds to an egocentric reading of the universe. When awakened, the mind \textit{is} at every moment a particular conscious state. \par\medskip

The goal of this second part is to show that, if the mind is surrounded by an adequate neuronal structure, its experience as we have just described it, corresponds effectively to our daily conscious experience, charged with our sensations, imaginations, emotions, decisions and especially with a new notion, the self-consciousness, which unifies the whole and which makes sense.\par
It is therefore a matter of developing a model, both quantum and neuronal, of the highest cognitive functions of the mind that explains human consciousness. Current neuroscience knowledge is unfortunately far from providing such a model. This is why I limit the goal to the construction of a plausible model which is coherent with the contemporary advances in neurosciences. Beyond these advances it will be summary and speculative. I will define the concepts of 
 q-neurons, BAM, imaginary maps, motor, emotional and attentional maps, 
parallel maps to the motor map and decision algorithms. We  have to keep in mind that these are fictions invented to build our model, even 
when for stylistic ease I use these concepts as if they were real.
They help to clarify the role of the random component of neuron behaviour
in the formation of self-consciousness.

\section{Neurons}
\subsection{Threshold Automaton\label{aut}}
The biological neuron, like any eukaryotic cell, is incredibly complex and diverse and communicates in many chemical ways [D.P.3]. Nevertheless, if we simplify at the extreme and study only the electrical signal transmission, a neuron has inputs, its synapses, and an output, the axon, which then ramifies into several branches containing a signal of the same intensity, which reach the synapses of other neurons.\par
Consider a neuron with $n$ synapses. 
 If its $j$-th synapse receives an electrical signal of intensity $s_j\in\Re$, it transmits the signal $s_je_j$  to the neuron; $e_j$ is called the synaptic efficiency. When positive, the synapse is called excitatory. Otherwise, the synapse is called inhibitory. These different synaptic contributions are summed at the trigger zone of the neuron, located at the beginning of the axon to form the signal 
$x=\di\sum_{j}s_je_j$. The neuron then emits an action potential $p$ if and only if $x$ is greater than a threshold $s$, characteristic of the neuron [D.P.3, page 103]. In turn, this action potential is propagated along the axon to some synapses of other neurons.
\par

For some neurons, called sensory, it is appropriate to add to $x$ electrical signals resulting from the interaction of the neuron with the external environment. We also distinguish motor neurons whose axons are plugged into a muscle to control its contraction. \par
The number $n$ of synapses, their synaptic efficiencies $e_j$ as well as the threshold $s$ of the neuron evolve according to the reception by the neuron of various chemical or electrical signals. 
This simplified  neuron is also called a threshold automaton [JP.N, page 10]. \par
To further simplify, we  assume that the action potential is $p=$1 and the resting  potential is $p=0$.
\par\medskip
In these conditions, if we choose $e_j=1$ for all $j$ and $s=1$, the neuron emits if and only if at least one of the $s_j$ is 1, so that the neuron output corresponds to the logical function ``or'' applied to the inputs. We then say it is an ``or'' neuron.  Similarly, if we choose $e_j=1$ for all $j$ and $s$ equal to the number of synapses, the neuron emits if and only if all $s_j$ are equal to 1. We  say it is an ``and'' neuron.

\subsection{Stochastic Neuron}
For threshold automaton, with  previous notations, $p=S_0(x)$, where $S_0$ (Figure 1) is the 
threshold function, equal to 0 when $x<s$ and 1 when $x\geq s$.  In order to incorporate some randomness into neuron behaviour, a common model [E.D, pages 35-40] uses a $C^{\infty}$ increasing function $S$ (Figure 2)  whose codomain is $[0,1]$, 
which approximates $S_0$ by requiring that $S(x)$ be close to 0 when $x\leq s-\epsilon$ and close to 1 when $x\geq s+\epsilon$, where $\epsilon$ is a given parameter close to 0.  The neuron is now stochastic: it emits an action potential $p=1$ with a probability $S(x)$, in the sense that
if we denote by $\bf{neu}$ the given neuron and $\bf{env}$ the rest of the universe, there is a coexistence of several scenarios, some leading to the emission of an action potential, others not;\par
$univ(t_2)=\di\sum_i\alpha_i neu_{i,emits}\otimes env_i+\sum_j\beta_j neu_{j,silent}\otimes env_j$, with 
$S(x)=\di\sum_i|\alpha_i|^2$.\par
In fact, $S(x)$ is only interpreted as a probability if this decomposition of $univ(t_2)$ into a sum of two terms is measured by a mind, included in $\bf{env}$. 
\par\medskip

\hspace{3cm}
\begin{minipage}{10cm}
 \psscalebox{2}{
\begin{pspicture}(-0.5,-1)(2.5,1.5)
\psset{labelFontSize=\scriptstyle}
\psaxes{->,xAxis=false}(0,0)(-0.5,-0.5)(2.5,1.5)
\psline{->}(-0.5,0)(2.5,0)
\psline[linecolor=red,linewidth=1pt]{-[}(0,0)(1,0)
\psline[linecolor=red,linewidth=1pt]{[-}(1,1)(2,1)
\psline[linestyle=dotted](0,1)(1,1)
\psline[linestyle=dotted](1,0)(1,1)
\rput[t](1,-0.25){$\scriptscriptstyle s$}
\rput[br](-0.2,1.3){$\scriptstyle p$}
\rput[tl](2.3,-0.2){$\scriptstyle x$}
\rput[tl](1.5,0.9){$\scriptstyle S_0$}
\rput[tl](0,-0.7){\scriptsize figure 1}
\end{pspicture}
}

\vglue 2cm

\ \ \ \ \ \ \psscalebox{2}{
\begin{pspicture}(-0.5,-1)(2.5,1.5)
\psset{labelFontSize=\scriptstyle}
\psaxes{->,xAxis=false}(0,0)(-0.5,-0.5)(2.5,1.5)
\psline{->}(-0.5,0)(2.5,0)
\psplot[linecolor=red,linewidth=1pt]{0}{2}
{1 1 2 1 x sub 0.06 div exp add div}
\psline[linestyle=dashed](0,1)(2,1)
\psline[linestyle=dotted](0.7,0)(0.7,1)
\psline[linestyle=dotted](1.3,0)(1.3,1)
\rput[br](-0.2,1.3){
\begin{minipage}{1.25cm}
\tiny firing\par
probability
\end{minipage}
}
\rput[tl](2.3,-0.2){$\scriptstyle x$}
\rput[tl](1.5,0.9){$\scriptstyle S$}
\rput[tl](0,-0.7){\scriptsize figure 2}
\rput[t](1,-0.15){$\scriptscriptstyle s$}
\psline(1,-0.075)(1,0.075)
\rput[t](0.6,-0.1){$\scriptscriptstyle s-\epsilon$}
\rput[t](1.4,-0.1){$\scriptscriptstyle s+\epsilon$}
\end{pspicture}
}

\vglue 2cm

\ \ \ \ \ \ \psscalebox{2}{
\begin{pspicture}(-0.5,-1)(2.5,1.5)
\psset{labelFontSize=\scriptstyle}
\psaxes{->,xAxis=false}(0,0)(-0.5,-0.5)(2.5,1.5)
\psline{->}(-0.5,0)(2.5,0)
\psline[linecolor=red,linewidth=1pt]{-[}(0,0)(0.7,0)
\psline[linecolor=red,linewidth=1pt]{[-[}(0.7,0.5)(1.3,0.5)
\psline[linecolor=red,linewidth=1pt]{[-}(1.3,1)(2,1)
\psline[linestyle=dotted](0,1)(1.3,1)
\psline[linestyle=dotted](0.7,0)(0.7,1)
\psline[linestyle=dotted](1.3,0)(1.3,1)
\rput[t](0.7,-0.15){$\scriptscriptstyle s$}
\rput[t](1.3,-0.05){$\scriptscriptstyle s'$}
\rput[Br](-0.1,0.5){$\scriptstyle P_0$}
\psline[linestyle=dotted](0,0.5)(0.7,0.5)
\rput[br](-0.2,1.3){
\begin{minipage}{1.25cm}
\tiny firing\par
probability
\end{minipage}
}
\rput[tl](2.3,-0.2){$\scriptstyle x$}
\rput[tl](1.5,0.9){$\scriptstyle S$}
\rput[tl](0,-0.7){\scriptsize figure 3}
\end{pspicture}
}
\end{minipage}
\par\medskip

It is often assumed [JP.N, page 16] that the information conveyed by a neuron is contained in the frequency  of its action potentials, which is on average proportional to the probability $S(x)$. Then $S(x)$ represents an approximate value of the signal generated by the neuron. The derivable character of the output $S(x)$ as a function of the inputs $s_j$ enhances the field of available mathematical tools, which is why this model is used in artificial intelligence, notably in the backpropagation algorithm
 [J.H.2, pages 75-81]. However, from a biological point of view, such coding is not very economical, so it does not explain the remarkable speed of reaction of the brain to the stimuli it receives.  It is likely that different kinds of neural coding are used in the brain\footnote{See Wikipedia's well-documented article entitled ``neural coding''.};
even though the intensity of a muscular contraction depends essentially on the frequency of firing of the associated motor neuron, we will see in paragraph \ref{pla} that certain synaptic efficiencies evolve according to the time interval between two action potentials. Under these conditions, the winning neural coding is the common denominator of these different codes, which considers that the instant of each action  potential of each neuron carries information. 

\subsection{Q-Neurons\label{qn}}
In chapters \ref{cog} and \ref{soi}, I give an important role to the random part of the neural signal in the construction of the ego, when this part is explicitly measured by the mind. To highlight the use of this randomness,
 even though in reality stochastic neurons are enough, we will assume that the population of neurons consists only of two distinct species: deterministic neurons, modelled by threshold automata, and q-neurons\footnote{``q'' for ``quantum''.}, that are simplified stochastic neurons, for which the $S$ function has the following form (Figure 3): it is
 defined using two thresholds $s$ and $s'$ such that $0\leq s<s'$ and using a number $P_0\in[0,1]$ which will play the role of a probability:
\begin{itemize}
\item When $x<s$, $S(x)=$0. That means that if the sum of neuron inputs weighted by synaptic efficiencies is less than the threshold $s$, then definitely the neuron remains inactive.
\item When $s\leq x<s'$, $S(x)=P_0$. This is the case where the neuron explicitly exhibits random behaviour.
\item When $x\geq s'$, $S(x)=1$. So, in this case, the neuron emits with certainty.
\end{itemize}
It is then easy for the neurons upstream of a q-neuron to impose on it a random behaviour.

When $x\in[s,s'[$, to ensure that the q-neuron fires with the probability $P_0$, we can imagine the following coin toss scenario: near the trigger zone, 
 two different protein species $p_a$ and $p_b$ are attached to the cell membrane. At a moment $t_0$, they rush towards the DNA inside the nucleus. The $p_a$ and $p_b$ proteins can only bind to a single site $\ell$ in DNA. If a $p_a$ protein does this, it triggers a cascade of intracellular reactions that induce such a polarization in the trigger zone that the q-neuron will emit an action potential as soon as we have $x\geq s$. On the contrary, in the case where
 it is a $p_b$ protein that manages to bind to $\ell$, a similar process imposes this time the condition $x\geq s'$ for the q-neuron activation.  When the event  ``$x\in[s,s'[$'' occurs at the trigger zone, it is handled by the previous coin toss and a new coin toss is initiated. To speed up  the working rhythm of a q-neuron, we can imagine that several pairs of different protein species $(p_{a,i},p_{b,i})$ make coin tosses.\par

Let us study one of them more precisely; 
the position of each protein, in the agitated intracellular medium, is a chaotic parameter whose wave function rapidly extends over a large region of space and thus becomes entangled with the environment in the form of a coexistence of positions. So between the initialization of the coin toss and the arrival of one of the proteins on the site $\ell$, the universe has became a superposition of a multitude of scenarios concerning these proteins; at the end of this process, the universe is in the state
$mind\otimes\Bigl(\di\sum_i \alpha_i [p_a\ wins]_i\otimes env_i+\sum_{j}\beta_j[p_b\ wins]_j\otimes env_j\Bigr)$.\par
We insist that $P_0$ can only be interpreted as a probability if this decomposition into a sum of two terms is measured by a mind. Then when $x\in[s,s'[$, the firing probability  is $P_0=\di\sum_i|\alpha_i|^2$.\par
By modifying the relative concentrations of the $p_a$ and $p_b$ proteins on the cell membrane, one can regulate this probability.  A retroactive control of the q-neuron is thus possible, for example according to the hormonal response of the organism to the consequences of the ``decision'' taken by the q-neuron.\par
If the mind then measures the possible activation of the q-neuron in random mode, the state of the universe becomes\par
$\bigl(mind_{[p_a\ wins]}\otimes\di\sum_i \alpha_i [p_a\ wins]_i\otimes env_i\bigr)+\bigl(mind_{[p_b\ wins]}
\otimes\sum_{j}\beta_j[p_b\ wins]_j\otimes env_j\bigr)$.\par

In this case, $P_0$ becomes a probability for $\bf{mind}$, but we will see later how such a random process is sometimes perceived by the mind as a ``self'' decision, i.e. a decision whose cause is a ``self'' that assumes full responsibility.
\par\medskip

So, after c-neuron, 
the random component of neural behaviour, whose concept of  
 q-neuron is a caricature, is the second ingredient allowing the construction of self-consciousness.
It places the notion of quantum measurement at the heart of the human decision-making structure. Quantum measurement is no longer confined to sophisticated experiments by physicists, it becomes an integral part of human psychology. 
\par\medskip

Using a q-neuron, it is simple to build a neural device behaving as an ``xor'' (Figure 4). \label{64}
 If the  thresholds and the synaptic efficiencies are chosen correctly, the following behaviour will occur: as long as axon 1 is inactive, the q-neuron remains inactive and nothing happens, whereas when axon 1 becomes active, it induces a random behaviour in the q-neuron. Thus, with a certain probability $P_0$, the q-neuron fires and with a probability $1-P_0$ it remains silent. When it fires, its axon  triggers action 1 and at the same time activates the inhibitory synapse of the classic neuron, preventing the trigger of action 2. When the q-neuron is inactive, axon 1 activates the classic neuron, because it is not inhibited by the synapse receiving the q-neuron signal, which triggers action 2, without action 1 being triggered. 
\par
\begin{pspicture}(0,0)(14,8)

\rput[t](6,6){\rnode{qn}{\psframebox{
q-neuron}}}

\rput[tl](10,4){\rnode{nc}{\psframebox{
\begin{minipage}{1.5cm}
classic\par 
neuron\end{minipage}}}}

\rput[tl](2,2){\rnode{a1}{ Action 1}}
\rput[tr](12,2){\rnode{a2}{ Action 2}}
\rput(6,8){\rnode{vide}{}}
\rput(6,4){\rnode{vide2}{}}
\rput(6,7){\rnode{vide3}{}}
\rput(0,8){\rnode{vers1}{}}
\rput(14,8){\rnode{vers2}{}}

\psset{arrows=->,arrowscale=2.5,linewidth=.05,linearc=.3}

\ncline{vide}{qn}
\nbput[npos=0.2]{ 1}
\ncangle[angleA=-90,angleB=0]{qn}{a1}
\ncangle[angleA=-90,angleB=180,nodesep=0.1cm]{-|}{qn}{nc}
\naput[npos=1.8]{\scriptsize inhibitory}
\nbput[npos=1.8]{\scriptsize synapse}
\ncangle[angleA=0,angleB=90]{vide3}{nc}
\ncline{qn}{vide2}
\ncline{nc}{a2}
\ncangle[angleA=-90,angleB=180,linecolor=red]{vers1}{a1}
\ncangle[angleA=-90,angleB=0,linecolor=red]{vers2}{a2}
\rput[t](7,1){
\begin{minipage}{7.5cm}
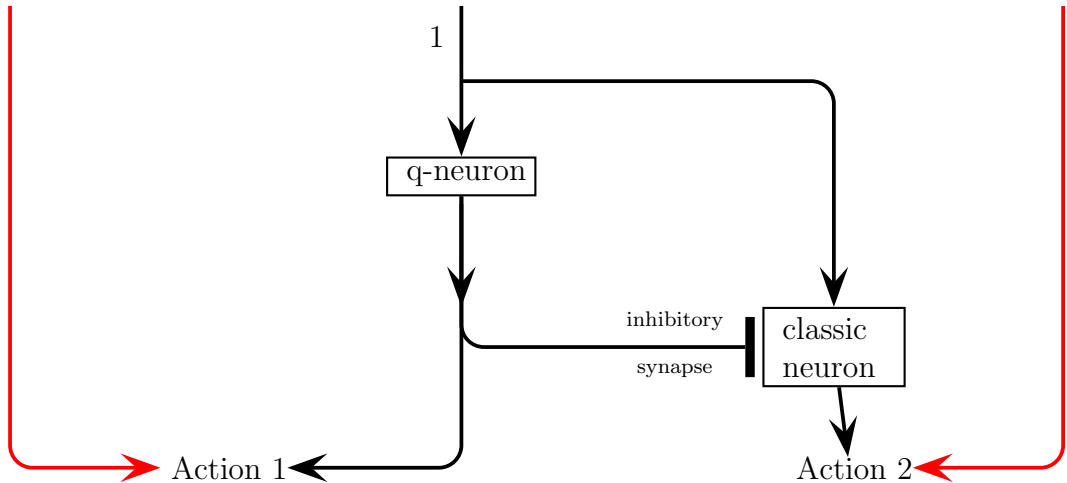
figure 4 :   ``xor'' \scriptsize(the black arrows).\par
The red arrows may be used 
to override this "xor".
\end{minipage}
}

\end{pspicture}

\section{Sensory Maps}

\subsection{The Different Senses}

The neurosciences count several sensory systems for the human body [D.P.2]:
\par
\begin{itemize}
	\item The vision; 
	\item The auditory system; 
	\item the chemical senses : 
	\begin{itemize}
		\item the taste;
		\item the olfactory system;
		\item the trigemal system, for the perception of irritating or noxious molecules;
	\end{itemize}
		\item the vestibular system (inner ear), for the perception of the overall position of the body and its movements;
	\item the somatic sensory system which has several subsystems: 
	\begin{itemize}
		\item the sense of touch, for the perception at every point of the body surface of temperature, pressure and vibration.;
		\item proprioception, to perceive the tension of muscles, tendons and joints;
		\item pain, which is a sense in its own right.
	\end{itemize}
\end{itemize}
Each system independently has a range of specific sensory receptors, nerve wiring to the brain, and specific processing areas in the brain.

\subsection{Cortical Maps}
In these areas, the raw information provided by one of the senses is filtered to extract the important components. In the primary visual cortex, some neurons are activated only when, in a specific region of the visual space, a small line segment oriented in a specific direction is present [D.P.3, page 266]. If one changes slightly the region or if one modifies a little the inclination of the line segment, these neurons become inactive, while other neurons are activated, close to the first ones in the cortex. The set of these different neurons form what is called a cortical map [JA.B], which in our example only takes into account the raw visual information for the presence of line segments and, in case of presence, for their locations and orientations. \par
Thus, a cortical map is a network of neurons sensitive to a limited number $k$  of sensory parameters denoted
$p=(p_1,\ldots,p_k)$ where $k\leq 5$. Each neuron $n$ on the map is activated only when the value of  $p$  is within a small domain denoted $D_n$, called the receptive field of the neuron. If $n$ and $n'$ are two physically distant neurons on the network, then $D_n\cap D_{n'}=\emptyset$.\par
There are a multitude of such cortical maps, which filter according to the position, the direction, the presence of a particular pattern, the speed of movement, the intensity of the luminosity, the color, by combining these various criteria by two, three or even more. It is valid for every sensory system.
\par\medskip

The simplest cortical maps, directly linked to sensory receptors, are created during fetal development likely using a genetically programmed process based on chemoaffinity [D.P.3, page 519].\par
These primary maps provide new parameters that feed secondary maps, which feed other maps in a hierarchical structure.\par
Knowledge about the development, organization and use of these maps is growing, however, we currently lack definitive and comprehensive answers.\par
Some of these maps are probably formed through a process of self-organization 
 analogous to Kohonen's maps [E.D], [JP.N], [J.H.2], to ensure that the detail level used to describe information is correlated with its frequency of occurrence.\par
On the other hand, the Hmax model and its refinements [M.R], [C.L.2] explain how, starting from primary sensory maps, 
it is possible to construct 
secondary maps that can recognize a certain conjunction of parameters (recognition of a square shape for example) independently of other parameters (position and size of the square).
 By caricaturing, the presence of a square corresponds to the existence of 4 correctly arranged line segments, therefore if each of these segments is coded by the activation of a neuron, by connecting these four neurons to an ``and'' neuron, the latter will be activated only in the presence of the square. Thus ``and'' neurons make it possible to recognize superimpositions of patterns, which leads to more sophisticated patterns. Moreover, if we connect to an ``or''  neuron all the neurons responding to the presence of a square, whatever its size and position, the ``or''  neuron will fire in the presence of a square, independently of its position and size.\par
Thus, when one goes up in the hierarchy of cortical maps, the criterion recognized by the map increases in complexity as well as in invariance;  at the top of the pyramid, in the visual cortex, one undoubtedly finds neurons sensitive to the presence of a smile on a face, independently of the size, the orientation, the luninosity etc..

\section{Hebbian Learning}

\subsection{Synaptic Plasticity\label{pla}}
Consider a synapse $s$ between a presynaptic neuron $n_0$ and a postsynaptic neuron $n_1$.
The efficiency $w$ of the synapse evolves according to the activity of the neurons $n_0$ and $n_1$ as well as the activity of the neighbouring synapses. This synaptic plasticity follows different rules depending on the neurotransmitters used by $s$ and its neighboring synapses and depending on the characteristics of $n_0$ and $n_1$.  
\par
We limit ourselves to the rule that seems to be the most common, guessed by Donald Hebb as early as 1949 [D.H], refined by actual synapse experiments [G.B]; $w$ is increased when an axon activity of $n_0$ in $s$ is followed after less than 40 ms by the emission of an action potential in the axon of $n_1$. On the contrary, $w$ is decreased when an activity of $n_0$ in $s$ occurs less than 40 ms after a firing of $n_1$. This decrease is however less strong than the previous increase, so if the synapse $s$ tends to be active at the same time as the neuron $n_1$, within a window of $\pm 10$ms, on average, $w$ will increase. \par
The increase is temporary if the simultaneity is only occasional, nethertheless, if it occurs more often it becomes permanent for several hours, even several days, or even all life, by binding particular proteins at specific sites of the $n_1$ DNA in order to modify the regulation of genetic transcription, thus the production capacities of certain proteins. The latter can even induce the creation of a new synapse close to $s$, still from the neuron $n_0$ to the neuron $n_1$.

\subsection{Hopfield Networks}
In artificial intelligence, to build a Hopfield network [JP.N], [J.H] we start from $N$ neurons which are zero threshold automata, and connect each of these neurons to all the others. Let $n_i$ be the $i$th neuron and let $w_{i,j}$ be the  efficiency of the synapse linking $n_i$ to $n_j$. Initially, we require $w_{i,j}=0$; 
The synapses are silent, which is biologically plausible [D.P.3, page 175].\par
A network configuration is given by a $N$-uple $\xi = (\xi_1,\ldots,\xi_N)\in\{0,1\}^N$, where $\xi_i$ represents the axonal activity of the neuron $n_i$.\par
During a first phase, called learning, we present to the network $p$ configurations $\xi^1,\ldots,\xi^p$, where $p$ is less than $\frac{N}{4\ln N}$ [JP.N, page 41]. During this phase, the synaptic efficiencies $w_{i,j}$ are modified according to the Hebbian rule, somewhat idealized as follows: after the presentation of the $k$-th configuration, for all $i,j\in\{1,\ldots,N\}$, $w_{i,j}$ is incremented by $\lambda (2\xi_i^k-1)(2\xi_j^k-1)$, where $\lambda$ is a strictly positive constant that corresponds to the network learning speed. This guarantees symmetry: $w_{i,j}=w_{j,i}$.\par
In a second phase, called network use, the synaptic efficiencies are no longer modified.  We present the network with some arbitrary configuration $\xi$ then we let this network of threshold automata work for a few iterations. One shows [JP.N, pages 34-41] that it then converges to one of the $\xi^k$, a priori the closest to $\xi$. So the network works like an associative memory; if we submit to the network a configuration close to one of the $\xi^k$, it recognizes $\xi^k$ in the sense that it adopts this configuration.\par

If we modify the learning rule somewhat to make it more realistic, or if we remove some connections between the $N$ neurons, the network continues to behave like an associative memory, at the cost of less storage capacity [JP.N, page 49]. Thus Hopfield networks can be considered as an idealization of certain strongly and symmetrically wired biological neural networks, which we  still call by extension Hopfield networks.
\par\medskip
Hopfield networks have in particular a content-addressable memory behaviour;
by equating a configuration with the set of neurons in the network that are active for that configuration, if we present part of one of the configurations $\xi^k$ to the network, it  converges to $\xi^k$, that is, it can retrieve the entire example $\xi^k$ from one of its parts. Of course, if we use a common part of two configurations  $\xi^k$, the network has to make a choice, and in practice, with a non-idealized network, it can also make convergence errors: to err is human.\par\medskip
In practice, the learning and use phases are often mixed:  the network learns as it is used. This requires that 
the learning rule is counterbalanced by a forgetting rule, which systematically decreases the absolute value of synaptic efficiencies with each iteration  
at a greater or lesser speed. When this speed and $\lambda$ are large, the Hopfield network is a short-term memory that forgets very quickly while it can learn every example from its first appearance.  On the contrary, when the speeds of forgetting and learning are low, the Hopfield network is a long-term memory, which forgets only after several days or several months, and in return learns only the examples that are repeated several times. 

\subsection{Single Sensory Objects}
For a human sense, the vision for example, let us consider all of its high-level cortical maps. Let us assume that the presence of your coffee cup in your  vision field is encoded by the activation of some neurons on some of these maps. We also assume that the outputs of these maps are the inputs of a  highly connected network of neurons, whose synaptic efficiencies are initially arbitrary.
This is a rather low hypothesis because we only assume the existence of a multitude of highly connected neurons with no initial organization, some of which having synapses that receive signals from the  neurons of these maps.\par 
Then this network will behave like a Hopfield network and learn the most commonly encountered objects in your vision field. Your cup of coffee in particular corresponds to a configuration learned by this network. If now your cup is half hidden by a water bottle, the Hopfield network still recognizes the cup by reproducing the corresponding complete configuration after a few iterations. Your brain produces augmented reality on a daily basis.
\par\medskip
It can be envisaged that several Hopfield networks are connected in this way to the upper vision maps, with different learning and forgetting speeds, in order to memorize visual objects in the more or less long term. It is the same for all other senses. 

\subsection{Bidirectional Associative Memories}
This is a variant of  Hopfield networks, also developed within the framework of artificial intelligence [B.K]. This variant only uses the Hebbian rule, so it is still plausible to assume that  brain contains similar mechanisms.\par
A BAM (Bidirectional Associative Memory) is made up of two layers of null threshold neurons, denoted by $E$ and $S$. Let $m_1,\ldots,m_N$ be the neurons of the layer $E$ and $n_1,\ldots,n_P$ those of the layer $S$. Each neuron in one layer is connected to all the neurons in the other layer, and there is no connection between neurons in the same layer. Let $w_{i,j}$ be the efficiency of the synapse joining $n_i$ to $m_j$ and 
$w'_{j,i}$ the efficiency of the reverse synapse joining $m_j$ to $n_i$.\par
A network configuration is now given by a couple $(A,B)$, \par
where $A=(A_1,\ldots,A_N)\in\{0,1\}^N$ represents the activity of the $E$ layer and \par
 where $B=(B_1,\ldots,B_P)\in\{0,1\}^P$ corresponds to the $S$ layer.\par
During the learning phase, the network is presented with $m$ configurations $(A^k, B^k)$, with 
$m<\min(N,P)$. The synaptic efficiencies, initially null, are modified according to the Hebbian rule; for all 
$i\in\{1,\ldots,N\}$ and $j\in\{1,\ldots,P\}$, the presentation of the $k$-th configuration increments $w_{i,j}$ by $\lambda (2A_i^k-1)(2 B_j^k-1)$ and increments 
$w'_{j,i}$ by the same value.\par
Then, during the use phase, if we present to the network a configuration $(A,B)$,  it is proved that it converges in a few iterations towards one of the learned configurations $(A^k,B^k)$, close to $(A,B)$.\par 
In particular, if we present to the network a configuration of the form $(A^k,B)$, where $B$ is arbitrary, and where $A^k$ is part of the examples learned, the network converges to $(A^k,B^k)$. Conversely a configuration of the form $(A,B^k)$ leads to $(A^k,B^k)$. Thus the network behaves like a hetero-associative memory, which can associate  the $A^k$ pattern with the $B^k$ pattern.

\subsection{Multisensory Objects}

For each sense, we have associated Hopfield networks
which can record
 single-sensory objects on different time scales. For each of the $p$ senses of a  human being, 
let us pick one of these networks, denoted by $H_i$, where $i\in\{1,\ldots,p\}$. Let us assume that the axon of each neuron of $H_i$ is connected
  to a new neuron. Let $H'_i$ be the set of these new neurons and suppose they are not 
 connected between them. They thus constitute an output layer of the network $H_i$. We  assume that a neural device guarantees that $H_i$ delivers its result at $H'_i$ only once convergence is reached. \par\medskip
When $i\not=j$, we assume that $H'_i$ and $H'_j$ are the two layers of a BAM.\par
The vision of a cup often occurs at the same time as the hearing of the word  ``cup'', so if
 $H_i$ corresponds to the vision and $H_j$ to the hearing, the BAM linking $H'_i$ and $H'_j$ can learn the couple $(A,B)$, where $A$ is the activity of $H_i$, transmitted to $H'_i$, generated by the vision of a cup, even partial, and where $B$ is the activity of $H_j$, transmitted to $H'_j$, generated by the hearing of the word ``cup'', even with low noise. 
Once the learning is completed, the mere sight of a cup will induce the imagination of hearing the word ``cup'' and vice versa. \par

The cup thus becomes a multisensory object, and its evocation according to one sense induces the corresponding imaginations according to the other senses\footnote{This is not systematic however because according to paragraph \ref{att}, it depends on how the attention is directed.}.\par
Thus, when a neural activity is induced in $H'_j$ by an $H'_i$ and not by the $j$-th sense maps, it is an imagined fact. For this reason, thereafter, $H'_j$ will be called an imaginary map of the $j$-th sense.\par\medskip

We assume that the higher level cortical maps are made of c-neurons. Their c-bits thus bring to the mind the experience directly produced by the senses. We also assume that the networks $H'_j$ are made of c-neurons, while the networks $H_j$ are made of classic neurons.  Then the c-bits of an imaginary map of the  $j$-th sense bring to the mind, on the one hand an augmented reality perceived by the $j$-th sense, consistent with the c-bits of high-level cortical maps of the same sense, and on the other hand imagined facts, relative to the $j$-th sense, from a real sensation produced by another sense.
\par\medskip

The conscious experience generated for example by the vision of the coffee cup  is reproduced identically with each vision of the cup. When this vision is only imagined after hearing the word ``cup'', it is also reproduced identically into the imaginary visual map, though without the presence of the corresponding activity pattern in the upper cortical visual maps.
\par\medskip
Each
 imaginary map is connected to other imaginary maps by BAM, so by transitivity, a first real perception can induce multiple imaginary facts. It already looks like an automatic sequence of thoughts. 

\subsection{Libet Experiments\label{Lib}}
In summary, each sensory system transmits information to the brain that is processed and stored in cortical maps. If they are low level, they are not made of c-neurons and therefore do not participate directly in conscious sensations. Conversely, we are conscious of the information contained in high-level cortical maps and imaginary maps.\par\medskip
C-neurons are threshold automata with c-bits. 
With the notations in paragraph \ref{aut}, inside a c-neuron, $x$ is compared at each moment with $s$ at its trigger zone, located at the beginning of the axon; when $x\geq s$, the c-neuron emits an action potential whose duration is of the order of 5 ms. 
According to the scenario described on page \pageref{32}, for this action potential to participate in consciousness, that is, for it to be encoded in the c-bit configuration of the c-neuron, 
one has to wait for a $p_2$ protein to travel from the trigger zone to the position  $\ell$ of the DNA inside the nucleus. A $p_2$ protein is therefore a motor protein [T.D.2], or it is carried by a motor protein. At best, such a protein, for example a dynein, moves along microtubules at a speed\footnote{see http://book.bionumbers.org/how-fast-do-molecular-motors-move-on-cytoskeletal-filaments/} of $6.10^{-6} m.s^{-1}$.  The size of a neuron varies between 5 and 120 micrometers. However, we can suppose that c-neurons have an eccentric nucleus, located near the trigger zone, which is rather common in  neurons of an adult brain. If 
the distance between the nucleus and the trigger zone is in the order of 2$\mu m$, 
   the time required for the protein to enter the nucleus is in the order of 300 ms. If we assume that the diameter of the nucleus is in the order of 2$\mu m$, the coding of the activation of an action potential in the c-bits will be done with a delay in the order of 500ms.\par

Such a delay corresponds to the experimental observations of Benjamin Libet [B.L]; by applying electrical stimulations directly in the cortex of consenting patients while they were undergoing surgery in the brain, Benjamin Libet observed that stimulations only reached the subject's consciousness after a duration of 500 milliseconds.\par
He also observed that a stimulation lasting less than 500 ms never reached consciousness. The transport of the $p_2$ protein towards the nucleus then towards the c-protein should therefore  be conditioned by the continuation of the emission of action potentials over 500 ms;  one can imagine that the motor protein moves only if it evolves in an environment where the electric potential $x$ is higher than the threshold  $s$. Then, as soon as $x<s$, the emission of action potentials stops and, for the same reason, the progression of the $p_2$ protein is stopped. So when the event ``$x\geq s$'' occurs for less than 500ms, the $p_2$ protein does not have time to reach the site $\ell$ and the action potentials are not encoded in the c-bits, so they do not participate in the observer's consciousness.\par

Consequently, the information contained in the c-bits does not fully reflect the axonal activity of the c-neurons. Moreover, some c-bits may code only the activity of one  synapse of the c-neuron, so conversely, the action potentials delivered or not by the c-neurons do not fully reflect the conscious experience of the observer. 
\par\medskip
A brief cutaneous stimulation of about ten milliseconds accesses the consciousness though,  because it generates a sequence of activities in the brain lasting more than 500 ms, but  it is coded in the mind with a delay of some 500 ms.\par
In another experiment [B.L, page 95], Benjamin Libet sends an electrical stimulation directly into the sensory cortex for more than 500 ms starting at a moment $t_0$.  At the moment $t_0 + 200 ms$, he sends a cutaneous stimulation of about ten milliseconds.  The cortical stimulation accesses consciousness at the instant $t_0+ 500 ms$ and the cutaneous stimulation at the instant $t_0+ 700 ms$. Yet the subject claims to be aware of the cutaneous stimulation \it{before} the cortical stimulation, precisely at the moment $t_0+200 ms$ when it was initialized. This remains true  even if we replace  200 ms by a duration $t_1$ lower than 500ms. When $t_1=500ms$, the two events are perceived as simultaneous and when $t_1>500ms$, the chronological order is restored. 

The hypothesis  H3 makes it possible to solve this paradox; during each period of consciousness, the mind remains in a constant quantum state during a duration $T$ of the order of 50 ms. It is therefore necessary that this state of mind encodes the state of the environment not only spatially, but also temporally. The coding of each conscious elementary event therefore includes temporal information: start time, end time, speed, etc. It is then possible to imagine processes of time correction which consist in modifying this information. Here, the time label of the cutaneous stimulation would be decremented by 500ms compared to that of the cortical stimulation. More generally, it is assumed that any real stimulation, i.e. coming from the primary sensory maps, is temporally labelled according to the approximate time of arrival of the signal in the primary maps, whereas signals deprived of such origin are by default temporally labelled according to the actual time. According to this mechanism, between\par
 $t_0+500ms$ and $t_0+700ms$, 4 conscious frames
 are produced which include the 
cortical stimulation but not the cutaneous stimulation, then beyond $t_0+700ms$, the conscious frames contain the cortical stimulation, temporally labeled by $t_0+500ms$ and the cutaneous stimulation labeled by $t_0+200ms$, therefore the second stimulation appears to consciousness as a former event than the first stimulation.  These new conscious frames  prevail when the patient testifies.

\section{Skeletal Muscle Contractions\label{motr}}
Each motor neuron, located in the spinal cord or the brain stem, sends its axon onto a few fibers of one of our 570 muscles. The firing of the motor neuron induces the contraction of fibers that triggers skeletal movement.\par
An elementary movement, such as bending an arm, requires a complex and coordinated sequence of contractions of the muscles involved. It is largely controlled without brain intervention, by  local neural circuits, again in the spinal cord or the brain stem, which cause motor neurons and certain sensory neurons to interact. Thus, the main functions of the motor cortex are movement planning, initiation, coordination and precise control.\par
Wilder Penfield showed that the motor cortex is organized as a motor map;  the excitation of a neuron on this map causes movement in a region of the body, the neighbouring regions corresponding to neighbouring neurons on the map. If one excites a neuron of this map, located in the part corresponding to  the right arm, while it is close to the part of the map corresponding to the  mouth [D.P.3, page 385], one obtains a movement of the arm bringing the hand towards the mouth. This map, often noted MM thereafter, thus provides a set of elementary movements to survive in nature and adapt to it. \par\medskip

We assume that between each imaginary map and the motor map we have
a  highly connected network of 
 symmetrically interconnected neurons. By Hebbian learning, this network can behave like a BAM which thus connects a sensation with a movement when they are often concomitant. This model is consistent with the approach developed by Christian Keysers and Valeria Gazzola in [C.K.].\par
In particular, before birth, the spontaneous movements \label{72} (cf paragraph \ref{spon}) of the foetus make it possible to link the triggering of a movement from  MM  with the somesthesic sensations that it generates into the imaginary maps concerned. Later, this learning allows an adult to connect each of his movements with the full range of corresponding sensations for each of the senses.\par
 When an individual $A$ observes the hand of a person $B$ grabbing an object, due to the invariance properties of the high-level sensory maps, the neural activity of the $A$'s higher visual maps is similar to what it would be if $A$ grabbed the object himself. The BAMs to the motor map then induce the same hand movement in $A$. That explains our tendency to imitate.

However, we often inhibit this inclination. 
To model this skill, 
we replace, for each neuron $n$ of  MM, the scheme $n\rightarrow \hbox{mn}$, which symbolizes the fact that the axon of  neuron $n$ is connected to some motor neurons, by the following more complex scheme: 
$\begin{array}
	{ccccc}
	n&\rightarrow&n'&\rightarrow&mn\\
	&&\top&&\\
	&&n''&&\\
\end{array}$, where
$n'$ and $n''$ are two more motor cortex neurons.
 The connection between $n$ and $n'$ is an excitatory synapse whereas the connection between $n''$ and $n'$ is inhibitory.  Thus the activation of $n''$ will prevent the activation of $n'$ and therefore the execution of the associated movement, even when $n$ is activated.
In these conditions, the activation of $n$ together with that of $n''$ inhibits the movement. Nethertheless via the BAMs, it triggers the imagination of the sensations related to the movement. That fits well with what $A$ feels inside when he observes $B$ grabbing an object.\par
Thus, when we observe the movements of another person, we feel largely the same sensations as when we actually perform the same movements and we even activate some neurons of type $n$ of the motor map, while inhibiting those of type $n'$, via the activation of neurons of type $n''$. This could explain the origin and usefulness of mirror neurons [A.K], [G.V].
\par\medskip

This model also explains how a baby can imitate facial expressions as early as the sixth week, whereas at this age he does not have the possibility of studying his face in a mirror [C.K]. 
His relatives imitate the baby firstly: 
when the baby pulls out his tongue, most often one of his parents does the same, so that the activation in the infant of the corresponding neurons on the motor map is associated via the BAM with the vision of a face that pulls out the tongue. Thereafter, baby can imitate her mother when she pulls out her tongue.
\par\medskip

Christian Keysers and Valeria Gazzola [C.K] show that this model can also explain the fine coordination of rapid movements; they take the example of a usual succession of movements: reaching for a biscuit, grasping the biscuit, bringing the biscuit to the mouth.  Let  $mvt_i$ be the $i$-th of these 3 movements. 
If at the moment $t$ the neurons of the motor map involved in $mvt_1$ are activated, this movement is actually performed at the moment $t+100ms$\label{key} then its perception is recorded in the sensory maps at the moment $t+200ms$. Then at $t+200ms$ the motor neurons involved in $mvt_2$ already become active. Thus the BAM  associates the perception of $mvt_1$ with the trigger of $mvt_2$. Indeed, we have seen that the simultaneity that allows synaptic reinforcement must take place within a time window of 40ms. Similarly, the perception of $mvt_2$ is associated with the trigger of $mvt_3$. After learning, the rapid succession of these 3 movements is programmed into the BAM;  the perception of $mvt_1$ triggers via the BAM $mvt_2$, then the perception of $mvt_2$ triggers $mvt_3$. Thus, the repetition of the sequence of these 3 movements makes it possible to automate its production.

\section{Emotions}
The different sensory systems with their cortical and imaginary maps thus closely interact with the somatic motor system. They form the overall sensory-motor system of the human being. \par\medskip
Besides, the human body is provided with the visceral  nervous system, which automatically controls  homeostasis of the organism, in other words its different balances. It regulates heart rate, gastric secretions, pupil dilation, tears, horripilation, etc. It also regulates hormone secretions which act in particular on neurotransmitters and thus on brain behaviour. \par\medskip
The visceral nervous system works according to a rather limited range of predefined behaviours that allow it to adapt to external circumstances. Each of these behaviours can be labelled by an emotion. Thus fear is associated with an acceleration of  heartbeat, shortness of breath, trembling of the legs, contraction of the bowels, etc., i.e. the mobilization of the body's resources to face a danger.
Besides fear, the most common emotions are happiness, surprise, anger, sadness, disgust and contempt. Each emotion has variants; anger can turn into rage or decrease to become annoyance. Different emotions can also be experienced at the same time. Antonio Damasio [AR.D] also defines background emotions such as well-being, discomfort, calm or tension, and  more social emotions such as shame, jealousy, pride, etc.\par
These emotions are triggered by the action on the visceral nervous system of neural networks in the brain, which I call emotional inducers. Each emotion is, probably innately, combined with a stereotyped configuration of certain skeletal muscles, which defines emotional facial expressions and body postures. 
\par\medskip

\hspace{0.3cm}
 \psscalebox{0.85}{
\begin{pspicture}(0,-2.5)(17.5,13.5)

\rput(4.5,13){\underline{Visceral nervous system}}

\rput(4.5,11){\rnode{ce}{\psframebox{Emotional map}}}
\rput(4.5,7){\rnode{ie}{\psframebox{Emotional inducers}}}
\rput(4.5,5){\rnode{ch}{\psframebox{Homeostatic control}}}
\rput(9,3){\rnode{m}{\psframebox{Muscles}}}
\rput(12,1){\rnode{e}{Outside}}
\rput(15,3){\rnode{s}{\psframebox{Senses}}}
\rput(13.5,6){\rnode{cm}{\psframebox{
\begin{minipage}{3.5cm}
Motor map\par
\scriptsize
including emotional facial expressions\par
 and body postures\par
\end{minipage}
}}}
\rput(13.5,9){\rnode{cs}{\psframebox{
\begin{minipage}{4cm}
Sensory maps\par
\scriptsize
including those dedicated to emotional facial expressions\par
 and body postures\par
\end{minipage}
}}}
\rput[b](13.5,11){\rnode{ci}{\psframebox{Imaginary maps}}}
\rput(13.5,13){\underline{Sensory-motor system}}
\rput{90}(0,10){Conscious}
\rput{90}(0,5){Unconscious}

\psline[linestyle=dashed, linewidth=0.5pt](0,8)(17.5,8)
\psline[linestyle=dashed, linewidth=0.5pt](9,4)(9,13)

\psset{arrows=->,arrowscale=2.5,linewidth=.05}

\ncline{ce}{ie}
\nbput{9}
\ncline{ie}{ch}
\nbput{7}
\ncbar[arm=1,angleA=180, angleB=180]{ch}{ce}
\nbput[npos=1.2]{8}

\psarc[linecolor=red,linewidth=.03]{<-}(3,9.5){1}{45}{315}
\rput(2.9,9.5){
\begin{minipage}{1.2cm}
\color{red}\tiny
Emotional\par
feedback\end{minipage}}

\ncangle[angleA=-90,angleB=180]{ch}{m}
\ncangle[angleA=-90,angleB=180]{m}{e}
\nbput{6}
\ncangle[angleA=0,angleB=-90]{e}{s}
\ncangle[angleA=-90,angleB=0]{cm}{m}
\nbput{5}

\psarc[linecolor=red,linewidth=.03]{->}(15,4.3){0.8}{45}{315}
\rput(15,4.3){
\begin{minipage}{1cm}
\color{red}\tiny
Sensory-\par
motor\par
loop
\end{minipage}}

\ncbar[arm=1]{s}{cs}
\naput[npos=1.2]{1}

\ncbar[arm=1.5]{<->}{cm}{ci}
\ncput*[npos=1.8]{\scriptsize BAM}
\nbput[npos=1.2]{4}

\ncbar[arm=0.5,angleA=180,angleB=180]{cs}{ci}
\naput[npos=1.9]{2}

\psarc[arrowscale=1.5]{<->}(14,11){0.8}{180}{0}
\rput(14,10.6){\scriptsize BAM\ \ 3
}

\ncbar[angleA=90,angleB=90, arm=0.75]{<->}{ce}{ci}
\ncput*{\scriptsize BAM}

\ncline{<->}{ce}{cm}
\ncput*[npos=0.2]{\scriptsize BAM}

\ncline{ci}{ie}
\ncline{cs}{ie}

\ncline[offsetB=-0.5,nodesepB=-0.3]{ie}{cm}
\nbput[npos=0.3]{10}

\rput[t](7,0){
\begin{minipage}{15cm}
figure 5 :\par
\small
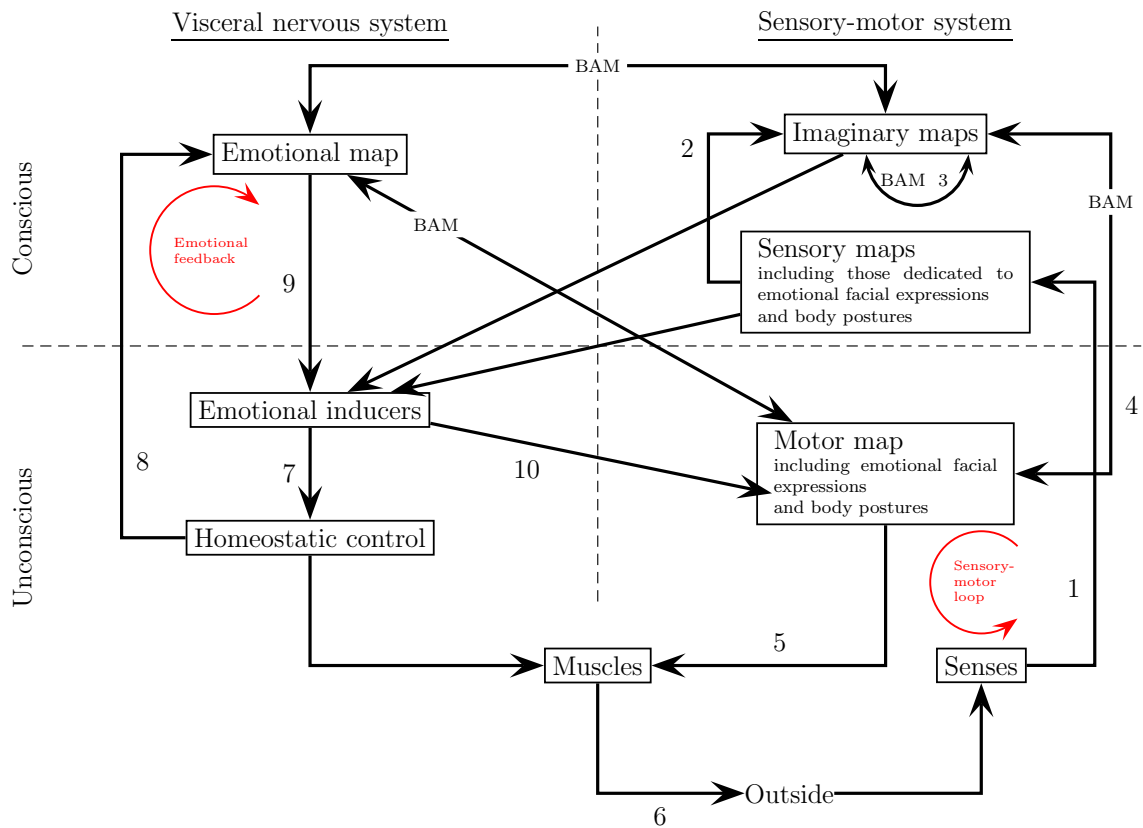
The sensory-motor loop corresponds to the cycle [1,2,3,4,5,6,1];\par
Emotional feedback corresponds to the cycle [7,8,9,7];\par
The arrow 10 controls the creation of emotional facial expressions.
 \end{minipage}
}

\end{pspicture}
}
\par

The visceral nervous system has many sensory receptors.  They are mostly integrated into neural loops of unconscious automatic activities. Some information still access the consciousness, via a map called emotional, that I suppose made up of c-neurons. In the mind, the elements of this conscious emotional component are by definition feelings.\par

The emotional map is connected to the imaginary maps and the motor map by BAMs (see Figure 5).
 This explains why the simple fact of forcing oneself to smile generates a vague feeling of happiness. 
It also explains our natural empathy for others: their emotional facial expressions that we perceive in our imaginary maps are translated via the BAMs into feelings in our emotional map.\par
The emotional inducers are innately activated by the joint action of certain neurons of some sensory or imaginary maps. Thus a young bird is frightened by the passage over the nest of a shape similar to that of a falcon. Emotions can be triggered indirectly, when a real or imagined event induces from BAM to BAM a certain activity in an imaginary map, which then activates emotional inducers. The emotional map probably also acts on the emotional inducers, which generates positive or negative emotional feedback phenomena. \par\medskip
Figure 5 summarizes our comments about these two sensory-motor and visceral nervous systems, distinguishing between conscious and unconscious parts.

\section{Higher Cognitive Functions\label{cog}}

\subsection{Spontaneous Actions\label{spon}}

Two neural maps are called \it{parallel} when they have the same number $k$ of neurons and, for every $i\in\{1,\ldots,k\}$, the $i$-th neuron on the first map is connected to the $i$-th neuron on the second map. 
The connections all go from the first map to the second, or they all go from the second to the first, or they are all symmetrical.
This symmetric binary relation over maps is extended by transitivity into  an equivalence relation. 
\par\medskip

We evoked page \pageref{72} the use of spontaneous movements to feed the BAMs linking the motor map and the imaginary maps. To produce such movements, we can imagine that MM is connected to a parallel map made of q-neurons that I call a q-map, denoted by $qM_1$. We assume that a neuron $d_1$ is connected to all q-neurons of $qM_1$ (see Figure 6). 
  $d_1$ collects signals at its synapses from certain sensory, imaginary or emotional maps. When they activate $d_1$, this neuron in turn activates the q-map so that each of its q-neurons goes into random mode, with however a rather low activation probability. This activates a few neurons on the motor map which trigger a structured movement of several parts of the body, unless the latter is impracticable or dangerous; it is assumed that an inborn neural device, downstream of the motor map, inhibits the movement in this case. So  human being is indeed able of spontaneous movements.

\subsection{Observing His Hand\label{main}}

When an infant studies one of his hands, I propose the following scenario inside his brain.
Firstly the particular posture he adopts is innately recognized by the neuron $d_1$, which triggers spontaneous movement.\par

\begin{pspicture}(0,-3)(10,10)

\rput(1,9){\rnode{se}{
\begin{minipage}{2cm}
External\par
 stimulus
\end{minipage}}}

\rput(4.5,9){\rnode{c}{\psframebox{Maps}}}
\rput(8,9){\rnode{d}{\psframebox{$d_1$}}}
\rput(8,7){\rnode{v}{}}

\rput(8,5){\rnode{qc}{\pspolygon(-2,-1)(1,-1)(2,0)(-1,0)}}
\rput(10.5,4.5){$qM_1$}
\rput(8,2){\rnode{cm}{\pspolygon(-2,-1)(1,-1)(2,0)(-1,0)}}
\rput(10.5,1.5){$MM$}
\rput(8,-1){\rnode{ms}{
\begin{minipage}{2.5cm}
Spontaneous\par
movement
\end{minipage}
}}

\psset{arrows=->,arrowscale=1.5,linewidth=.05}

\ncline[nodesepB=-0.2]{v}{qc}
\ncline[nodesepB=-0.2,offsetB=0.4]{v}{qc}
\ncline[nodesepB=-0.2,offsetB=0.8]{v}{qc}
\ncline[nodesepB=-0.2,offsetB=1.2]{v}{qc}
\ncline[nodesepB=-0.2,offsetB=-0.4]{v}{qc}
\ncline[nodesepB=-0.2,offsetB=-0.8]{v}{qc}

\ncline[nodesepB=-0.5,offsetB=0.1]{v}{qc}
\ncline[nodesepB=-0.5,offsetB=0.4]{v}{qc}
\ncline[nodesepB=-0.5,offsetB=0.8]{v}{qc}
\ncline[nodesepB=-0.5,offsetB=1.2]{v}{qc}
\ncline[nodesepB=-0.5,offsetB=-0.5]{v}{qc}
\ncline[nodesepB=-0.5,offsetB=-1]{v}{qc}

\ncline[nodesepB=-0.8,offsetB=-0.1]{v}{qc}
\ncline[nodesepB=-0.8,offsetB=0.4]{v}{qc}
\ncline[nodesepB=-0.8,offsetB=0.8]{v}{qc}
\ncline[nodesepB=-0.8,offsetB=1.2]{v}{qc}
\ncline[nodesepB=-0.8,offsetB=-0.6]{v}{qc}
\ncline[nodesepB=-0.8,offsetB=-1.3]{v}{qc}

\ncline[nodesepA=1,nodesepB=-0.2,offset=-0.8]{qc}{cm}
\ncline[nodesepA=1,nodesepB=-0.2,offset=-0.4]{qc}{cm}
\ncline[nodesepA=1,nodesepB=-0.2,offset=0]{qc}{cm}
\ncline[nodesepA=1,nodesepB=-0.2,offset=0.4]{qc}{cm}
\ncline[nodesepA=1,nodesepB=-0.2,offset=0.8]{qc}{cm}
\ncline[nodesepA=0.8,nodesepB=-0.2,offset=1.2]{qc}{cm}

\ncline[nodesepA=1,nodesepB=-0.5,offset=-0.9]{qc}{cm}
\ncline[nodesepA=1,nodesepB=-0.5,offset=-0.5]{qc}{cm}
\ncline[nodesepA=1,nodesepB=-0.5,offset=-0.1]{qc}{cm}
\ncline[nodesepA=1,nodesepB=-0.5,offset=0.3]{qc}{cm}
\ncline[nodesepA=1,nodesepB=-0.5,offset=0.7]{qc}{cm}
\ncline[nodesepA=0.9,nodesepB=-0.5,offset=1.1]{qc}{cm}

\ncline[nodesepA=1,nodesepB=-0.8,offset=-1]{qc}{cm}
\ncline[nodesepA=1,nodesepB=-0.8,offset=-0.6]{qc}{cm}
\ncline[nodesepA=1,nodesepB=-0.8,offset=-0.2]{qc}{cm}
\ncline[nodesepA=1,nodesepB=-0.8,offset=0.2]{qc}{cm}
\ncline[nodesepA=1,nodesepB=-0.8,offset=0.6]{qc}{cm}
\ncline[nodesepA=1,nodesepB=-0.8,offset=1]{qc}{cm}

\psset{arrows=->,arrowscale=2,linewidth=.08}

\ncline{se}{c}
\ncline{c}{d}
\ncline{d}{v}
\ncline[nodesepA=1,nodesepB=0.2]{cm}{ms}

\rput[t](7,-2){figure 6 }

\end{pspicture}
	
	\par

 In Figure 7, the mechanism generating a spontaneous move is reproduced. It corresponds to arrows 1, 2 and 3. \par
If it is an observed hand movement, 
then the sensory and imaginary maps of the somesthesic and visual senses are activated together. We assume that a $test$ neuron can recognize this configuration using its synapses, correctly connected to the previous maps (arrow 6).

Let us assume that each neuron $n$ of $MM$ is symmetrically connected to a neuron $n'$ of a parallel map denoted $HM_2$ .  Suppose the $test$ neuron is also connected to all the neurons of $HM_2$ and $n'$ is activated  if an only if $n$ and $test$ are both active. Thus, when the spontaneous movement does not correspond to a hand movement, $HM_2$ remains inactive, whereas in the opposite case, the activation of $MM$ corresponding to this movement is copied into the parallel map $HM_2$ (arrows 4 to 7). 
If now $HM_2$ is a Hopfield network, it can memorize among the spontaneous movements those that correspond to an observed hand movement.\par\medskip

After this learning phase, for example during a sleep period, let us assume that a parallel q-map denoted $qM_2$ is connected to $HM_2$ (arrow b) and that a neuron $d_2$ is connected to all neurons of $qM_2$ (arrow a).
So, once our baby wakes up, he can move on to an active phase, which triggers the neuron $d_2$.
Via the arrow a, all q-neurons of $qM_2$ switch to random mode. They send on $HM_2$ a random signal which in a few iterations converges towards one of the memorized movements. This neural activity pattern is transmitted via the arrow c to the motor map that performs it (arrows d and e).\par

\begin{pspicture}(0,0)(15,12)

\rput(2,11){\rnode{ce}{\psframebox{
\begin{minipage}{1.9cm}
Emotional\par
map
\end{minipage}
}}}

\rput(6,11){\rnode{d1}{\psframebox{$d_1$}}}
\rput(6,9){\rnode{qc1}{\psframebox{$qM_1$}}}
\rput(6,7){\rnode{cm}{\psframebox{$MM$}}}
\rput(6,5){\rnode{ch2}{\psframebox{$HM_2$}}}
\rput(6,3){\rnode{qc2}{\psframebox{$qM_2$}}}
\rput(6,1){\rnode{d2}{\psframebox{$d_2$}}}
\rput(9,7){\rnode{m}{\psframebox{muscles}}}
\rput(13,7){\rnode{cs}{\psframebox{
\begin{minipage}{1.5cm}
Sensory\par
maps
\end{minipage}
}}}
\rput(9,5){\rnode{t}{\psframebox{test}}}

\psset{arrows=->,arrowscale=2,linewidth=.05}

\ncline{ce}{d1}
\naput{1}
\ncline{d1}{qc1}
\nbput{2}
\ncline{qc1}{cm}
\nbput{3}
\ncline[offset=-0.4]{cm}{ch2}
\nbput{7}
\ncline[offset=-0.4]{ch2}{cm}
\nbput{c}
\ncline{qc2}{ch2}
\nbput{b}
\ncline{d2}{qc2}
\nbput{a}
\ncline{cm}{m}
\naput{4,d}
\ncline{m}{cs}
\naput{5,e}
\ncangle[angleA=-90,angleB=0]{cs}{t}
\naput{6}
\ncline{t}{ch2}
\naput{7}

\rput[t](7,0){Figure 7}

\end{pspicture}

\par\bigskip

The repeated use of the neuron $d_2$ thus teaches the BAM placed between the imaginary visual maps and the motor map  to associate the movements of the hand with their vision, if we suppose that our baby continues to observe his hand.\par
The correct working of this device implies that the arrow c is inhibited during the learning phase and that the Hopfield network stops learning when the active phase is switched on.
\par\medskip

If we generalize, when $HM_i$ is a parallel Hopfield map symmetrically connected to $MM$, it can memorize certain movements during a learning phase. So
if we add a neuron $d_i$ and a q-map $qM_i$, according to the arrangement $(d_2,qM_2,HM_2,MM)$ in Figure 7, we obtain a device for random generation of the movements stored in $HM_i$.\par
Figure 8
 then shows how to use $HM_i$ to select from its stored movements those that satisfy another $test$ and store them in a second Hopfield map $HM_j$. By adding a similar device $(d_j,qM_j)$ to $HM_j$, we can then use $HM_j$ as a new random  generator of selected movements. For example, if $HM_i$ generates any hand movements, $HM_j$ may only retain those that coincide in the visual maps with movements made by baby's relatives and that he had memorized in an imaginary visual map. It could be gripping movements, particularly interesting because this is how his mother sometimes waved a toy in front of him. 
\par\medskip
By varying the part of the body concerned and the test used, we thus obtain the construction of several random generators of movements that the child can study and use at will.\par

By definition, the \it{decision map} is made up of the neurons $d_i$ as well as other neurons that will be specified later.
The neurons on this map are called decisional neurons.
\par
Assuming BAMs exist between this new map and the other maps, the use phase also makes it possible to link a movement of a relative's hand with the decisional neuron corresponding to the movements of the hand. So when mom moves her hands, baby is inclined to make a hand movement himself. A communication is setting up.\par\medskip

\begin{pspicture}(0,-1.5)(15,8)

\rput(0,7){\rnode{di}{\psframebox{$d_i$}}}
\rput(2,7){\rnode{qci}{\psframebox{$qM_i$}}}
\rput(4,7){\rnode{chi}{\psframebox{$HM_i$}}}
\rput(6,7){\rnode{cm}{\psframebox{$MM$}}}
\rput(6,5){\rnode{chj}{\psframebox{$HM_j$}}}
\rput(6,3){\rnode{qcj}{\psframebox[linecolor=red]{$qM_j$}}}
\rput(6,1){\rnode{dj}{\psframebox[linecolor=red]{$d_j$}}}
\rput(9,7){\rnode{m}{\psframebox{muscles}}}
\rput(13,7){\rnode{cs}{\psframebox{
\begin{minipage}{1.6cm}
Sensory\par
map
\end{minipage}
}}}
\rput(9,5){\rnode{t}{\psframebox{test}}}

\psset{arrows=->,arrowscale=2,linewidth=.05}

\ncline{di}{qci}
\naput{1}
\ncline{qci}{chi}
\naput{2}
\ncline{chi}{cm}
\naput{3}
\ncline[offset=-0.4]{cm}{chj}
\nbput{7}
\ncline[linecolor=red,offset=-0.4]{chj}{cm}
\nbput{\textcolor{red}{c}}
\ncline[linecolor=red]{qcj}{chj}
\nbput{\textcolor{red}{b}}
\ncline[linecolor=red]{dj}{qcj}
\nbput{\textcolor{red}{a}}
\ncline[offset=0.2]{cm}{m}
\naput{4}
\ncline[linecolor=red,offset=-0.2]{cm}{m}
\nbput{\textcolor{red}{d}}

\ncline[offset=0.2]{m}{cs}
\naput{5}
\ncline[linecolor=red,offset=-0.2]{m}{cs}
\nbput{\textcolor{red}{e}}

\ncangle[angleA=-90,angleB=0]{cs}{t}
\naput{6}
\ncline{t}{chj}
\naput{7}

\rput[t](7,0){
\begin{minipage}{12cm}
figure 8 : \par
\small
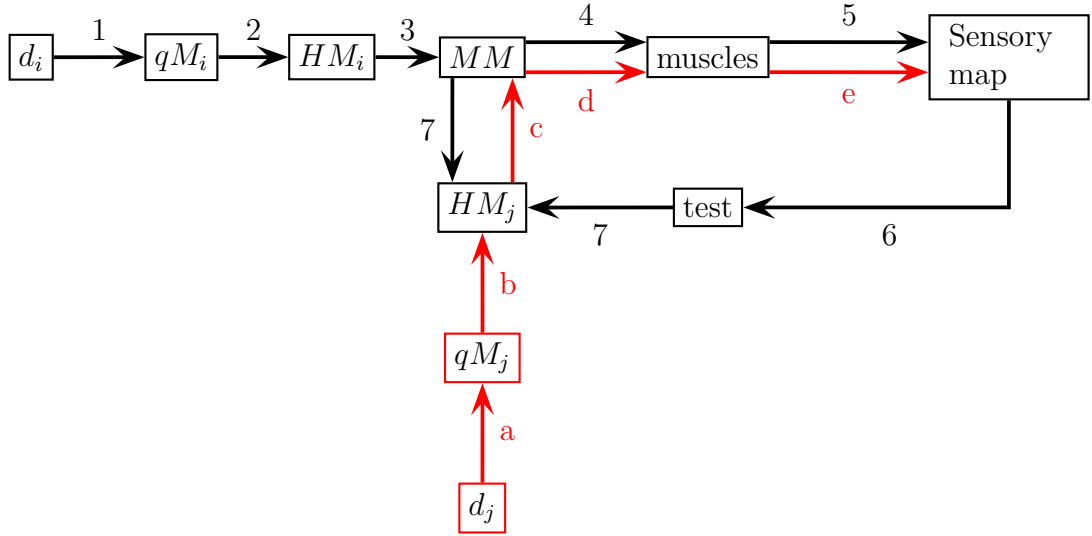
Arrows 1 to 7 : $HM_j$ learning phase.\par
\textcolor{red}{Arrows a to e} :  $HM_j$ use phase.
\end{minipage}}

\end{pspicture}
\par
\par\medskip

\subsection{Decision Algorithms}
Consider a family of decisional neurons $(d_i)_{1\leq i\leq k}$, each provided with a q-map denoted  $qM_i$ and a Hopfield map denoted $HM_i$, according to the previous paragraph. \par
Let us build a first decision algorithm which, if activated at the same time as one  $d_i$, imagines a movement chosen in $HM_i$ by $qM_i$, scores it positively or negatively according to a certain criterion, then actually executes it only if the score is positive. Figure 9 shows this algorithm, where 
we assume that $d_i=d_1$. It is assumed that an upstream device guarantees the impossibility of activating two neurons $d_j$ at the same time. Thus at time 1, only $d_1$ is activated, as well as $a_1$ which is the root neuron of the decision algorithm. \par
The description of $MM$ seen on page \pageref{72} is reproduced, where 
for each neuron $n$ of  MM, the connection  $n\rightarrow \hbox{mn}$ from $n$ to some motor neurons is replaced by the more complex form
$\begin{array}
	{ccccc}
	n&\rightarrow&n'&\rightarrow&mn\\
	&&\top&&\\
	&&n''&&\\
\end{array}$. This amounts to adding to $MM$ a map $MM'$ parallel to $MM$, consisting of neurons $n'$ and a map $MM''$ containing  neurons $n''$ (see Figure 9).  Let $MN$ be the set of  all motor neurons.  Thus, when $MM''$ is activated, it inhibits the  map $MM'$ which blocks the transmission of the $MM$ activity to the motor neurons and then to the muscles. In this case, the map $MM$  only activates imaginary maps via the BAMs.
\par
\begin{pspicture}(0,-2)(15,14)

\rput(5,1){\rnode{ce}{\psframebox{
\begin{minipage}{2.1cm}
Emotional\par
map
\end{minipage}
}}}

\rput(5,3){\rnode{ci}{\psframebox{
\begin{minipage}{2.1cm}
Imaginary\par
map
\end{minipage}
}}}

\rput(1,13){\rnode{v1}{}}
\rput(3,13){\rnode{v2}{}}
\rput(7,13){\rnode{v3}{}}

\rput(1,11){\rnode{d1}{\psframebox{$d_1\ \textcolor{red}{\bullet}$}}}
\rput(1,9){\rnode{qc1}{\psframebox{$qM_1$}}}
\rput(1,7){\rnode{ch1}{\psframebox{$HM_1$}}}

\rput(3,11){\rnode{dk}{\psframebox{$d_k\ \textcolor{red}{\bullet}$}}}
\rput(3,9){\rnode{qck}{\psframebox{$qM_k$}}}
\rput(3,7){\rnode{chk}{\psframebox{$HM_k$}}}

\ncline[linestyle=dashed,nodesep=1.5pt]{d1}{dk}
\ncline[linestyle=dashed,nodesep=1.5pt]{qc1}{qck}
\ncline[linestyle=dashed,nodesep=1.5pt]{ch1}{chk}

\rput(5,5){\rnode{cm}{\psframebox{$MM$}}}
\rput(7,5){\rnode{cmp}{\psframebox{$MM'$}}}
\rput(7,7){\rnode{cms}{\psframebox{$MM''$}}}

\rput(7,11){\rnode{a1}{\psframebox{$a_1\ \textcolor{red}{\bullet}$}}}
\rput(9,5){\rnode{mn}{\psframebox{$MN$}}}
\rput(9,1){\rnode{t}{\psframebox{$test$}}}
\rput(11,7){\rnode{i}{\psframebox{$I$}}}

\psset{arrows=->,arrowscale=2,linewidth=.05}

\ncline{v1}{d1}
\nbput{1}
\ncline{d1}{qc1}
\nbput{2}
\ncline{qc1}{ch1}
\nbput{3}

\ncline{v2}{dk}
\ncline{dk}{qck}
\ncline{qck}{chk}

\ncangle[angleA=-90,angleB=180,offsetB=-0.2]{ch1}{cm}
\nbput{4}

\ncangle[angleA=-90,angleB=180,offsetB=0.2]{chk}{cm}

\ncline{cm}{ci}
\nbput{6}
\ncline{ci}{ce}
\nbput{7}

\ncline{cm}{cmp}
\naput{5}

\ncline{cm}{ci}
\nbput{6}

\ncline[nodesep=1pt]{->|}{cms}{cmp}
\nbput{5}

\ncline{cmp}{mn}
\nbput{10}
\ncline{ce}{t}
\naput{8}

\ncline{v3}{a1}
\naput{1}
\ncangle[angleB=90]{a1}{i}
\naput{2}

\ncline{i}{cms}
\naput{3,4}

\ncangle[angleB=-90,nodesep=1pt]{->|}{t}{i}
\naput{9}

\rput(1.5,-1.3){\rnode{v4}{}}
\rput(2.5,-1.3){\rnode{v5}{}}

\ncline{->|}{v4}{v5}

\rput[tl](1.5,0){
\begin{minipage}{17cm}
figure 9 : \par
\small
$\textcolor{red}{\bullet}$ : indicates a neuron of the decision map;\par\medskip
\hspace{1cm} : inhibitory synapse.
\end{minipage}
}

\end{pspicture}
\par

In Figure 9, the times 1,2,3,4 correspond to the random generation of a movement of $HM_1$ and its transmission to the map $MM$. At the same time, $a_1$ activates a neuron $I$ that activates $MM''$.  Thus, at time 5, access to the motor neurons is blocked. At time 6, $MM$ communicates with the imaginary maps, which eventually generates a certain emotional situation. The neuron $test$ fires if and only if the emotion is positive, according to a criterion specific to the neuron $test$.  If the emotion is considered negative, nothing happens because access to $MN$ is still blocked, whereas, if the emotional score is positive, the neuron  $test$ is activated. It inhibits the neuron $I$ and that silences the map $MM''$. Then the map $MM'$  is no longer inhibited and in this case, the initially selected movement is transmitted to the motor neurons for execution. \par\medskip

This decision algorithm consists only of the three neurons $a_1$, $I$ and $test$. It is wired once for all decisional neurons $d_1,\ldots,d_k$. 
\par\medskip
\hspace{1.5cm}
\begin{minipage}{16cm}
\begin{pspicture}(0,-2)(15,17)

\rput(5,1){\rnode{ce}{\psframebox{
\begin{minipage}{2.1cm}
Emotional\par
map
\end{minipage}
}}}

\rput(5,3){\rnode{ci}{\psframebox{
\begin{minipage}{2.1cm}
Imaginary\par
maps
\end{minipage}
}}}

\rput(1,12){\rnode{v1}{}}
\rput(3,12.3){\rnode{v2}{}}
\rput(9,12){\rnode{v3}{}}

\rput(1,11){\rnode{d1}{\psframebox{$d_1\ \textcolor{red}{\bullet}$}}}
\rput(1,9){\rnode{qc1}{\psframebox{$qM_1$}}}
\rput(1,7){\rnode{ch1}{\psframebox{$HM_1$}}}

\rput(3,11){\rnode{d2}{\psframebox{$d_2\ \textcolor{red}{\bullet}$}}}
\rput(3,9){\rnode{qc2}{\psframebox{$qM_2$}}}
\rput(3,7){\rnode{ch2}{\psframebox{$HM_2$}}}

\rput(5,5){\rnode{cm}{\psframebox{$MM$}}}
\rput(7,5){\rnode{cmp}{\psframebox{$MM'$}}}
\rput(7,7){\rnode{cms}{\psframebox{$MM''$}}}

\rput(9,11){\rnode{a2}{\psframebox{$a_2\ \textcolor{red}{\bullet}$}}}
\rput(13,3){\rnode{d}{\psframebox{$d\ \textcolor{red}{\bullet}$}}}
\rput(10,5){\rnode{mn}{\psframebox{$MN$}}}
\rput(9,1){\rnode{t}{\psframebox{$test$}}}
\rput(9,9){\rnode{i}{\psframebox{$I$}}}

\rput(1,13){\rnode{at1}{\psframebox{
\begin{minipage}{1.7cm}
Wait for\par
200ms
\end{minipage}
}}}

\rput(5,11){\rnode{at2}{\psframebox{
\begin{minipage}{1.7cm}
Wait for\par
200ms
\end{minipage}
}}}
\rput(6,15){\rnode{at3}{\psframebox{
\begin{minipage}{1.7cm}
Wait for\par
100ms
\end{minipage}
}}}
\rput(6,13){\rnode{at4}{\psframebox{
\begin{minipage}{1.7cm}
Wait for\par
100ms
\end{minipage}
}}}

\rput(4,12){\rnode{e1}{\psframebox{$e_1$}}}
\rput(8,12){\rnode{e2}{\psframebox{$e_2$}}}
\rput(9,14){\rnode{qn}{\psframebox{$qn$}}}
\rput(9,3){\rnode{ch}{\psframebox{$HM$}}}
\rput(11,3){\rnode{qc}{\psframebox{$qM$}}}

\psset{arrows=->,arrowscale=2,linewidth=.05}

\ncline[offset=-0.2]{v1}{d1}
\nbput{1}
\ncline{d1}{qc1}
\nbput{2}
\ncline{qc1}{ch1}
\nbput{3}

\ncline[linestyle=dashed]{v2}{d2}
\ncline[offset=-0.2]{d2}{qc2}
\ncline{qc2}{ch2}

\ncangle[angleA=-90,angleB=180,offsetB=-0.2]{ch1}{cm}
\nbput{4}

\ncangle[angleA=-90,angleB=180,offsetB=0.2]{ch2}{cm}

\ncline{cm}{ci}
\nbput{6}
\ncline{ci}{ce}
\nbput{7}

\ncline{cm}{cmp}
\naput{5}

\ncline[nodesep=1pt]{->|}{cms}{cmp}
\nbput{5}

\ncline{cmp}{mn}
\naput{21}
\ncline{ce}{t}
\naput{8}

\ncline{v3}{a2}
\naput{1}
\ncline[offset=0.2]{a2}{i}
\naput{2}

\ncloop[angleA=-90,angleB=90,loopsize=1.2,arm=.5,linearc=.2,offset=-0.2]{i}{i}

\ncangle[angleA=-90,offsetA=0.2]{i}{cms}
\naput{3,4}

\ncline{t}{ch}
\naput{9}

\ncline{cm}{ch}
\ncput*{9}

\ncline{ch}{cmp}
\ncput*{20}
\ncline{qc}{ch}
\naput{19}
\ncline{d}{qc}
\naput{18}

\rput(13,9){\rnode{v6}{}}
\ncangle[angleB=90]{qn}{v6}
\naput{17}
\ncline{v6}{d}
\naput{17}
\ncline[nodesep=1pt]{->|}{v6}{i}
\naput{17}

\ncangles[angleB=90,arm=1]{a2}{at3}
\nbput[npos=2.5]{10}

\ncangle[angleA=90,nodesep=1pt]{->|}{qn}{at3}
\nbput{16}
\ncangle[angleB=-90]{at4}{qn}
\nbput{15}
\ncline{at3}{at4}
\naput{11}
\ncbar[angleA=180,angleB=180,arm=1]{at4}{at3}
\naput{12}

\ncbar[angleA=180, angleB=180,arm=1]{d1}{at1}
\ncline[offsetA=0.5,nodesepA=-0.2]{at1}{e1}
\ncangle[angleA=180,angleB=90,offsetB=0.2]{e1}{d1}
\naput{14}

\ncline{d2}{at2}
\ncline[offsetA=-0.3,nodesepA=-0.1]{at2}{e2}
\ncbar[angleA=-90,angleB=90,offsetB=-0.2,arm=0.7]{e2}{d2}

\ncangle[angleA=-90,offsetA=-0.2]{at4}{e1}
\ncput*{\scriptsize 13}
\ncangle[angleA=-90,angleB=180]{at4}{e2}
\ncput*{\scriptsize13}

\rput(1.5,-1.3){\rnode{v4}{}}
\rput(2.5,-1.3){\rnode{v5}{}}

\ncline{->|}{v4}{v5}

\rput[tl](1.5,0){
\begin{minipage}{17cm}
figure 10 : \par
\small
$\textcolor{red}{\bullet}$ : indicates a neuron of the decision map;\par\medskip
\hspace{1cm} : inhibitory synapse.
\end{minipage}
}

\end{pspicture}
\end{minipage}

\par

The second decision algorithm is a little more sophisticated. We admit 
again that an upstream device ensures that at most one $d_i$ is activated. 
This algorithm then reactivates $d_i$ a random number of times. At each iteration,
a movement is generated by $HM_i$, with some consequences in the imaginary maps and then in the emotional map, which makes it possible to assign a score to the movement. If positive, the algorithm 
copy the movement into a Hopfield map  denoted $HM$.  After this testing period, the algorithm randomly selects one of the movements recorded in $HM$ and actually executes it.\par

This decision algorithm is closer to our daily behaviours than the previous one;
we weigh different possibilities, we retain only a few, among which we execute one without really knowing why we chose it.
Figure 10 shows the algorithm structure, while restricting ourselves to $k=2$ and assuming again that $d_i=d_1$. The root neuron of this algorithm is denoted  by $a_2$.\par
The arrows 1 to 8 are identical to those of the first algorithm.
At time 9, if the test is positive, the Hopfield map $HM$ stores the movement from 
$HM_1$. Otherwise, in the absence of synaptic activation from the neuron $test$, the map $HM$ is not activated by $MM$ and the movement is not memorized.\par
The time 10 starts at the same moment as the time 1, however, the neurons ``wait 100ms'' cause a delay. These neurons, which may even be networks of neurons, pick up an input signal and output it 100 milliseconds later.
 Suppose firstly that the q-neuron $qn$ remains inactive. Then, with a period of 200ms, via 13, the neurons $e_j$ all receive a first synaptic activation. However, only $e_1$ receives a second synaptic activation, induced by the axonal activity of $d_1$ 200ms earlier, while $d_2$ was inactive at the same time. The thresholds of the neurons $e_j$ are such that only $e_1$ is activated, which again triggers the activation of $d_1$, thus the generation of a movement in $HM_1$, every 200ms.\par

Via the arrow 15, $qn$ is  stimulated every 200ms. This switches on its random behaviour every 200ms. Now suppose that $qn$ actually becomes active. Then the arrow 16 turns off the periodic activation of $d_1$ and the arrow 17 activates the neuron $d$ which activates the q-map $qM$, which causes $HM$ to randomly pick one of its recorded movements. The arrows 20 and 21 ensure the muscular achievement of the movement.\par
The neuron $I$ activates itself because a part of its axon returns to one of its synapses. Thus, when activated at  the time 2, it remains activated indefinitely and continuously blocks the passage from the motor map to the motor neurons, which ensures that the movements generated during the test period are only imagined. However, when $qn$ becomes active, the arrow 17 brings a synaptic inhibitory component to the neuron $I$, which extinguishes its activity.\par
On average, the number of movements selected in $HM$ is equal to $n=1+\di\frac1{P_0}$, where $P_0$ is the firing probability of $qn$. We can consider that $P_0$ is
 adjusted according to the emotion of the moment, so that $n$ is about ten in case of serenity and only about 3 or 4 in case of anger or fear. \par
This decision algorithm consists of about ten neurons and two maps $HM$ and $qM$ parallel to $MM$.
Like the first algorithm, it is wired once for all decisional neurons $d_1,\ldots,d_k$.  There is no difficulty to add a neuron $d_{k+1}$ as well as its maps 
$qM_{k+1}$ and $HM_{k+1}$ to this second algorithm. 

\par\medskip
One can thus imagine multiple such decision algorithms. For each of them, we assume that its root neuron belongs to the decision map. Possibly other neurons that articulate the algorithm, such as the neuron $d$ in the second algorithm, also belong to the decision map.\par
We can also consider that other decision algorithms control which of the previous root neurons is activated and therefore which of the previous decision algorithms is used, and that they also control the choice of the neuron $d_i$, so the choice of the library $HM_i$ of movements that is used. 
For example, by linking the root neurons of several decision algorithms by ``xor'' q-neurons (see page \pageref{64}), we can choose randomly one of them.\par

\subsection{Abilities of the Decision-Making System}

\subsubsection{Adaptability}
The appearance in one imaginary map of an object likely to provide a food, tactile, financial (and so on) reward induces an emotion of desire, when the high-level sensory maps do not perceive the object, that is, when it is only imagined. This desire, encoded in the emotional map, activates a decision algorithm acting on one or more libraries of movements. If it succeeds in fulfilling this desire, the chosen algorithm is reinforced, by modifying the probabilities of the q-neurons which led to this choice, and the chosen movements are reinforced by reproducing them several times in the Hopfield map where they are memorized, for example during a period of sleep. Conversely, defective choices are undermined for the future.\par
The decision-making system thus adapts to its environment. It optimizes  rewards and minimizes punishments. 

\subsubsection{Learning and Imitation}
Thanks to the different BAMs linking his cortical maps,  a human being can  spot moves of others that lead to rewards or avoid inconveniences. He memorizes the perceptions of these moves in his imaginary maps, then learns to reproduce them in his libraries of movements.
For example, by observing his relatives grasp toys, a baby visualizes some gripping moves and then learns to reproduce them.

\subsubsection{Mixing Thought and Impulse}
A Hopfield map $HM_i$  parallel to the motor map is not only used to generate a random movement among all those it has memorized; we can consider that some decision algorithms generate a movement in $HM_i$ by providing, at the same time as a random signal from $qM_i$, the signal present in $MM$. We have seen that the latter is the sum of influences from the imaginary maps and that it often contains the move statistically best adapted to previous actions, i.e. the move that  would be done impulsively. By properly weighting the signal from 
$qM_i$ and the signal from  $MM$ with a correct setting of  the synaptic efficiencies,
the movement chosen by $HM_i$ is more or less close to the thoughtless movement contained in $MM$.

\subsubsection{Creating}

One can also consider decision algorithms that imagine, evaluate and then execute not a single elementary movement, but a succession of such movements selected in several $HM_i$. 
In front of a problem requiring the implementation of several moves, chance can thus provide to the imagination of the observer solutions more or less good, and sometimes an amazing discovery.

\subsubsection{Storing a Succession of Elementary Movements}
One can also consider a triple $(d_i,qM_i,HM_i)$ which memorizes a succession of $k$ elementary movements, where $2\leq k\leq 5$. When $k=3$ for example, $qM_i$ and $HM_i$ are composed of 3 layers of maps parallel to $MM$. This complicates the wiring of learning phases of $HM_i$ and of  reading phases  of $HM_i$ by $MM$. It is not insurmountable however.

\subsection{Attention\label{att}}
Thereafter, when we refer to a \it{map}, without specifying anything else, it may be  indifferently
 a high-level sensory map, 
an imaginary map, the emotional map, the motor map, the decision map or finally the attention map that we  now define. \par\medskip

In accordance with the neural architecture that we have just set up, a single initial sensation activates an imaginary map and then using different BAMs activates other maps. It thus can trigger a decision algorithm with the generation of several imagined movements which themselves  lead to multiple other configurations of activities in the different maps. So, without an additional mechanism
that can inhibit some of these activities, the maps would most often fall into an incessant buzzing. Among the countless information coming from the environment, they would not succeed in filtering the relevant information, notably for the survival of the organism.\par
 To avoid such overheating, we assume that for any region of any map, all neurons in the region receive a synaptic inhibitory contribution from one single  neuron. So the activation of this neuron 
cancels most of the activities in that region; attention is then diverted from that region.\par
We also assume the existence of a second neuron that provides an exciting synaptic contribution to all neurons in the same region. It focuses attention on this region.\par
The set of these new neurons constitutes by definition the attention map. 
\par\medskip

An attentional inhibitor neuron must not, however, prevent emergency signals from activating certain neurons in certain maps, which is  in fact the mark of the emergency.  In this case, it is likely that the attention map detects the emergency and in turn stimulates the region where the emergency came from to better manage it. Figure 11 shows one way to implement such a mechanism; 
$n_i$ is an attentional inhibitory neuron whose region of influence contains the neuron $n_u$ (arrow 1). Arrows 2 and 3 indicate an emergency signal that activates $n_u$ despite the inhibition of $n_i$. The neuron $n_l$ then receives two synaptic excitations from $n_i$ and $n_u$, which activate $n_l$. Via the arrow 4, $n_l$ inhibits the activity of $n_i$, which removes the inhibition that $n_i$ had on $n_u$ and on the other neurons under its influence. 
Still via the arrow 4, $n_l$ activates the excitatory neuron $n_e$ on the attention map that has the same region of influence as $n_i$. So attention is now turned to the region containing the neuron $n_u$. 

\par\medskip

\begin{pspicture}(0,-1)(12,9)

\rput(6,1){\rnode{gu}{
\begin{minipage}{3cm}
Handling of\par
the emergency
\end{minipage}
}}

\rput(0,5){\rnode{v1}{}}
\rput(6,5){\rnode{v2}{}}
\rput(10,5){\rnode{v3}{}}

\rput(4,5){\rnode{nu}{\psframebox{$n_u$}}}
\rput(8,3){\rnode{ne}{\psframebox{$n_e$}}}
\rput(8,5){\rnode{nl}{\psframebox{$n_{l}$}}}
\rput(8,7){\rnode{ni}{\psframebox{$n_i$}}}

\psset{arrows=->,arrowscale=2,linewidth=.05}

\ncline{v1}{nu}
\naput[npos=0.4]{emergency}
\nbput[npos=0.8]{2}

\ncangle[angleA=180,angleB=-90]{ne}{nu}
\naput{5}
\ncangle[angleA=180,angleB=90,nodesep=1.5pt]{->|}{ni}{nu}
\nbput{1}

\ncline[linecolor=red]{nu}{v2}
\naput{\textcolor{red}{3}}
\ncline[linecolor=red]{v2}{gu}
\naput[npos=0.3]{\textcolor{red}{3}}
\ncline[linecolor=red]{v2}{nl}
\naput{\textcolor{red}{3}}

\ncline{ni}{nl}
\naput{3}
\ncline{nl}{v3}
\naput{4}

\ncangle[angleA=-90]{v3}{ne}
\naput{4}
\ncangle[angleA=90,nodesep=1.5pt]{->|}{v3}{ni}
\nbput{4}

\rput[t](7,0){figure 11}

\end{pspicture}

\par

In emergency situations, sensory and imaginary maps can therefore act on the attention map. It is also controlled by the emotional map because each emotion has its own  pattern of attention.\par
It is still controlled by the decision map;  for example, when a decision algorithm imagines a movement to which it needs to assign a score, it acts on the attention map to focus attention on the adapted area of the emotional map and on the areas of the sensory maps that matter for scoring. 
  \par\medskip
The decision map can also act directly on the attention map, which we denote thereafter by $AM$, the same way it acts on $MM$; 
If $d$ is a neuron that can activate a q-map $qM$ parallel to $AM$ and connected to $AM$, then the activation of $d$ induces a random attention profile. By adapting Figures 7 and 8, we can then construct  some Hopfield maps $HM$ parallel to $AM$ which store some attention profiles adapted to certain tasks. For example, the fundamental exercise of mindfulness meditation is to learn to actively focus one's attention on breathing [C.T], [M.L.C, page 40]. According to my modelling, this learning builds a Hopfield map $HM$ containing several attention profiles; one profile that privileges the chest muscles and the diaphragm, another that privileges the breathing rhythm, one profile that focuses on inhalation and exhalation, another that studies the temperature variations of the nasal air etc. This map $HM$  is associated with 
a q-map and a $d$ neuron, which belongs to the decision map. To practice this meditation exercise is to activate $d$ to trigger one of these attention profiles, then activate $d$ again  relentlessly to change the attention profile. 

\subsection{Objects and Their Methods\label{meth}}
After extensive learning, an adult has, for each object $o$ in his environment, a decision neuron $d_o$ which can generate randomly one of the movements that act on this object, using the maps $qM_o$ and $HM_o$. In object-oriented programming terminology, these movements are the methods of the object. 
When the individual feels or imagines the presence of the object $o$, the neuron $d_o$ is stimulated and if the other synaptic contributions of $d_o$ allow it, it emits an action potential, which generates a method acting on the object.  At the same time, a decision algorithm is chosen to define the strategy for using this method. For example, the presence of a door in my vision field
stimulates $d_{door}$, whose methods are ``act on the handle to open'', ``act on the handle to close'',
``slam the door'', ``wait for opening if it is an automatic door'' etc. 
 \par\medskip
Let us estimate the number of neurons involved in all the maps $qM_o$ and $HM_o$. It is reasonable to approximate the number of objects associated with such maps by the number of words an individual uses to talk, taking into account proper names. This number is between 1000 and 20000 depending on whether the individual is more or less cultured.
 Moreover, to manage the 570 skeletal striated muscles, let us assume that the number of neurons on the motor map is between 2000 and 50000. Then, for each object, the union of its q-map and Hopfield map contains a number of neurons between $1.2\ 10^4$ and $3\ 10^5$, if we use maps 
that can code three successive movements on average. 
The number of neurons mobilized by all these maps
is between $1.2\ 10^7$ and $6\ 10^9$, that is between $0.012\%$ and $6\%$ of all the neurons of our  brain. So it is not beyond his capabilities.

\subsection{Language}

Pronouncing an elementary phoneme of our language is a movement among others. A baby can learn it by a process similar to that of paragraph \ref{main}; while he is making spontaneous gestures, the baby focuses his attention on his auditory map. The device in Figure 7, with an adapted neuron $test$, then feeds a Hopfield map $HM_p$ with the different phonemes he pronounces.
During the $HM_p$ using phase, the BAM linking $MM$ and the auditory map learns to associate the hearing of a phoneme with its pronunciation.\par
Then, when a close relative pronounces a phoneme in front of him, the baby is inclined to reproduce this phoneme and repeat it many times. This reinforces the phonemes actually used in the language of his relatives, while the others are gradually forgotten by $HM_p$. 
\par\medskip
If in the sentences used to communicate with the baby, a certain phoneme is often followed by another phoneme, the BAM between $MM$ and the auditory map will associate the hearing of the first phoneme with the pronunciation of the second, according to the process detailed at the end of  chapter \ref{motr}. So when the baby says ``mum'', he is inclined to continue with ``my''. \par\medskip

Naming an object is usually one of the methods of that object, which a child can learn when one shows him the object while distinctly uttering its name. The child understands that he is in a learning situation. This activates the root neuron of an appropriate decision algorithm; the child repeats the name, or simply imagines that he repeats the name. This inserts successively in $MM$ the phonemes of the name which are then stored in the Hopfield map of the object.

\section{Self-Consciousness\label{soi}}

\subsection{The Ego}
We have already assumed that 
high level sensory maps, imaginary maps and emotional maps are made of c-neurons.  We now do the same assumption for the decision map. On the contrary we assume that the motor and attention maps are made of classic neurons.\par
So, $\bf{mind}=\bf{sense}\otimes\bf{emo}\otimes\bf{dec}$, where $\bf{sense}$ is the tensor product of the  c-bits on the sensory and imaginary maps, where $\bf{emo}$ is the product of the c-bits 
 of the emotional map and where $\bf{dec}$ is the tensor product of the c-bits of the decision map. \par
The state of $\bf{sense}$ is the set of sensations perceived and imagined by the mind.  The state of $\bf{emo}$ is the set of feelings experienced by the mind. What is the state of $\bf{dec}$?
\par\medskip

A decision algorithm acts on q-neurons, often structured in q-maps, to make random choices, including choices of movements or attention profiles in a Hopfield map. The neurons that act on the synapses of q-neurons to set them in random mode are part of the decision map.\par
A decision algorithm also has a specific strategy for using these different random choices. Some key neurons in the algorithm are also part of the decision map. \par
The latter therefore informs, in the context of a decision-making strategy, of the presence of  random choices made by q-neurons and how their results are used for the final decision. \par

So there is a difference in nature between the information carried by the state of \par
$\bf{sense}\otimes \bf{emo}$ and that carried by the state of $\bf{dec}$;
 In the first case, changes in c-bits are interpreted as consequences, real or only imagined, of changes of objects located outside or inside the body according to a flow of activity going from these objects to the c-bits. In the second case, on the contrary,  changes of the c-bits of the decision map inform on the changes that their c-neurons cause downstream. 
Decisional neurons are focused on motor and attention neurons; decisional neuron activity is perceived as a cause of motor and attention neuron activity.

\par\medskip

To be more precise, let us take again the example of  hand movements. Let us suppose that we decide at the moment $t_0$  to imagine a hand movement and execute it immediately.  According to our model, a neuron $d$ on the decision map acts on the q-neurons of a q-map and sets them in random mode at time $t_1$.  This random activity of the q-map is transmitted to a Hopfield map that has memorized some hand movements. After a few iterations, from the signal provided by the q-map, the Hopfield map converges to one of the stored movements. According to the algorithm we have chosen, this movement is 
transmitted to the motor map for execution. The BAMs directly linking $MM$ to the  imaginary maps provides first the mind at time $t_2$ with the imagination of the movement, quickly followed at time $t_3$ by the sensations related to the muscular actions of the movement.  $t_3>t_2$ because these real sensations come from a more complex path than the imagined sensations: precise neuronal shaping of the movement, muscle activation, activation of the sensory neurons linked to this movement, which then feed up the sensory maps.\par
Let us\label{83} denote $\bf{mind}=\bf{mind'}\otimes \bf{D}$, where $\bf{D}$ is the tensor product of  c-bits of the c-neuron $d$.  Let us also denote by $\bf{Q}$ the q-map concerned and $\bf{env}$ the rest of the universe.
At time $t_0$, from our mind's  point of view
\footnote{According to the linearity principle FP2, we can limit ourselves to such a separable form for $\bf{Q}\otimes\bf{env}$.},\par
$univ(t_0)=[mind'\otimes D_{inactive}]\otimes Q_{inactive}\otimes env$. 
Then\par
 $univ(t_1)=[mind'\otimes D_{active}]\otimes \di\sum_{i}\alpha_iQ_i\otimes env_i$,\par
 where the family $(Q_i)_i$ contains the
different possible states of the q-map.\par
 Then, at the moment $t_2$, from our mind's  point of view, \par
$univ(t_2)=[mind'_i\otimes D_{active}]\otimes Q_i\otimes env_i$, where
$mind'_i$ contains the imagined sensation of the hand movement.
At last, at time $t_3$,\par
 $univ(t_3)=[mind''_i\otimes D_{active}]\otimes Q_i\otimes env_i$, where
$mind''_i$ contains  the imagined sensation followed by the actual sensation of the hand movement.
\par
Thus, the mind perceives the activation of the c-neuron $d$, then the imagination of a hand movement, then the sensation of its actual execution. We can decide, using a suitable decision algorithm, to repeat this experience several times and the chronology will always be the same;
the mark left in $\bf{dec}$ by the activation of $d$ at the time $t_1$
 is systematically followed by the imagination of a variable movement of one hand then the sensations of the actual execution of the same movement.\par
Between the event $D_{active}$ that occurs in $\bf{dec}$ and the sequence of events $(mind'_i,mnid''_i)$ that occurs in $\bf{sense}\otimes\bf{emo}$, we find all the characteristics of a causal link: \par
\begin{itemize}
\item the anteriority;
\item the necessity,
because the presence of $D_{active}$ is systematic before a hand movement occurs;
\item the specificity, because the event $D_{active}$  occurs only before the appearance of a hand movement.
\end{itemize}
\par
This is the notion of causality as conceived by David Hume in his
``Enquiry Concerning Human Understanding'', that is, causality based on experience and habit.\par
Thus, from the mind's point of view, the event $D_{active}$ is perceived as the cause of the events 
$(mind'_i,mind''_i)$.
More generally, when an individual uses a decision algorithm, the state of $\bf{dec}$ is, from his
 mind's point of view, the cause of this algorithm and its consequences, of which the movement actually decided among the various possible movements is part.
This cause is therefore responsible for the execution of the movement. 
 This approach is consistent with the theory of Daniel M. Wegner presented in
 ``Pr\' ecis of The Illusion of Conscious Will''  [DM.W].
\par
 In short,
$\bf{dec}$ being the cause of the choices made by the brain in the context of a decision making,
this is the ego,  the soul, ``I'', the self and so on.
Thereafter, I thus replace the notation $\bf{dec}$ by $\bf{ego}$. 

\subsection{Free Will, Libet and Soon Experiments}
\subsubsection{Protocol and Modelling of Libet Experiment}
After studying the delays of conscious perception, Benjamin Libet conducted experiments concerning free will [B.L, pages 141-157]; he asks a seated subject to flex his wrist as soon as the subject has decided to, avoiding any pre-planning of his gesture.
The experimenter measures several times:
\begin{itemize}
	\item 
 the moment $S$ of the wrist flexion is measured with an electrode that can detect the muscle activity corresponding to flexion.
\item 
The moment $P$ of the beginning of the motor preparation performed by the brain is evaluated by electrodes applied to the subject's scalp, near motor or premotor areas. $P$ therefore indicates the moment when some neurons in these areas fire to trigger a wrist flexion.
\item 
The moment $W$ when the subject becomes aware of his will to act is evaluated by asking the subject, during the experiment, to watch a clock indicating the time with an accuracy better than 50ms. When he becomes aware of his will to act, the subject must at the same time memorize the  clock position.
\end{itemize}
Note that $W$ is a subjective time, derived from the subject's conscious experience, whereas $S$ and $P$ are objective times. 
\par
Benjamin Libet obtains on average $W=S-150ms$, i.e. the subject becomes aware of his willingness to act 150ms before his action, and $P=W-350ms$ which is more surprising because it means that some neurons on the motor map initiate the flexion \it{before} it is consciously decided. This experience seems to prove that free will is an illusion, that the decisions of a human being are not decided by him but only by subliminal parameters that influence him unconsciously.
\par\medskip

In the context of our modelling, this experiment can be described at the neural level by Figure 12.
When the subject is informed of the experiment protocol, he places himself in a learning situation. 
Using a mechanism similar to that of Figure 8,  he can memorize in a Hopfield map $HM$ a wrist flexion, and even several wrist flexions, more or less fast. $HM$ is controlled by a q-map and a decisional c-neuron $d$, so that the activation of $d$ triggers a wrist flexion according to arrows 7 to 10. \par
To conform to the experiment protocol, the subject chooses a decision algorithm among the different algorithms available.  Its root neuron is denoted by $a$. At the beginning of the experiment, he activates $a$.
 As long as the q-neuron $qn$ remains inactive, loop 3-4 is executed which produces nothing for 50ms. At each iteration of the loop, the q-neuron is set in random mode. Its firing probability is tuned so that the average waiting time is about 20 seconds. When the q-neuron finally fires at the moment $t$, arrow 5 inhibits loop 3-4 and arrow 6 activates $d$. 

\par

\begin{pspicture}(0,-2)(10,14)

\rput(9,1){\rnode{mi}{\psframebox{
\begin{minipage}{3.5cm}
memorize the\par
clock position
\end{minipage}
}}}

\rput(2,3){\rnode{at2}{\psframebox{
\begin{minipage}{1.7cm}
Wait for\par
25 ms
\end{minipage}
}}}

\rput(2,5){\rnode{at1}{\psframebox{
\begin{minipage}{1.7cm}
Wait for\par
25 ms
\end{minipage}
}}}

\rput(9,4){\rnode{at3}{\psframebox{
\begin{minipage}{1.7cm}
Wait for\par
500 ms
\end{minipage}
}}}

\rput(6,4){\rnode{d}{\psframebox{$d\ \textcolor{red}{\bullet}$}}}
\rput(2,7){\rnode{a}{\psframebox{$a\ \textcolor{red}{\bullet}$}}}

\rput(6,7){\rnode{qc}{\psframebox{q-map}}}
\rput(4,4){\rnode{qn}{\psframebox{qn}}}

\rput(6,9){\rnode{ch}{\psframebox{$HM$}}}
\rput(6,11){\rnode{cm}{\psframebox{$MM$}}}
\rput(6,13){\rnode{m}{\psframebox{muscles}}}
\rput(2,8.5){\rnode{v1}{}}

\psset{arrows=->,arrowscale=2,linewidth=.05}

\ncline{v1}{a}
\nbput{1}
\ncline{a}{at1}
\nbput{2}

\ncangle[angleA=90,nodesep=1pt]{->|}{qn}{at1}
\nbput{5}
\ncangle[angleB=-90]{at2}{qn}
\nbput{4}
\ncline{at1}{at2}
\naput{3}
\ncbar[angleA=180,angleB=180,arm=1]{at2}{at1}
\naput{4}

\ncline{qn}{d}
\naput{6}
\ncline{d}{qc}
\nbput{7}
\ncline{qc}{ch}
\nbput{8}
\ncline{ch}{cm}
\nbput{9}
\ncline{cm}{m}
\nbput{10}
\ncline{d}{at3}
\nbput{11}
\ncline{at3}{mi}
\nbput{12}

\rput(1.5,-1.3){\rnode{v4}{}}
\rput(2.5,-1.3){\rnode{v5}{}}

\ncline{->|}{v4}{v5}

\rput[t](7,0){
\begin{minipage}{10cm}
figure 12 : \par
\small
$\textcolor{red}{\bullet}$ : indicates a neuron of the decision map;\par\medskip
\hspace{1cm} : inhibitory synapse.
\end{minipage}
}

\end{pspicture}
\par

\subsubsection{Free Will, a Mind's Point of View}
Let us take the notations from page \pageref{83}, where $\bf{Q}$ now stands for  $qn$. So at the beginning of the experiment, $univ(t_0)=[mind'\otimes D_{inactive}]\otimes Q_{inactive}\otimes env$, then after about 30 seconds, \par
$\eqalign{univ(t_1)=&\di\sum_{i\in I}\alpha_i[mind'_i\otimes D_{inactive}]\otimes\sum_{j} \beta_{i,j}Q_{inactive, i,j}\otimes env_{i,j}
\hfill\cr
&+\di\sum_{i\in J}\alpha_i[mind'_i\otimes D_{active,i}]\otimes \sum_j \beta_{i,j}Q_{active,i,j}\otimes env_{i,j} : \hfill\cr}$
\par
The first sum represents a coexistence of states of the universe in which  $qn$ has not yet fired. In the second sum, the term containing
 $mind'_i\otimes D_{active,i}$ represents a universe in which the q-neuron fired at time $t=\tau_i$, which is encoded in the mind by $D_{active,i}$, because the activation of $qn$ causes the activation of the c-neuron $d$. $mind'_i$ represents the state of the rest of the mind after  the wrist flexion.

From the subject's   point of view, $mind(t_0)=mind'\otimes D_{inactive}$ is a mind aware of the start of the experiment thanks to the c-bits of $a$. Then, in the case where $qn$ fires at the moment $t=\tau_i$, after  the wrist flexion,  
$mind(t_1)=mind'_i\otimes D_{active,i}$.
 Chronologically, we will see that the  state change of $\bf{D}$ occurs just before the state changes of 
$\bf{mind'}$  that result from the activation of the motor map by $d$. 
So, from the perspective of the subject's mind, which only perceives one of the terms at the moment $t_1$, some person chose, between $t_0$ and $t_1$, that the moment $t$ would be $\tau_i$. And everything indicates that that person is $\bf{ego}$, of which $\bf{D}$ is a factor: anteriority, necessity, specificity.\par
It is a free decision, because the subject is aware that he could just as easily have chosen another value of $t$. In fact, this is what happens in the other terms of the second sum of $univ(t_1)$. The feeling of free will is consistent with the fact that the collapse of the wave function is not real but only perceived by the mind in each term. 

\subsubsection{Strengthening of Causality}
The interpretation to be given to $W$ is delicate. If we apply a cutaneous stimulation to the subject and ask him to spot the time $M$ indicated by the clock when he becomes aware of it, he indeed indicates  the actual moment of the stimulation [B.L, page 148].
Thus, the time $M$ is calculated by the brain according to a non detailed mechanism that memorizes the position of the clock as it is coded by the neuronal activity of the visual imaginary map, at the moment when the neurons of the sensory map sensitive to the cutaneous stimulation are activated. When testifying about the time $M$, the subject  uses this memorized image to construct his report. In reality, according to paragraph \ref{Lib}, the subject becomes aware of the cutaneous stimulation at time $M+500ms$, however this conscious event is backdated and has indeed $M$ as time label in the mind. \par

Nethertheless, I show now that to reinforce the feeling of causality, it is preferable that the c-bits of $d$ do not backdate its action potential. Under these conditions, the subjective time $W$ is equal to $t+500ms$. It is assumed that the brain performs this calculation correctly, according to arrows 11 and 12; the action potential of $d$ is delayed 500ms before being used as an input to the mechanism that previously calculated $M$.

Thus, the decision to flex the wrist is made at the moment $t$, but the awareness of this decision by the mind occurs only at the moment $W=t+500ms$ (see Figure 13).
\par\medskip

\hspace{3cm}
\begin{minipage}{10cm}
\begin{pspicture}(0,-1.5)(8,2)
\psset{arrowscale=2, linewidth=.05}

\psline{->}(-1,0)(8,0)
\rput(0,1){$t$}
\psline(0,-0.5)(0,0.5)

\rput(1.5,1){$P$}
\psline(1.5,-0.5)(1.5,0.5)
\rput(0.75,-0.5){\small 150 ms}

\rput(5,1){$W$}
\psline(5,-0.5)(5,0.5)
\rput(3.25,-0.5){\small 350 ms}

\rput(6.5,1){$S$}
\psline(6.5,-0.5)(6.5,0.5)
\rput(5.75,-0.5){\small 150 ms}

\rput[t](4,-1){figure 13}

\end{pspicture}
\end{minipage}

\par
\par\medskip

The relationship $P=W-350ms$ can be written $P=t+150ms$ which restores a coherent chronology: the neurons of $MM$ are activated only \it{after} the decision is made.\par
The relationship $W=S-150ms$ can be written $S=P+500ms$: a rather long delay of $500ms$ is thus necessary between the activation of the motor map and the muscular activity.
 It corresponds to signal propagation time, motion control and muscle activation, but it is 5 times slower than the rapid succession of movements studied on page \pageref{key}
at the end of the chapter \ref{key}.  
In order to explain this relative slowness of these  thoughtful movements which result from
 of a decision algorithm, we can provide two reasons:
\begin{itemize}
\item The map $HM$  is part of the motor areas, so its activity is perceived by the electrodes. Then the moment $P$ corresponds to the beginning of the $HM$ activity, which takes some time to converge towards one of its learned movements.
\item When a movement results from a decision algorithm, one can suppose that a mechanism of temporization intervenes at the level of the motor map to possibly allow a counter-decision.
\end{itemize}

The neural circuit that produces the imagination of the movement is faster because it uses BAMs that connect the motor map to the imaginary maps.
Let us assume that the flexion is imagined 150ms before being executed and 
that it still takes 150ms for the actual movement to be perceived in the sensory maps. 
Assuming that both the imagined movement and the actual movement are backdated by 500ms when encoded in c-bits, their conscious time labels coincide with real time. Then the imagined movement reaches consciousness with time $S-150ms=W$ and the real movement is consciously perceived with time $S+150=W+300ms$.
This reinforces the feeling that the ego is the cause of the decision;  the moment he becomes aware of his decision to act, the mind imagines the movement he is going to make, and then it is realized. Thus, from the mind's point of view, the ego has not only decided the moment of the wrist flexion but has also decided the parameters of this movement, including its speed. \par
 If the awareness of the decision to act had been backdated, one would have obtained $S-150=W+500ms$ : the movement would consciously be imagined $500ms$ after the ego decided to act. The mind would perceive an ego that has decided to act, 
but without knowing if the flexion he has decided is more or less rapid, 
so without knowing the choices he is supposed to have made.

\subsubsection{Soon Experiment}
Chun Siong Soon and his colleagues studied a variant of the previous experiment [CS.S]. The seated subjects have two buttons on which they can press, respectively with the right or left index finger. The subject is asked, whenever he wishes, to decide which button to press and to execute this action as soon as it is decided, avoiding any pre-planning of the time and choice of button.
The results of the experiment indicate that the decision is consciously made at the time $W=S-750$ on average, but here $S$ indicates the moment when the button is actually pressed, which probably explains the delay of some $600ms$ compared to Libet experiment.\par
On the neural level, the decision algorithm of the previous experiment can be used again, the only difference being that the map $HM$ now stores right button pressure movements and left button pressure movements. Thus, the above explanations remain valid with this new protocol.
\par
However, instead of electrodes, the experimenters analyze the subjects' brains  with functional magnetic resonance imaging. Each region of the brain is analyzed to determine the extent to which its activity predicts the subject's decision, using statistical pattern recognition methods. The authors obtained predictions with a reliability between 55\% and 60\%, and that nearly 10 seconds \it{before} the subjective time $W$  of the decision. Such a result seems a priori to invalidate the previous model according to which decision making is done at the moment $t=W-500ms$ and give the coup de grâce to the notion of free will.
\par
Nethertheless, in the framework of my model, it is likely that during the 10 seconds before the decision is made, the subject imagines different ways to press the right or left button, and that the map $HM$ memorizes these imagined movements. Depending on whether more right or left index movements are imagined during this period, the result of the convergence in $CM$ of the random signal provided by the q-map will be more or less likely to be a right or left index movement. This preliminary brain activity would explain why statistical techniques can detect bias in the subject's future decision, as Eddy Nahmias [E.N] suggests. The reliability of predictions does not exceed 60\%, which is not far from the 50\% that one would obtain by making purely random predictions. They would therefore only be the detection of a bias in the subject's decision, a bias introduced into the map $HM$ by a right/left imbalance during a preliminary process of memorization of the movements. Despite this bias, the decision is actually made when the neuron $d$ fires.\par
In the subject's mind, the moment  this decision is made is defined by a change of the $\bf{ego}$ state, who is perceived as the cause. The imagination of the decided movement arrives in the mind at the same time. So the ego seems to be the one who freely decided which button to press and when to do it. 

\subsection{Consistence and Inconsistence of the Ego}

The conscious perception of the ego is real, thanks to the c-bits of the decision map. 
The decisions made by the ego have consequences for the future of the individual and for the society in which he evolves. These consequences inevitably induce positive emotions of success or self-confidence, or negative emotions of failure, shame or guilt. They induce a sense of responsibility that gives the ego a very real consistence.
\par\medskip
However, if the ego decides to analyze this consistence, if he wants to clarify his nature, beyond the evidence of his existence as a perception of the mind, beyond the consequences of his decisions, he becomes inconsistent, unlike other objects perceived by the mind, which can be analyzed, decomposed, compared, categorized, etc.
Indeed, the ego is never really the cause of a decision, it is only the indication that a decisional c-neuron has been activated, but this activation itself results from upstream neuronal activations, conscious or unconscious. Moreover, when he activates q-neurons, he does not decide what result they produce; indeed, the other possible results are actually realized because, from an external point of view, the universe is a coexistence of universes and each of the possible decisions is indeed made in one of them. \par
Still from an external point of view, quantum mechanics is deterministic, so the only cause of the future is the past. And if we reverse the arrow of time,  the future  becomes the cause of the past. Thus, the notion of causality is no longer relevant. The ego is not at the origin of the decisions he believes he is making, he is only the indication in the mind that the mind has made a quantum measurement on some q-neurons. This does not absolve us of our responsibility; the consequences of our decisions are very real as well as the emotions they induce and, morally, we are supposed to change our future behaviors to reduce their negative consequences, therefore we are supposed to have a plasticity that strengthens the mechanisms that lead to good decisions and weakens those that lead to bad decisions. 

\subsection{The Meaning of 
Conscious Perceptions
\label{sens}}

By the mere fact of existing, the mind's perceptions are conscious. Through their reproducibility, they acquire meaning because this enables them to be associated with each other in fairly flexible causal relationships. For example, the imagination of the word ``dog'' leads to the visualization of a dog. When the ego does not intervene, if attention is diverted from the decision map, the word ``dog'' evokes various memories stored in the imaginary maps and by association, our mind can become aware of the conversation we had with a neighbor who was walking his dog. When the ego intervenes, he can decide to widen the class of  dogs to that of animals and study what this evokes to him, he can on the contrary restrict his attention to a particular dog. Thus, through multiple associations and multiple choices, my brain can explain to the mind what a dog is, by reference to other notions, which he can just as well explain using other notions. The mind contains the perception of an ego that actively maintains the network of his knowledge of the world.  By deciding to remember them, to analyze their components, to compare them, by seeking out analogies, causal links, similarities, generalizations etc.,
 the ego becomes the owner of this knowledge and acquires in the mind a central place of  conductor. \par
The fundamental motivation that leads the ego to study his knowledge relentlessly is the instinct for survival; natural selection has favoured organisms that seek good control of their environment, to avoid dangers and maximize rewards. 
\par\medskip
On the neural level, a child's relatives are associated, in his brain, with objects within the meaning of paragraph \ref{meth}, equipped with methods that permit him to name his relatives and communicate with them in multiple ways.\par
A child also has objects corresponding to each part of his body, which he manages to aggregate into a single object.  This object corresponds to that relative who appears in the mirror, to his body and to the ego, because each decision of the ego, in definitive,  concern a movement of the body. Thus the ego is personified in an object, one of whose methods is to say ``I''. And each decision of the ego is accompanied by the activation of this object, which for example induces the imagination of the body or a part of the body.
This object ``I''  has specific algorithms. In particular, 
the triggering of a decision algorithm causes another algorithm to be activated in the object ``I'' to analyze the decision and its consequences, which induce a certain emotion.
\par
Thus, the object ``I'' becomes central in the good functioning of the organism during the waking periods. It is essentially controled by the judgment of the ego facing his own decisions:  The emotions generated by these judgments largely control the long-term behaviour of the individual. 
\par
On the sensory level, the ego becomes the owner of the body; it is not only the one who decides, who judges his decisions, who chooses where to focus his attention, but also the one who feels. He is the conductor of every perception in the mind.\par
Objects that represent relatives are also equipped with specific algorithms. One of them is to imagine what we would do in someone else's place. The choices we imagine then are the result of the activation of decisional neurons which induce in the mind the awareness of ``self'' as the cause responsible for decisions, but the concomitant presence of c-bits specific to the algorithm we describe indicate that the cause must be shifted to others.  This is the case for all supposedly sensitive objects. Then, from a single soul felt in the mind, we switch to a multitude of spiritual entities, which sometimes catch our full attention. 
\par
Moreover, we understand that others evaluate and judge us according to the same mechanisms, which reinforces the need to pay attention to our ego, to its actions, to our judgment of its actions as well as to the judgments we assume others make about our actions.

\par\medskip
This is how our soul and those of others become essential objects of our daily life. Our survival instinct requires us to search for their deep nature in order to better understand and master them. Nethertheless, if ``I'' can decide to analyze and dissect each of my sensory and emotional perceptions, according to the previous paragraph, I cannot return this capacity of analysis on the heart of the ego, his decision-making component, whose true nature is beyond my senses. 
I think that this is the source of a fundamental dissatisfaction that is related to what Buddhism calls ``Dukkha'' [W.R]. 

\subsection{Uniqueness of Mind}

\subsubsection{Parts of the Mind}\label{90}
According to hypothesis H3, $\bf{mind}$ 
is the tensor product of $N$ c-bits $(\bf{c}_1,\ldots,\bf{c}_N)$.\par
If $J$ is a subset of $\{1,\ldots,N\}$, then $\bf{mind}_J=\di\bigotimes_{i\in J}\bf{c}_i$ also satisfies
 the constraints of H3. 
So $\bf{mind}_J$ is also an observer.  We would get $2^N$ observers, which is a lot for one body.
\par\medskip
These partial consciousnesses do exist, but they do not have a self-consciousness who analyses and maintains relentlessly the sense of his
perceptions.
The existence of such self-consciousness requires not only that $J$ contain the c-bits of the decisional  c-neurons, but also all the c-bits needed for the perception of the ego as the conductor of a global network of knowledge of the world. \par
It is likely that among all the $mind_J$, 
$\bf{mind}=\di\bigotimes_{1\leq i\leq N}\bf{c}_i$ is the only mind with a working self-consciousness. 

\subsubsection{Union of Several Minds}
Let $\bf{mind}^1$ and $\bf{mind}^2$ be  the minds of two observers.
Let $(\bf{c}_1,\ldots,\bf{c}_{N_1})$ be the c-bits of $\bf{mind}^1$ and 
$(\bf{c}_{N_1+1},\ldots,\bf{c}_{N_1+N_2})$ those of $\bf{mind}^2$. 
So $\bf{mind}^1\otimes \bf{mind}^2=\di\bigotimes_{1\leq i\leq N_1+N_2} \bf{c}_i$
satisfies
 the constraints of H3. He is also an observer.\par
However, if we consider a conscious state $mind^1\otimes mind^2$, it corresponds to the conjunction of a consciousness state of $\bf{mind}^1$ and a consciousness state  of $\bf{mind}^2$. If we change the c-bit states of $\bf{mind}^2$, the consciousness of $\bf{mind}^1$ is not (immediately) affected. Similarly, changing the c-bit states of $\bf{mind}^1$ does not affect the consciousness state of $\bf{mind}^2$, so this ``united'' observer does exist, but he does not have a self-consciousness. He has two of them, each experimenting the universe independently of the other, in accordance with the relationships presented at the end of paragraph \ref{coex} page \pageref{30}. 
\par
More generally, the union of several observers is indeed a consciousness, but it is devoid of a unified self-consciousness. 

\newpage

\section*{Conclusion}
\addcontentsline{toc}{section}{Conclusion}
Starting from the principles of non-relativistic quantum mechanics, without an a priori definition of probabilities, we have developed a particular conception of the latter as well as the notions of observer, existence and consciousness:\par
An object exists when it is in a pure state, however such an existence can also be read as the simultaneous coexistence of several existences which do not interact between them and which interfere. This is particularly the case of the universe in its entirety, which can be read as a coexistence of states of the form $mind_i\otimes env_i$, where $mind_i$ is a pure state of a mind and where $env_i$ is the coexistence of environments that set the mind in the state $mind_i$.  His consciousness becomes meaningful when his states are reproducible memories and he has a suitable neural environment. Before such a mind measures the state of an object, the probability for that object to be after the measurement in the state $obj_i$ is defined as the proportion of the number of minds that will experience the reality $obj_i$ after the measurement. 
\par\medskip
However, the assumptions we have imposed on an observer are constraining. Practically, are they actually implanted in some of our neurons? Theoretically, are they only possible?     Can these assumptions be relaxed in less restrictive forms? 
The last chapter of the first part explored these questions but did not exhaust them. 
\par\medskip
I finish by mentioning a few points that should be addressed or developed.
\begin{itemize}
\item The second part provides a too sketchy model of higher cognitive functions and self-consciousness. It should be developed significantly. 
\item In particular, the action that the mind exerts on its environment is hardly mentioned. 
According to what mechanisms do c-neurons read the state of their c-bits and how does this reading intervene in neuronal activity, in decision-making? What are the selective advantages for the organism provided by c-bits?

\item
What are the metaphysical consequences of our approach?

\item 
 On a more technical level, how our interpretation
harmonize with the use of probabilities in statistical physics, particularly with the notion of entropy?  What does it tell us about irreversibility and the arrow of time?

\item
What becomes of this article in the framework of a quantum gravity theory? 

\item Do some unicellular organisms and plants possess a mind in the form of a tensor product of c-bits? This would allow these brainless organisms to have a form of consciousness.
\end{itemize}

\newpage
\section*{Acknowledgements}
\addcontentsline{toc}{section}{Acknowledgements}
I am very grateful to 
Maxim Kontsevich who allowed me to present this work to him at IHES, in the presence of Vasily Pestun, Henri Epstein, Yan Soibelman and Thibault Damour. I thank them for their attentive and benevolent listening. Their comments and questions led me to strengthen several ideas.\par
My warmest thanks go to my friends Eric Platon, Daniel Ivanier, Michel Gar\c{c}on, Pascal Boulnois, Luc Albert and Eric S\'er\'e for their advice, re-reading and relevant comments.\par
I would like to thank Anne Burban and all the ``Inspection G\'en\'erale de Math\'ematiques'' who allowed me to take a sabbatical year to complete this work. \par
But this article would not exist without C\'eline, my wife, Florette and Philippe, my children, Alla Putintseva and Daniel L\'eopold, my friends. Thank you for your unfailing and invaluable support. \par

\newpage

\section*{Bibliography}
\begin{itemize}

\item\textbf{[A.A.1]} : Andreas Albrecht, Daniel Philips. Origin of Probabilities and Their Application to the Multiverse.
arxiv : 1212.0953v3. 2014.\par\medskip

\item\textbf{[A.A.2]} : Anthony Aguirre, Max Tegmark. 
Born in an Infinite Universe: a Cosmological Interpretation of Quantum Mechanics.\par
arXiv : 1008.1066v2 [quant-ph]   2012.\par\medskip

\item\textbf{[A.D.O]} : A. D. O'Connell, M. Hofheinz, M. Ansmann, Radoslaw C. Bialczak, M. Lenander, Erik
Lucero, M. Neeley, D. Sank, H. Wang, M. Weides, J. Wenner, John M. Martinis, A.N 
Cleland. Quantum Ground State and Single Phonon Control of a
Mechanical Resonator.
\textit{Nature 464, 697-703}. 2010.\par\medskip

\item\textbf{[A.E]} : A. Einstein, B. Podolsky, N. Rosen. Can Quantum-Mechanical Description of Physical Reality Be 
Considered Complete?\par
 \textit{Physical Review, vol 47 pg 777-780}. 1935.\par\medskip

\item\textbf{[A.H]} : Alan Hajek, ``Mises Redux''- Redux : Fifteen Arguments Against Finite Frequentism, Kluwer Academic Publishers, Erkenntnis
45 : pg 209-227. 1997. \par\medskip

\item\textbf{[A.K]} : Alexander Kraskov, Numa Dancause, Marsha M.Quallo, Samantha Sherpherd, Roger N.Lemon, 
Corticospinal Neurons in Macaque Ventral Premotor Cortex with Mirror Properties : A Potential Mechanism for Action
Suppression? \textit{Neuron 64, 922-930, }2009.\par\medskip

\item \textbf{[AK.S]} : Anja K.Sturm, Peter K\"{o}nig (2001) Mechanisms to Synchronize Neuronal Activity. \textit{Biological Cybernetics 84, 153-172.}
\par\medskip

\item\textbf{[A.M]} : Andr\'e L.G. Mandolesi. Analysis of Wallace's Decision Theoretic Proof of the Born Rule in Everettian
Quantum Mechanics.\par
 arXiv : 1504.05259v1 [quant-ph]. 2015.\par\medskip

\item\textbf{[A.M.G]} : 
A. M. Gleason, Measures on the Closed Subspaces of a Hilbert Space, \textit{J. Math. Mech. 6, 885-894}
(1957).\par\medskip

\item\textbf{[A.N]} : The Wave Function : Essays on the Metaphysics of Quantum Mechanics, de Alyssa Ney et David Z. Albert, Oxford University Press. 2013.\par\medskip

\item\textbf{[AR.D]} : Antonio R. Damasio. Le sentiment m\^eme de soi, corps, \'emotions, conscience, Editions Odile Jacob, 1999.\par\medskip

\item\textbf{[B.G]} : This is a translation of \it{The Hidden Reality}:\par
 La r\'ealit\'e cach\'ee, les univers parall\`eles et les lois du cosmos, de Brian Greene, Editions Robert Laffont. 2011.\par\medskip

\item\textbf{[BH.T]} : 
Toyama BH, Savas JN, Park SK, et al. Identification of Long-Lived Proteins Reveals Exceptional Stability of Essential Cellular Structures. \par
\textit{Cell. 154(5):971-982}. 2013
doi:10.1016/j.cell.2013.07.037.\par\medskip

\item\textbf{[B.K]} : Bart Kosko, Bidirectional Associative Memories, \textit{IEEE Transactions on syst\`emes,Vol 18 No 1,} 1988.\par\medskip

\item \textbf{[B.L]} : 
This is a  translation of  \it{Mind time : The temporal factor in consciousness}:\par
L'esprit au-del\`a des neurones, de Benjamin Libet,  \'editions Dervy, 2012.\par\medskip

\item \textbf{[B.M]} : Bradley Monton (2001) Mechanisms to Synchronize Neuronal Activity. \textit{Biological Cybernetics 84, 153-172.}
\par\medskip

\item\textbf{[C.A.F]} : Christopher A.Fuchs. Quantum Mechanics as Qantum Information (And Only a Little More).
arXiv : quant-ph/0205039v1, 2002.\par\medskip

\item\textbf{[C.C]} : Craig Callender. The Emergence and Interpretation of Probability in Bohmian Mechanics. Studies in History and Philosophy of Science Part B: Studies in History and Philosophy of Modern Physics
Volume 38, Issue 2, June 2007, Pages 351-370.\par\medskip

\item\textbf{[C.K]} : 
Keysers C, Gazzola V. 2014,
Hebbian Learning and Predictive Mirror Neurons
for Actions, Sensations and Emotions. \textit{Phil.
Trans. R. Soc. B 369: 20130175.}
http://dx.doi.org/10.1098/rstb.2013.0175
\par\medskip

\item \textbf{[C.L.1]} : Introduction \`a la chimie quantique, de Claude Leforestier, Editions Dunod, 2012.\par\medskip

\item\textbf{[C.L.2]} : Chang Liu, Fuchun Sun. Hmax Model : a Survey. \textit{IEEE 978-1-4799-1959-8.} 2015.\par\medskip

\item\textbf{[CS.S]} : Chun Siong Soon, Marcel Brass, Hans-Jochen Heinze, John-Dylan\par
 Haynes. Unconscious Determinants of Free Decisions in the Human Brain. \par
\textit{Nature neuroscience. }2008.
doi : 10.1038/nn.2112.\par\medskip

\item\textbf{[C.T]} : Respirer la vie, de Catherine Ternaux, \'Editions de la table ronde, 2003.\par\medskip

\item\textbf{[CW.G]} : 
Colin W. Garvie, Cynthia Wolberger. Recognition of Specific DNA Sequences. \textit{Molecular Cell, Vol. 8, 937-946,} November, 2001.\par\medskip

\item\textbf{[D.D]} : David Deutsch, Adriano Barenco, et Artur Ekert.
Universality in Quantum Computation, 1995.\par
Available on the internet, on arXiv : quant-ph/9505018v1\par\medskip

\item\textbf{[D.H]} : Hebb D. 1949 The Organisation of Behaviour. New
York, NY: John Wiley and Sons.\par\medskip

\item\textbf{[D.M]} : Cours complet Pr\'epa MP \& MP* de Denis Monasse, Vuibert Sup\'erieur. 1998.\par\medskip

\item\textbf{[DM.W]} : Pr\'ecis of The Illusion of Conscious Will, de Daniel M. Wegner, Cambridge University Press, 2005.\par\medskip

\item \textbf{[D.P.1]} :
Purves D., Paydarfar J. A., Andrews T. J. (1996). The Wagon Wheel Illusion in Movies and Reality. \par
\textit{Proc. Natl. Acad. Sci. U.S.A. 93, 3693-369710.1073/pnas.93.8.3693}
\par\medskip

\item \textbf{[D.P.2]} :
Purves D, Augustine GJ, Fitzpatrick D, et al., editors. Neuroscience. third edition. Sunderland (MA): Sinauer Associates; 2004. 
\par\medskip

\item \textbf{[D.P.3]} :
Purves, Augustine, Fitzpatrick, Hall, Lamantia, White. Neuroscience (french translation). 5i\`eme edition. de boeck  2015. 
\par\medskip

\item \textbf{[D.R]} : El\'ements de biologie cellulaire, 4i\`eme \'edition, de Daniel Robert et Brigitte Vian, Doin \'editeurs, 2013.\par\medskip

\item \textbf{[D.W.1]} : David Wallace (Avril 2005).  Quantum Probability from Subjective Likelihood:  Improving on Deutsch's Proof of the Probability Rule. \par
Available on the internet, on arXiv : quant-ph/0312157v2\par\medskip

\item \textbf{[D.W.2]} : The Emergent Multiverse, de David Wallace, Oxford university press, 2012.\par\medskip

\item\textbf{[E.D]} : Des r\'eseaux de neurones, de Eric Davalo et Patrick Na\"{i}m. Editions Eyrolles.1990.\par\medskip

\item\textbf{[E.N]} : Eddy Nahmias. Why We Have Free Will. \textit{Scientific American,} janvier 2015. \par\medskip

\item\textbf{[F.B]} : Florian Boge, On Probabilities in the Many Worlds Interpretation of Quantum Mechanics.\textit{University of Cologne 
Bachelor Thesis}. 2016.\par\medskip

\item\textbf{[G.B]} : Bi G, Poo M. 2001 Synaptic Modification by
Correlated Activity: Hebb's Postulate Revisited. \textit{Annu.
Rev. Neurosci. 24, 139-166.} (doi:10.1146/annurev.
neuro.24.1.139)\par\medskip

\item \textbf{[G.T]} : Giulio Tononi, Melanie Boly, Marcello Massimini, Christof Koch. Integrated Information Theory : From Consciousness to Its Physical Substrate. (2016) \textit{Nature reviews neuroscience, advance online publication.}
\par\medskip

\item\textbf{[G.V]} : Ganesh Vigneswaran, Roland Philipp, Roger N.Lemon, Alexander Kras\-kov, 
M1 Corticospinal Mirror Neurons and Their Role in Movement Suppression During Action Observation, 
\textit{Current Biology 23, 236-243, }2013.\par\medskip

\item\textbf{[H.B]} : 
La conscience de la vie,  de 
Henri Bergson, Conf\'erence Huxley donn\'ee \`a  l'Universit\'e de Birmingham, le 29 mai 1911.
\par\medskip

\item \textbf{[HD.Z]} :
H.D. Zeh. The Problem of Conscious Observation in Quantum Mechanical Description. arXiv:quant-ph/9908084v3 (2000).
\par\medskip

\item\textbf{[H.E.1]} : Hugh Everett, III. 
``Relative State'' Formulation of Quantum Mechanics. 
\textit{Reviews of Modern Physics, Vol. 29, No. 3, 454-462}, July 1957.
\par\medskip

\item\textbf{[H.E.2]} : Hugh Everett, III. The Theory of the Universal Wave Function. Princeton University Press. 1973.\par\medskip

\item \textbf{[H.MH]} :
Herzog MH, Kammer T, Scharnowski F. Time Slices: What Is the Duration of a Percept? \textit{PLoS Biol. 2016;14:e1002433. }doi: 10.1371/journal.pbio.1002433
\par\medskip

\item \textbf{[I.C]} : Itai Carmeli, Karuppannan Senthil Kumar, Omri Heifler, Chanoch Car\-me\-li, Ron Naaman : 
Spin Selectivity in Electron Transfer in Photosystem.
\textit{Chem Int Ed Engl. Aug 18;53(34):8953-8.} 2014. doi: 10.1002/anie.201404382. Epub 2014 Jul 2.
\par\medskip

\item\textbf{[JA.B]} : James A. Bednar, Stuart P.Wilson : Cortical Maps.  \textit{White Rose university consortium}. 2015.\par\medskip

\item\textbf{[J.C]} : Joshua Combes, Christopher Ferrie, Matthew S. Leifer, Matthew F. Pusey. Why Protective Measurement Does Not Establish the Reality of the Quantum State. arXiv : 1509.08893v1 [quant-ph] 2015.\par\medskip

\item\textbf{[J.D.C]} : Quantum Physics Notes, J.D. Cresser, Macquarie university. 2009.\par
\textit{Available on the internet, on }\par
\verb#http://physics.mq.edu.au/~jcresser/Phys301.html#\par\medskip

\item\textbf{[JG.C]} : John Cramer.  The Transactional Interpretation of Quantum Mechanics.  \textit{Reviews of Modern Physics 58, 647-688,} July (1986)\par\medskip

\item\textbf{[J.H]} : John J.Hopfield, David W.Tank, Computing with Neural Circuits : A Model,\textit{ Science, New series, Vol 233 No 4764,} 1986.\par\medskip

\item \textbf{[J.H.1]} : Les spineurs en physique, de Jean Hladik, Editions Masson, 1996.\par\medskip

\item \textbf{[J.H.2]} : R\'eseaux neuronaux et traitement du signal, de Jeanny H\'erault et Christian Jutten, Editions Hermes, 1994.\par\medskip

\item\textbf{[J.N]} : Mathematical Foundations of Quantum Mechanics, de John Von Neumann, Princeton University Press, 1955.\par\medskip

\item \textbf{[JP.N]} : R\'eseaux de neurones de la physique \`a la psychologie, de Jean-Pierre Nadal, Editions Armand Colin, 1993.\par\medskip

\item \textbf{[J.Z]} : J.Zihl, D. Von Cramon, N. Mai ; Selective Disturbance of Movement Vision After
Bilateral Brain Damage, \textit{Brain, Volume 106, Issue 2,} 1 June 1983, Pages 313-340, 
https://doi.org/10.1093/brain/106.2.313
\par\medskip

\item\textbf{[K.I]} : 
Karolinska Institutet. How Epigenetic Changes in DNA Are Interpreted. \textit{ScienceDaily,} 2017.\par\medskip

\item\textbf{[L.L]} :  Liaofu Luo. Protein Folding as a Quantum Transition Between Conformational States.\textit{ Front. Phys. 6(1), 133-140 }(2011). doi: 10.1007/s11467-010-0153-0 \par\medskip

\item\textbf{[L.V]} : L. Vaidman. Protective Measurements of the Wave Function of a Single System. arXiv : 1401.6696v2 [quant-ph] 2014.\par\medskip

\item\textbf{[M.F]} : Matthew P.A. Fisher. Quantum Cognition: The Possibility of Processing with Nuclear Spins in the Brain. (2015)
\textit{Available on the internet, on } https://arxiv.org/abs/1508.05929
\par\medskip

\item\textbf{[M.F.P]} : Matthew F. Pusey, Jonathan Barrett, Terry Rudolph. On the Reality of the Quantum State. 
arXiv : 1111.3328v3 [quant-ph] 2012.\par\medskip

\item\textbf{[M.K]} : Karen Michaeli, Nirit Kantor-Uriel, Ron Naaman, David Waldeck : 
The Electron's Spin and Molecular Chirality - How Are They Related and How Do They Affect Life Processes? Michaeli K, Kantor-Uriel N, Naaman R, Waldeck DH.
\textit{Chem Soc Rev.  Nov 21; 45(23):6478-6487.} 2016.
\par\medskip

\item\textbf{[M.L.C]} : Vive la m\'ediation !, de Marie-Laurence Cattoire. Leduc's \'Editions, 2017.\par\medskip

\item\textbf{[M.R]} : Maximilian Riesenhuber, Tomaso Poggio. Hierarchical Models of Object Recognition in Cortex. \textit{Nature neuroscience, vol 2, no 11.} 1999.\par\medskip

\item\textbf{[M.S.1]} : Decoherence and the Quantum-to-Classical Transition, de Maximilian Schlosshauer, Editions Masson,  2007.\par\medskip

\item\textbf{[M.S.2]} : Maximilian Schlosshauer, Tangereen V.B. Claringbold. Entanglement, Scaling, and the Meaning of the Wave Function in Protective Measurement. arXiv : 1402.1217v2 [quant-ph] 2014.\par\medskip

\item\textbf{[M.S.L]} : Matthew Saul Leifer. Is the Quantum State Real? An Extended Review of $\psi$-Ontology Theorems.
arXiv : 1409.1570v2 [quant-ph] 2014.\par\medskip

\item\textbf{[M.T]} : Max Tegmark. The Importance of Quantum Decoherence in Brain Processes. (1999)
\textit{Available on the internet, on }
arXiv: quant-ph/9907009v2\par\medskip

\item\textbf{[MV.LB]} : Michael V.L Bennett, R.Suzanne Zukin. 
Electrical Coupling and Neuronal Synchronization in the Mammalian Brain.
\textit{Neuron Volume 41, Issue 4,} 19 February 2004, Pages 495-511.\par
https://doi.org/10.1016/S0896-6273(04)00043-1
\par\medskip

\item\textbf{[N.M]} :  
Quantum Computation and Quantum Information, de Michael A. Niel\-sen  et Isaac L. Chuang,
Cambridge university press, 2010.
\par\medskip

\item\textbf{[O.C]} : Omid Charrakh. On the Reality of the Wavefunction.\par
arXiv : 1706.01819v1 [physics.hist-ph]. 2017.\par\medskip

\item\textbf{[PF.C]} : Pierre-Fran\c{c}ois Carton, Romain Pacaud, Gilles Salbert. \par
M\'ethylation/d\'e\-m\'e\-thy\-la\-tion de l'ADN et expression du g\'enome. \textit{Revue francophone des laboratoires. num\'ero 473.} 2015.\par\medskip

\item\textbf{[PJ.U]} :
Uhlhaas PJ, Pipa G, Lima B, et al. Neural Synchrony in Cortical Networks: History, Concept and Current Status. \textit{Frontiers in Integrative Neuroscience.} 2009;3:17. doi:10.3389/neuro.07.017.2009.\par\medskip

\item\textbf{[P.M]} : The Interpretation of Quantum Mechanics and the Measurement Process, de Peter Mittelstaedt, Cambridge university press, 1998.\par\medskip

\item\textbf{[P.VG]} : Biologie g\'en\'erale, 4\`eme \'edition, de Paulette Van Gansen et Henri Alexandre, Editions Springer, 2007.\par\medskip

\item\textbf{[RB.G]} : Robert B.Griffiths. Consistent Quantum Measurements.\par
  	arXiv:1501.04813 [quant-ph]. 2015.\par\medskip

\item\textbf{[R.C.1]} : Ramakrishna Chakravarthi, Rufin VanRullen. Conscious Updating Is a Rhythmic Process. \textit{PNAS Early Edition}, 2012.\par
www.pnas.org/cgi/doi/10.1073/pnas.1121622109.\par\medskip

\item\textbf{[R.C.2]} : R. Cooke, M. Keane, and W. Moran, An Elementary Proof of Gleason's Theorem, \textit{Math. Proc.
Camb. Phil. Soc. 98, 117-128} (1981); I. Pitowsky, Infinite and Finite Gleason's Theorems and the
Logic of Indeterminacy, J. Math. Phys. 39, 218-228 (1998).\par\medskip

\item\textbf{[R.F]} : Roman Frigg, Carl Hoefer. Probability in GRW theory. Studies in History and Philosophy of Science Part B: Studies in History and Philosophy of Modern Physics.
Volume 38, Issue 2, June 2007, Pages 371-389.\par\medskip

\item\textbf{ [R.R]} :  Richard Rolleigh. The Double Slit Experiment
and Quantum Mechanics. Available on the internet. 2010. \par\medskip

\item\textbf{[R.W]} : Richard Webb. First Quantum Effects Seen in Visible Object. \textit{New scientist,} 17 March 2010.
https://www.newscientist.com/article/dn18669-first-quantum-effects-seen-in-visible-object/\
\par\medskip

\item\textbf{[R.W.S]} : Robert W. Spekkens. In Defense of the Epistemic View of Quantum States : A Toy Theory.
arXiv : quant-ph/0401052v2. 2005.\par\medskip

\item \textbf{[S.A.1]} : 
Atasoy, S., Donnelly, I., Pearson, J. (2016). Human Brain Networks Function in Connectome-Specific
Harmonic Waves.\par
 \textit{Nature Communications, 7, 1-10}. http://doi.org/10.1038/ncomms10340\par\medskip

\item \textbf{[S.A.2]} : 
Selen Atasoy, Gustavo Deco, Morten L Kringelbach, Joel Pearson (2017).
Harmonic Brain Modes: A Unifying Framework for Linking Space and Time in Brain Dynamics.
bioRxiv 162040; doi: https://doi.org/10.1101/162040\par
Now published in \textit{The Neuroscientist} doi: 10.1177/1073858417728032
\par\medskip

\item\textbf{ [S.E]} : Sandra Eibenberger, Stefan Gerlich, Markus Arndt, Marcel Mayor,
Jens T\"{u}xen. Matter-Wave Interference of Particles Selected from a
Molecular Library with Masses Exceeding 10 000 Amu. \textit{Phys.Chem.}  2013,
15, 14696.\par\medskip

\item \textbf{[S.G.1]} : Shan Gao. Protective Measurement : A Paradigm Shift in Understanding Quantum Mechanics. (2013) \textit{Available on the internet : }\par
philsci-archive.pitt.edu/9627/4/PMPS.pdf
\par\medskip

\item \textbf{[S.G.2]} : Shan Gao. An Ontological Interpretation of the Wave Function. (2013) \textit{Available on the internet :}
\par
http://philsci-archive.pitt.edu/10129/1/mwf\_2013\_v3.pdf
\par\medskip

\item \textbf{[S.G.3]} : Shan Gao. An Argument for $\psi$-Ontology in Terms of Protective Measurements. 
arXiv : 1508.07684v1.
(2015) 
\par\medskip

\item\textbf{[S.H]} : Stuart Hameroff et Jack Tuszynski, Quantum States in Proteins Assemblies : The Essence of Life?
\textit{Proc. of SPIE Vol 5467, pg 27-41,} 2015.
\par\medskip

\item\textbf{[S.O]} : 
Ostojic, S., Brunel, N. \& Hakim, V. 
Synchronization Properties of Networks of Electrically Coupled Neurons in the Presence of Noise and Heterogeneities.
\textit{J Comput Neurosci  26: 369.} (2009) https://doi.org/10.1007/s10827-008-0117-3
\par\medskip

\item\textbf{[S.P]} :
Pockett S, Brennan BJ, Bold GEJ, Holmes MD. A Possible Physiological Basis for the Discontinuity of Consciousness. \textit{Frontiers in Psychology.} 2011;2:377. doi:10.3389/fpsyg.2011.00377.
\par\medskip

\item\textbf{[T.D.1]} : Le Myst\`ere du Monde Quantique, par Thibault Damour et Mathieu Burniat, Editions Dargaud, 2016. \par\medskip

\item \textbf{[T.D.2]} : Thomas Duke. Les prot\'eines motrices : la main-d'oeuvre de la cellule. (2000) \textit{Available on the internet :}
http://www.cnrs.fr/publications/\par
imagesdelaphysique/couv-PDF/ip2000/10a.pdf
\par\medskip

\item \textbf{[W.R]} : L'enseignement du Bouddha, de Walpola Rahula, Editions du seuil.
\par\medskip

\item \textbf{[W.Z.1]} : Probabilities from Entanglement, Born's Rule $p_k=
|\psi_k|^2$ from Envariance, de Wojceiech Hubert Zurek, Physical Review A71, 2005.\par
Available on the internet.
\par\medskip

\item\textbf{[W.Z.2]} : Wojceiech Hubert Zurek. Decoherence Einselection and the Existential Interpretation. 
arXiv:quant-ph/9805065. (2008) \par\medskip

\item\textbf{[W.Z.3]} : Wojceiech Hubert Zurek. 
Entanglement Symmetry, Amplitudes, and Probabilities: Inverting Born's Rule.
arXiv:1105.4810v1[quant-ph]. (2011) \par\medskip

\end{itemize}

\end{document}